\documentclass[12pt,a4paper]{report}

\usepackage{epsfig,amstext,fancyhdr,setspace,color,hyperref,url} 
\usepackage{caption,subcaption,amsmath,amssymb,enumerate,lscape,slashed}
\usepackage{dsfont,cancel,psfrag,array,textcomp,booktabs,shuffle}
\usepackage[sort&compress,numbers]{natbib}
\usepackage[utf8]{inputenc}

\DeclareFontFamily{OT1}{pzc}{}
\DeclareFontShape{OT1}{pzc}{m}{it}{<-> s * [1.350] pzcmi7t}{}
\DeclareMathAlphabet{\mathpzc}{OT1}{pzc}{m}{it}

\def\be{\begin{equation}}
\def\ee{\end{equation}}
\def\nn{\nonumber}
\def\nnl{\nonumber \\ }

\def\cT{\mathcal{T}}

\def\cA{\mathcal{A}}
\def\cN{\mathcal{N}}
\def\cO{\mathcal{O}}
\def\cD{\mathcal{D}}
\def\cQ{\mathcal{Q}}
\def\tf{\tilde{f}}
\def\eps{\epsilon}
\def\d{\mathrm{d}}

\def\tr{\operatorname*{tr}}

\def\trm{\text{tr}_-}
\def\trp{\text{tr}_+}
\def\trfive{\text{tr}_5}

\def\bra#1{\langle #1|}

\def\sqket#1{|#1]}
\def\braket#1{\langle #1 \rangle}

\def\SpDenom5{\braket{12}\braket{23}\braket{34}\braket{45}\braket{51}}

\def\eqn#1{eq.~\eqref{#1}}
\def\eqns#1#2{eqs.~\eqref{#1} and~\eqref{#2}}

\def\tab#1{Table~{\ref{#1}}}

\def\Tab#1{Table~{\ref{#1}}}

\def\sec#1{section~{\ref{#1}}}

\definecolor{myblue}{rgb}{0,0,0.7}

\def\usedelta#1{\includegraphics[scale=1.0,trim=0 8 0 0]{graphs/#1.pdf}}
\def\usegraph#1#2{\includegraphics[scale=1.0,trim=0 #1 0 0]{graphs/#2.pdf}}

\begin{document}
%

\pagestyle{empty}

\begin{center}
\LARGE
\vspace*{1cm}
{\bf A Tale of Two Loops: Simplifying All-Plus Yang-Mills Amplitudes}

\vspace{1cm}

\begin{figure}[ht]
\begin{center}
\epsfig{file=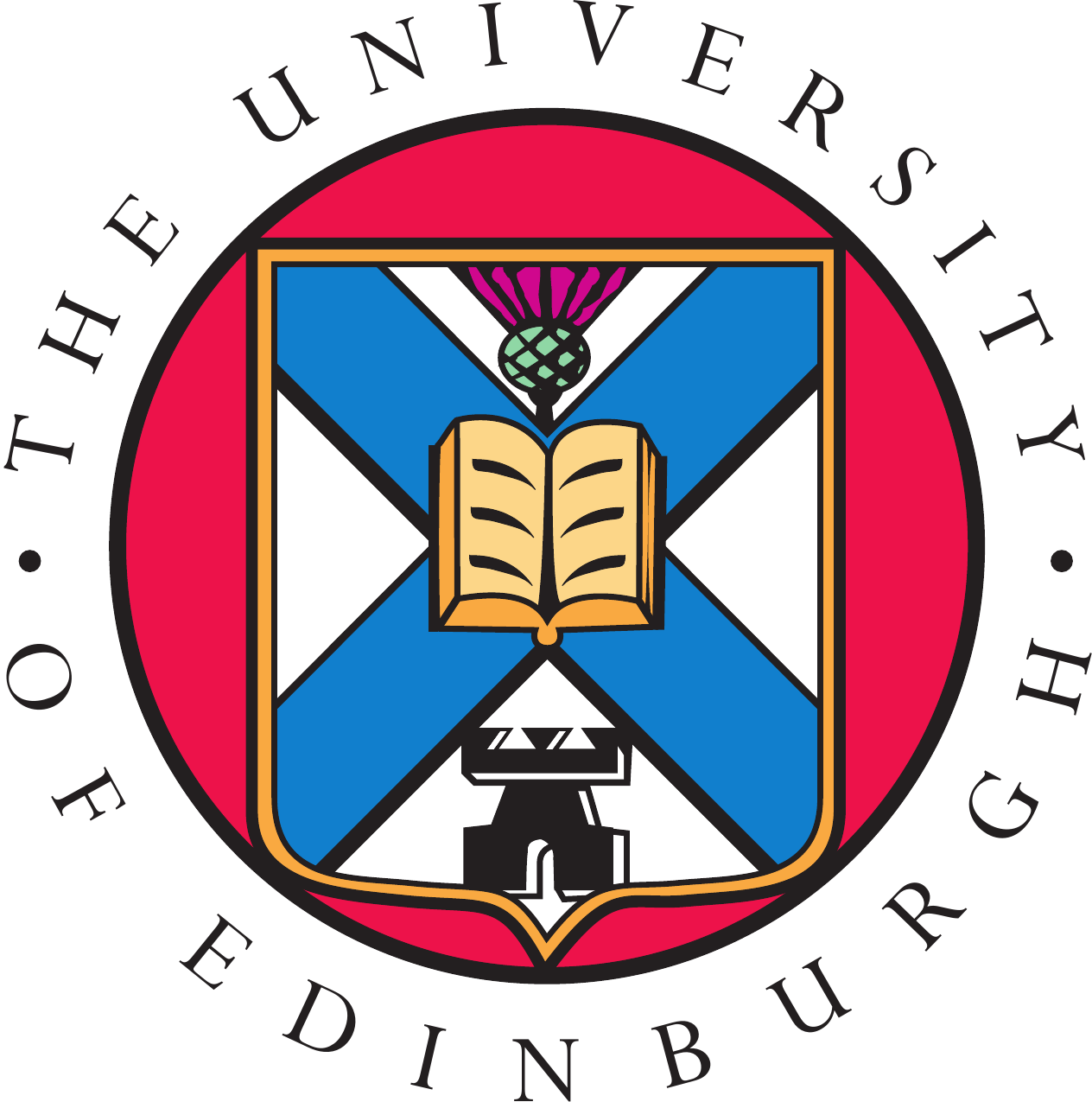,width=0.3\linewidth}
\end{center}
\end{figure}
\vspace{1cm}
\large
{\it David Gustav Mogull}

\vspace{7cm}
\large
Doctor of Philosophy\\
\vspace{.2cm}
University of Edinburgh\\
\vspace{.2cm}
July 2017
\end{center}




\pagestyle{fancy}
\fancyhead{} 
\renewcommand{\headheight}{28pt}
\cfoot{}
\fancyfoot[LE,RO]{\thepage}


\newpage
\clearpage\thispagestyle{empty}
\vspace*{8cm}
\begin{center}
  \Large {\em Till Helge Persson}
\end{center}

\ \newpage

\pagenumbering{roman}
\setcounter{page}{1}

\onehalfspacing

\addcontentsline{toc}{chapter}{Lay Summary}

\chapter*{Lay Summary}

Scattering amplitudes are a cornerstone of modern particle physics.  In a scattering process between fundamental particles, a set of incoming particles, carrying individual energies and momenta, becomes a new, outgoing set of particles.  In this context, loosely speaking, the scattering amplitude is the square root of the probability for such an interaction to occur.  Scattering amplitudes are a crucial ingredient for making predictions at collider experiments such as the Large Hadron Collider (LHC); the need to calculate them to ever-increasing accuracy is one of the main bottlenecks for making new predictions.

\begin{figure}[t]
\centering
\begin{subfigure}{0.25\textwidth}
  \centering
  \includegraphics[width=0.4\textwidth]{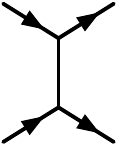}
  \caption[a]{Tree level}
\end{subfigure}
\begin{subfigure}{0.25\textwidth}
  \centering
  \includegraphics[width=0.47\textwidth]{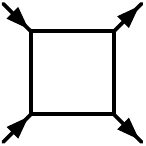}
  \caption[b]{One loop}
\end{subfigure}
\begin{subfigure}{0.25\textwidth}
  \centering
  \includegraphics[width=0.8\textwidth]{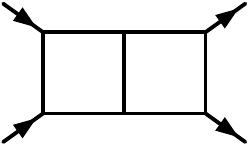}
  \caption[c]{Two loops}
\end{subfigure}
\caption{\small Example Feynman diagrams for a $2\to2$ scattering process, involving only cubic vertices, at different loop orders.\label{fig:loopexamples}}
\end{figure}

Scattering amplitudes are usually calculated using Feynman diagrams.  There are three steps involved: drawing all allowed diagrams, assigning algebraic expressions to them, and summing to obtain the amplitude.  Three example diagrams for a $2\to2$ scattering process, involving only cubic vertices and organised by loop order, are drawn in Figure~\ref{fig:loopexamples}.  Scattering processes can involve infinitely many diagrams, so categorisation in terms of loop order allows for a useful approximation: simply cap the number of internal loops allowed.  More complex diagrams, containing more loops, are associated with less favourable interaction paths.  By raising this cap on the loop order one can achieve better accuracies; however, the higher loop orders become increasingly difficult to calculate.

Part of the problem is that the number of diagrams involved grows factorially with the number of loops.  Also, each loop is associated with intermediate particles for which one should integrate over all possible values of their energies and momenta.  These integrals can be very challenging to perform analytically, and often give rise to infinities.  However, as the expressions one assigns to individual diagrams are not generally unique, there is usually freedom in how one distributes terms between the diagrams.  The hope is that this freedom can be exploited to reduce the number of non-vanishing diagrams while simplifying the job of integrating them.

In this thesis we explore various methods of simplifying Yang-Mills amplitudes up to two loops.  Yang-Mills theory describes interactions between the force carriers, such as photons, carriers of the electromagnetic force, and gluons, carriers of the strong nuclear force.  More specifically, we focus on the all-plus helicity sector, in which all incoming and outgoing particles must share the same polarisation.  These amplitudes are especially simple because, in a classical setting, these states do not scatter; it is only in the quantum regime where scattering occurs.  We see the amplitudes as a useful testing ground for the development of new techniques which, we believe, are generalisable to a wider class of amplitudes.


\addcontentsline{toc}{chapter}{Abstract}

\chapter*{Abstract}

Pure Yang-Mills amplitudes with all external gluons carrying positive helicity, known as all-plus amplitudes, have an especially simple structure.  The tree amplitudes vanish and, up to at least two loops, the loop-level amplitudes are related to those of $\cN=4$ super-Yang-Mills (SYM) theory.  This makes all-plus amplitudes a useful testing ground for new methods of simplifing more general classes of amplitudes.  In this thesis we consider three new approaches, focusing on the structure before integration.

We begin with the planar (leading-colour) sector.  A $D$-dimensional local-integrand presentation, based on four-dimensional local integrands developed for $\mathcal{N}=4$ SYM, is developed.  This allows us to compute the planar six-gluon, two-loop all-plus amplitude.  Its soft structure is understood before integration, and we also perform checks on collinear limits.

We then proceed to consider subleading-colour structures.  A multi-peripheral colour decomposition is used to find colour factors based on underlying tree-level amplitudes via generalised unitarity cuts.  This allows us to find the integrand of the full-colour, two-loop, five-gluon all-plus amplitude.  Tree-level BCJ relations, satisfied by amplitudes appearing in the cuts, allow us to deduce all the necessary non-planar information for the full-colour amplitude from known planar data.

Finally, we consider representations satisfying colour-kinematics duality.  We discuss obstacles to finding such numerators in the context of the same five-gluon amplitude at two loops.  The obstacles are overcome by adding loop momentum to our numerators to accommodate tension between the values of certain cuts and the symmetries of certain diagrams.  Control over the size of our ansatz is maintained by identifying a highly constraining, but desirable, symmetry property of our master numerator.


\addcontentsline{toc}{chapter}{Declaration}

\chapter*{Declaration}

I declare that this thesis was composed by myself, that the work contained herein is my own except where explicitly stated otherwise in the text, and that this work has not been submitted for any other degree or professional qualification except as specified.  The results of chapters \ref{ch:localintegrands}, \ref{ch:allplusfull} and \ref{ch:allplusbcj} were produced in collaboration and appear in the following publications:
\begin{description}
	\item [{\bf Chapter \ref{ch:localintegrands}}]\hfill\
	\begin{description}
		\item [\cite{Badger:2016ozq}]
		S.~Badger, G.~Mogull, T.~Peraro,
		\emph{Local Integrands for Two-Loop All-Plus Yang-Mills Amplitudes},
		JHEP {\bf 08} (2016) 063, [1606.02244]
		\item [\cite{Badger:2016egz}]
		S.~Badger, G.~Mogull, T.~Peraro,
		\emph{Local Integrands for Two-Loop QCD Amplitudes},
		PoS {\bf LL2016} (2016) 006, [1607.00311]
	\end{description}
	\item [{\bf Chapter \ref{ch:allplusfull}}]\hfill\
	\begin{description}
		\item [\cite{Badger:2015lda}]
		S.~Badger, G.~Mogull, A.~Ochirov, D.~O'Connell,
		\emph{A Complete Two-Loop, Five-Gluon Helicity Amplitude in Yang-Mills Theory},
		JHEP {\bf 10} (2015) 064, [1507.08797]
	\end{description}
	\item [{\bf Chapter \ref{ch:allplusbcj}}]\hfill\
	\begin{description}
		\item [\cite{Mogull:2015adi}]
		G.~Mogull, D.~O'Connell,
		\emph{Overcoming Obstacles to Colour-Kinematics Duality at Two Loops},
		JHEP {\bf 12} (2015) 135, [1511.06652]
	\end{description}
	
\end{description}

\vspace{20mm}

\hfill {\it D.G.~Mogull}

\hfill {\it July 2017}


\addcontentsline{toc}{chapter}{Acknowledgements}

\chapter*{Acknowledgements}

First, I'd like to offer my utmost thanks to Donal O'Connell, my supervisor.  Thank you for believing in me so early on.  Thank you for being so patient in our (sometimes very long!) meetings.  And thank you for showing me the path to postdoctoral research.  You have offered me a future in theoretical physics.

Next, I'd like to thank my collaborators in Edinburgh for all their brilliant work: Simon Badger, Alexander Ochirov and Tiziano Peraro.  A special thanks also goes to Einan Gardi for his help and advice over the years.

My three months in Uppsala last year were a phenomenal experience; I am hugely indebted to Henrik Johansson for offering it to me.  Also to his collaborators, Marco Chiodaroli and Oluf Engelund, and his student, Gregor K\"{a}lin.  I look forward to continued collaboration when I join you all later this year!  And a big thank you to my family in Sweden: Erik, Klara, Malin, Mats, Olle, Samuel and Ulla.  \emph{Tusen tack f\"{o}r allt!}

Edinburgh has been a truly welcoming environment; I've had so many good friends that it seems silly to mention you all, but I'll do it anyway!  James Cockburn, Susanne Ehret, James Gratrex, Mark Harley, Joshua Hellier, Xanthe Hoad, Stephen Kay, Ava Khamseh, Siobh\'{a}n MacInnes, Isobel Nicholson, Chay Paterson, Laura Sawiak, Jennifer Wadsworth, Andries Waelkens.  I'll miss all of you.

The last four years would not have been possible without the core team that is the Mogull family: Marc, Kerstin, Alex and Philip.  Words cannot express how important you are to me, so I won't try.  But I will say: \emph{semper conor}.

\singlespacing



\pagestyle{fancy}
\fancyhead{} 
\renewcommand{\headheight}{28pt}
\cfoot{}
\fancyfoot[LE,RO]{\thepage}

\renewcommand{\chaptermark}[1]{\markboth{\chaptername \ \thechapter.\ #1}{}}
\renewcommand{\sectionmark}[1]{\markright{\thesection.\ #1} {}}

\fancyhead[LE]{\small \slshape \leftmark}      
\fancyhead[RO]{\small \slshape \rightmark}     
\renewcommand{\headrulewidth}{0.3pt}    


\addcontentsline{toc}{chapter}{Contents}
\tableofcontents






\clearpage 


\cfoot{}
\fancyfoot[LE,RO]{\thepage}

\onehalfspacing
\pagenumbering{arabic}
\setcounter{page}{1}


\chapter{Introduction}

\begin{figure}[t]
\centering
\begin{subfigure}{0.25\textwidth}
  \centering
  \includegraphics[width=0.8\textwidth]{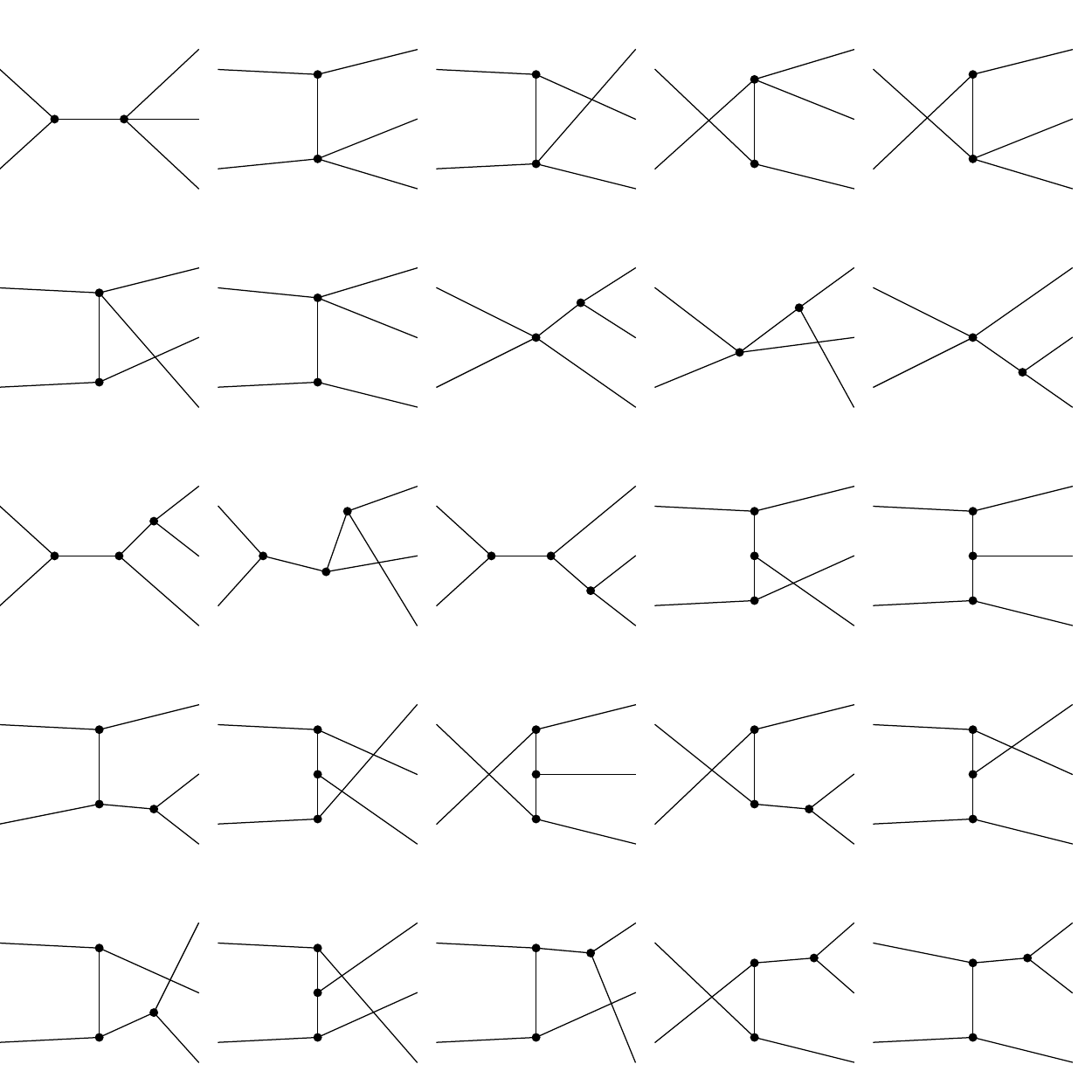}
  \caption[LO]{LO}
\end{subfigure}
\begin{subfigure}{0.25\textwidth}
  \centering
  \includegraphics[width=0.8\textwidth]{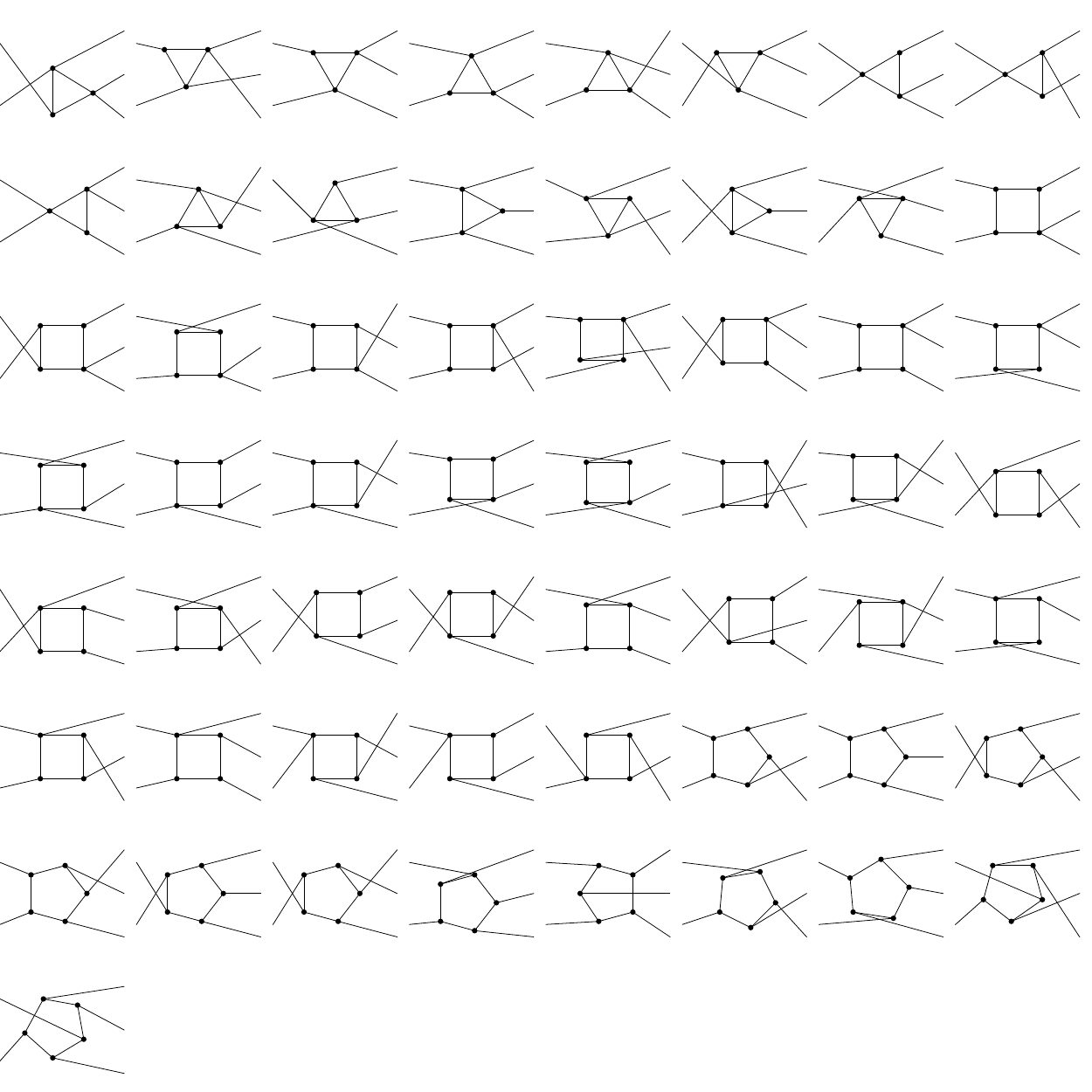}
  \caption[NLO]{NLO}
\end{subfigure}
\begin{subfigure}{0.25\textwidth}
  \centering
  \includegraphics[width=0.8\textwidth]{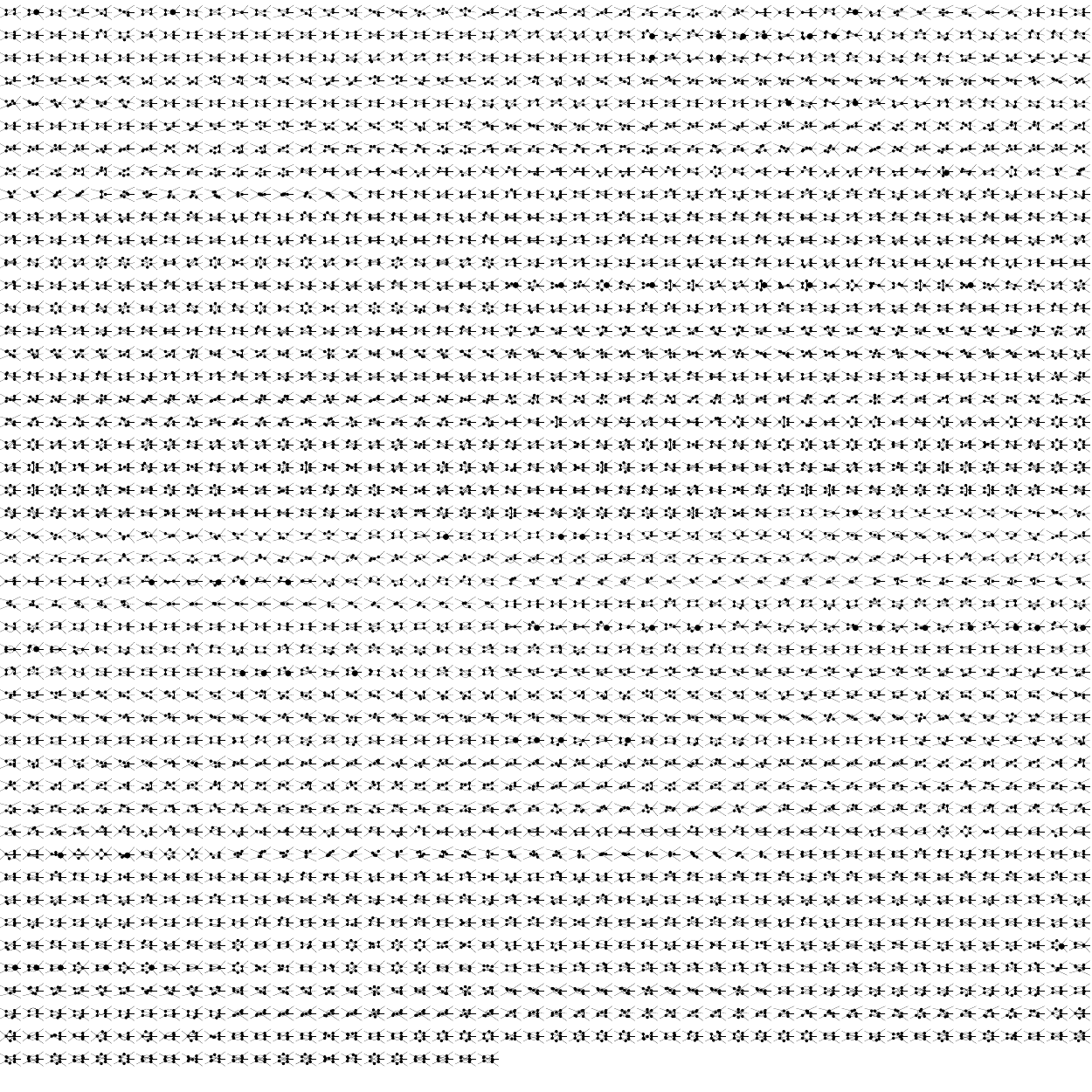}
  \caption[NNLO]{NNLO}
\end{subfigure}
\caption{\small Three-gluon jet production events.  At leading order (LO) there are 25 diagrams, at next-to-leading order (NLO) there are 57 diagrams and at next-to-next-to-leading order (NNLO) there are 2230. \label{fig:jetprod}}
\end{figure}

What is the best way of writing two-loop scattering amplitudes before integration?  In a traditional quantum-field-theory calculation, amplitudes are expressed as sums of Feynman diagrams, expressions for which are deducible from a set of Feynman rules.  This intuitive approach, while naturally lending itself to algorithmic implementation in a wide variety of cases, has severe drawbacks.  Individual Feynman diagrams are gauge-dependent objects, so their expressions can easily grow to be large and unmeaningful.  Meanwhile, the number of diagrams involved grows factorially with the number of scattered objects and number of loops.  For instance, in Figure~\ref{fig:jetprod} we see how for just $2\to3$ gluon scattering the number of two-loop diagrams is very large indeed.  Is there a better way?

Our motivation for asking this question is clear: amplitudes are vital for making precision predictions at collider experiments.  The problem of calculating one-loop amplitudes has now largely been solved, so two-loop amplitudes in particular represent the next big hurdle.\footnote{We refer the reader to the Les Houches working group report and references therein for a review of known and desired phenomenological predictions~\cite{Badger:2016bpw}.}  Also, the experimental data now being collected at Run II of the LHC allows the study of many observables with percent-level uncertainties. This level of precision represents a serious challenge for current perturbative techniques where a minimum of NNLO precision is desirable (QCD cross sections at NNLO carry two more powers of the strong coupling $\alpha_s\approx0.119$ than at LO).  This is particularly true for higher-multiplicity final states where the current bottleneck lies in the unknown two-loop matrix elements.

When combined with the technology of integration-by-parts (IBP) identities~\cite{Chetyrkin:1981qh,Tkachov:1981wb}, traditional Feynman diagram-based approaches have been quite successful for two-loop calculations of $2\to2$ scattering processes~\cite{Anastasiou:2000ue,Anastasiou:2000kg,Anastasiou:2001sv,Glover:2001af,Garland:2002ak,Garland:2001tf,Gehrmann:2011aa,Gehrmann:2015ora,vonManteuffel:2015msa,Caola:2015ila,Caola:2014iua}.  In these processes the number of diagrams involved is manageable; the main focus has been in the evaluation of the resulting master integrals.  But for higher multiplicities, the rapid growth in the complexity of the Feynman diagram representation motivates the development of alternative approaches for computing amplitudes.

One such alternative approach is generalised unitarity, of which many excellent reviews are now available~\cite{Bern:1994zx,Bern:1994cg,Carrasco:2011hw,Bern:2011qt,Ita:2011hi,Britto:2010xq}.  The idea is to build the integrands of loop-level amplitudes from knowledge of the underlying tree amplitudes, which are of course gauge-independent.  By working with gauge-independent objects throughout, one overcomes to a large extent the gauge-dependent messiness of Feynman diagrams.  A number of two-loop $2\to2$ scattering amplitudes have been computed this way~\cite{Bern:2000dn,Bern:2001df,Bern:2002zk,Bern:2002tk,Bern:2003ck}.

Generalised unitarity is not a perfect solution though: it only provides partial information about the amplitudes one may wish to calculate.  To reconstruct full amplitudes it is usually necessary to fit a basis of integrals, carrying unknown coefficients, to data provided by unitarity cuts (IBP identities can then be used to help evaluate these integrals).  When scattering coloured particles a basis of colour factors is also required.  The ability to make different valid choices of these bases means that the resulting coefficients can be large, unwieldy functions of external kinematics containing unphysical poles.  The gauge-dependent arbitrariness of Feynman diagrams seems to have re-emerged in another form, hampering efforts to go beyond $2\to2$ scattering in two-loop calculations.

Improved phenomenological predictions are not the only motivation for studying scattering amplitudes.  In recent years it has become clear that scattering amplitudes are fascinating mathematical objects in their own right, and worthy of independent study; several excellent reviews now exist~\cite{Roiban:2010kk,Ellis:2011cr, Bern:2011qt, Carrasco:2011hw, Dixon:2013uaa, Elvang:2013cua, Carrasco:2015iwa}.  In this domain more exotic ways of calculating them have led to remarkably compact expressions at higher points.  For instance, the tree-level Parke-Taylor formula for scattering of all-but-two positive-helicity gluons~\cite{Parke:1986gb},
\begin{align}
A^{(0)}(1^+,2^+,\cdots,i^-,\cdots,j^-,\cdots,n^+)=
i\frac{\braket{ij}^4}{\braket{12}\braket{23}\cdots\braket{n1}},
\end{align}
is a strikingly compact example of an $n$-point amplitude (our spinor-helicity conventions are defined below).

At loop level, much of this recent progress towards understanding the mathematical aspects of amplitudes has focused on supersymmetric theories, especially $\cN=4$ maximally-supersymmetric Yang Mills (SYM).  The focus has also often been on the planar sector, which arises in the limit of a large number of colours, $N_c\to\infty$.  The $n$-point planar all-loop integrand for $\cN=4$ scattering amplitudes is an excellent example: at two loops this takes the form~\cite{ArkaniHamed:2010kv,ArkaniHamed:2010gh}
\vspace{-.15in}
\begin{align}
A^{(2),[\cN=4]}(1,2,\cdots,n)=
\frac{1}{2}\sum_{i<j<k<l<i}\includegraphics[scale=0.3,trim=0 90 0 0]{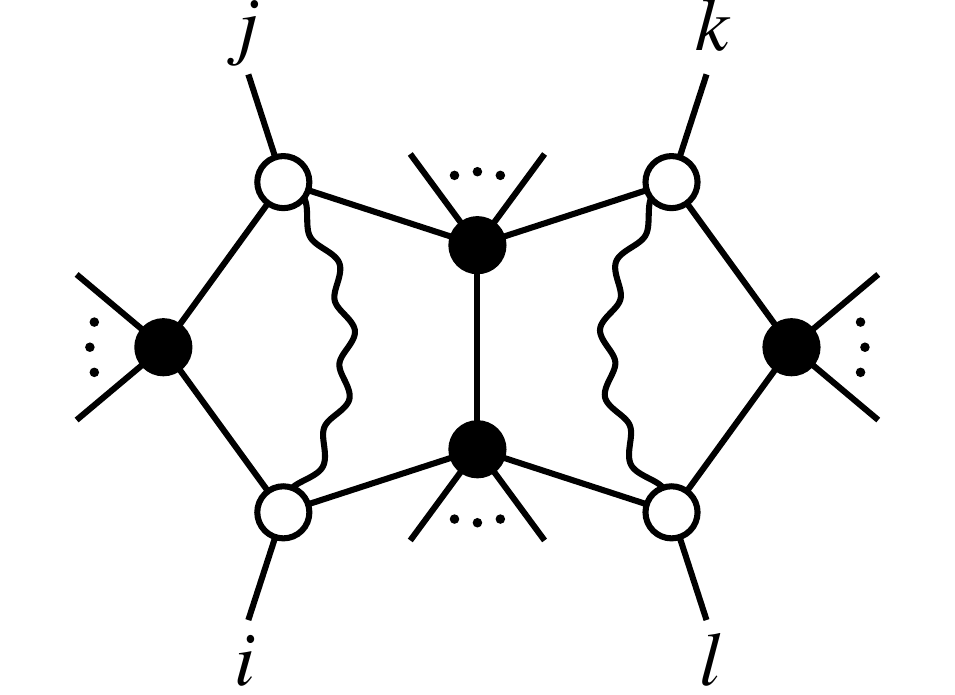},
\end{align}
which uses local integrands --- these will be discussed in chapter~\ref{ch:localintegrands}.  From a phenomenological standpoint $\cN=4$ SYM may be considered a toy model: an interesting mathematical construct that shares some, but not all, of the aspects of physical scattering processes.  Nevertheless, one naturally questions to what extent these new ideas may be generalisable to non-supersymmetric and non-planar theories.

In this thesis we will explore different ways of organising higher-point (at least five-point) two-loop integrands of pure Yang-Mills amplitudes for which all external gluons have positive helicity (all-plus).  The physical data for the integrands will come from generalised unitarity cuts; arrangements of both the kinematic and colour structure will be considered.  While all-plus amplitudes are involved in QCD processes, they only become relevant at N$^3$LO (a consequence of the tree amplitudes vanishing, which we will demonstrate in chapter~\ref{ch:review}).  In this sense we see the all-plus sector as its own toy model, but one which brings us much closer to QCD predictions.  Lastly, and most importantly, all-plus integrands are related to those of $\cN=4$ SYM up to at least two loops~\cite{Bern:1996ja,Badger:2013gxa}.  This will enable us to import techniques which, until now, have only found use in supersymmetric theories.

We will begin with a review of all-plus amplitudes in chapter~\ref{ch:review}.  In our first new presentation, to be discussed in chapter~\ref{ch:localintegrands}, the focus will be kinematic structure: we will consider up to six-gluon planar amplitudes.  Our starting point will be the previously-mentioned supersymmetric all-loop integrand; we will show how it can be translated to a Dirac-trace-based language applicable in dimensions $D\neq4$ (in contrast with the necessarily four-dimensional momentum twistors used in the original all-loop integrand).  This will give us a (partial) basis of integrals onto which we will fit all-plus amplitudes.  The presentation exposes infrared singularities and dependence on physical poles.  It is also remarkably compact.

In chapter~\ref{ch:allplusfull} our focus will shift to the colour structure.  We will develop a method for understanding loop-level colour structures based on a Del Duca, Dixon and Maltoni's (DDM's) colour decomposition~\cite{DelDuca:1999rs} of the tree amplitudes underlying the colour-dressed cuts.  The full-colour information can be recycled from leading-colour alone using tree-level amplitude relations found by Bern, Carrasco and Johansson (BCJ)~\cite{Bern:2008qj}.  It is therefore unnecessary to calculate nonplanar unitarity cuts.  Using this method we will determine an integrand for the (previously unknown) full-colour five-point two-loop all-plus amplitude.

Finally, in chapter~\ref{ch:allplusbcj} we will find a presentation of the five-gluon amplitude satisfying BCJ's colour-kinematics duality~\cite{Bern:2008qj}.  So-called colour-dual presentations place colour and kinematics on an equal footing, and are of particular interest for aiding study of (supersymmetric) gravity amplitudes.  While this five-point presentation is less compact than the other two presentations, it does have some interesting new properties, and tightens the link to $\cN=4$ SYM.  In chapter~\ref{ch:conclusions} we will conclude.

\section{Notation and conventions}

As a summary for the reader the following section lists most of the conventions that we will adopt throughout this thesis.

\subsection{Dimensionality}\label{sec:introdim}

We will work in dimensional regularisation with different dimensions in play simultaneously:
\begin{description}
\item[$D$] The number of dimensions in dimensional regularisation, $D=4-2\epsilon$.
\item[$D_s$] The spin dimension of internal gluons.
\item[$\cD$] The embedding dimension for $D$-dimensional momenta; usually we will choose $\cD=6$.
\end{description}
One can obtain results in the 't Hooft Veltman (tHV) scheme by setting $D_s=4-2\eps$ and the the four-dimensional helicity (FDH) scheme by setting $D_s=4$ \cite{Bern:2002zk}.

\subsection{Spinor and Lorentz products}

In any number of dimensions we will use the mostly-minus metric $\eta^{\mu\nu}=\text{diag}(+,-,-,-,\ldots)$.  External momenta, which we will denote as $p_i$ and always take outgoing, will always be taken in four dimensions; we will use the shorthands
\begin{align}
p_{ij\cdots k}=p_i+p_j+\cdots+p_k,&&
s_{ij\cdots k}=p_{ij\cdots k}^2.
\end{align}
Spinor products will be constructed from holomorphic ($\lambda_a$) and anti-holomorphic ($\tilde{\lambda}_{\dot{a}}$) two-component Weyl spinors, such that $\braket{ij}=\lambda_{i,a}\lambda_j^a$ and $[ij]=\tilde{\lambda}_i^{\dot{a}}\tilde{\lambda}_{j,\dot{a}}$.  At four points we will also use the usual Mandelstam invariants $s=(p_1+p_2)^2$, $t=(p_2+p_3)^2$ and $u=(p_1+p_3)^2$, and the permutation-invariant prefactor
\begin{align}\label{eq:cT}
\cT=\frac{[12][34]}{\braket{12}\braket{34}}.
\end{align}
Dirac traces are defined as
\begin{align}
\text{tr}_\pm(ij\cdots k)
=\frac{1}{2}\tr((1\pm\gamma_5)\slashed{p}_i\slashed{p}_j\cdots\slashed{p}_k),&&
\trfive(ij\cdots k)=\tr(\gamma_5\slashed{p}_i\slashed{p}_j\cdots\slashed{p}_k),
\end{align}
where $\trfive(ijkl)=4i\epsilon_{\mu\nu\rho\sigma}p_i^\mu p_j^\nu p_k^\rho p_l^\sigma$ and, for instance, $\trp(ijkl)=[ij]\braket{jk}[kl]\braket{li}$.  We will also occasionally abbreviate
\begin{align}
\trfive=\trfive(1234).
\end{align}
Finally, the spurious directions are
\begin{align}\label{omega}
   \omega_{ijk}^\mu = \frac{\braket{jk}[ki]}{s_{ij}}
                      \frac{\bra{i}\gamma^\mu\sqket{j}}{2}
                    - \frac{\braket{ik}[kj]}{s_{ij}}
                      \frac{\bra{j}\gamma^\mu\sqket{i}}{2},
\end{align}
which satisfy $p_i\cdot\omega_{ijk}=p_j\cdot\omega_{ijk}=p_k\cdot\omega_{ijk}=0$.

\subsection{Colour algebra}\label{sec:colourconventions}

When using the SU($N_c$) colour algebra we will normalise the fundamental-representation generators ${(T^a)_i}^{\bar\jmath}$ as $\tr(T^aT^b)=\delta^{ab}$.  The indices $a$, $i$ and $\bar\jmath$ belong to the adjoint, fundamental and anti-fundamental representations of SU($N_c$) respectively.  We also have the usual Fierz identity
\begin{align}\label{eq:fierz}
{(T^a)_{i_1}}^{\bar\jmath_1}{(T^a)_{i_2}}^{\bar\jmath_2}=
\delta_{i_1}^{\bar\jmath_2}\delta_{i_2}^{\bar\jmath_1}-
\frac{1}{N_c}\delta_{i_1}^{\bar\jmath_1}\delta_{i_2}^{\bar\jmath_2}.
\end{align}
Finally, we introduce the adjoint vertices $\tf^{abc}=\tr([T^a,T^b]T^c)=i\sqrt{2}f^{abc}$, where $f^{abc}$ are the structure constants more commonly found in the literature.  The colour algebra is then $[T^a,T^b]=\tf^{abc}\,T^c$.\footnote{Good reviews of the colour algebra and its practical applications may be found in refs.~\cite{Dixon:1996wi,Dixon:2011xs}.}

When writing colour factors that depend only on the adjoint vertices $\tf^{abc}$ we will often find it convenient to use a diagrammatic notation.  For instance,
\begin{align}
c\bigg(\usegraph{10}{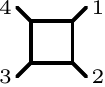}\bigg)=\tf^{b_1a_1b_2}\tf^{b_2a_2b_3}\tf^{b_3a_3b_4}\tf^{b_4a_4b_1},
\end{align}
where the external colour indices corresponding to outgoing partons $i$ are $a_i$.

\subsection{Loop momenta}

$D$-dimensional loop momenta $\ell_i$ will sometimes be separated into their four-dimensional and $(-2\epsilon)$-dimensional parts $\ell_i=\bar{\ell_i}+\mu_i$.  Rotational invariance in the extra dimensions forces $\mu_i$
to appear in the combinations $\mu_{ij}=-\mu_i\cdot\mu_j$.  For one-loop integrals we will write $\mu^2=-\mu\cdot\mu$.  In chapter~\ref{ch:localintegrands} we will often include these loop momenta in Dirac traces, necessitating the use of a $D$-dimensional Clifford algebra.  Our approach is formally outlined in appendix~\ref{app:traces} but in practice we will overcome such ambiguities by decomposing into four-dimensional traces:
\begin{align}
\text{tr}_\pm(i_1\cdots i_k\ell_x\ell_yi_{k+1}\cdots i_n)
=\text{tr}_\pm(i_1\cdots i_k\bar{\ell}_x\bar{\ell}_yi_{k+1}\cdots i_n)
-\mu_{xy}\text{tr}_\pm(i_1\cdots i_n).
\end{align}
This decomposition will allow us to evaluate $D$-dimensional traces using
\begin{align}\label{eq:trstacking}
\text{tr}_\pm(i_1\cdots i_k\ell_x\ell_yi_{k+1}\cdots i_n)
=\frac{\text{tr}_\pm(i_1\cdots i_n)}{s_{i_k,i_{k+1}}}\text{tr}_\pm(i_k\ell_x\ell_yi_{k+1}),
\end{align}
where $k$ should in this case be odd.

\subsection{Integrals}

For a given $L$-loop topology $T$, defined by a set of massless propagators $\{\cQ_\alpha\}$, the $L$-loop integration operator will be
\begin{align}
I^D_T\left[\mathcal{P}(p_i,\ell_i,\mu_{ij})\right]\equiv
\int\!\left(\prod_{k=1}^L\frac{\d^D\ell_k}{(2\pi)^{D}}\right)
\frac{\mathcal{P}(p_i,\ell_i,\mu_{ij})}
{\prod_{\alpha\in T}\cQ_{\alpha}(p_i,\ell_i)},
\end{align}
where we will generally identify topologies $T$ by explicitly drawing them (as we do for colour factors).  For instance, the box integral in our notation is
\begin{align}
I^D\bigg(\usegraph{10}{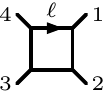}\bigg)[\mathcal{P}]=
\int\!\frac{\d^D\ell}{(2\pi)^D}\frac{\mathcal{P}}{\ell^2(\ell-p_1)^2(\ell-p_{12})^2(\ell+p_4)^2}.
\end{align}
When the integration operator acts directly on the numerator of a diagram, it should be implied that the relevant propagators to integrate with are those of the diagram in question:
\begin{align}
 I^D[\Delta_T]\equiv
\int\!\left(\prod_{k=1}^L\frac{\d^D\ell_k}{(2\pi)^{D}}\right)
 \frac{\Delta_T(p_i,\ell_i,\mu_{ij})}{\prod_{\alpha\in T}\cQ_{\alpha}(p_i,\ell_i)}.
\label{IntMeasure}
\end{align}
Finally,
\begin{align}\label{eq:cGamma}
c_\Gamma=\frac{\Gamma(1+\eps)\Gamma^2(1-\eps)}{(4\pi)^{2-\eps}\Gamma(1-2\eps)}
\end{align}
is the standard loop prefactor in dimensional regularisation.

\chapter{Review: all-plus helicity amplitudes}\label{ch:review}

\section{Introduction}

In this chapter we review the necessary aspects of all-plus helicity amplitudes up to two loops.  After introducing the trace-based colour decomposition we will work our way up through the loop orders, starting from tree level.  Then we will review infrared divergences.  As we shall see, the lower-loop structure has a strong influence on the higher-loop structure.

\section{Colour structure}

Often in this thesis we will make use of the trace-based colour decomposition.  An $L$-loop pure Yang-Mills amplitude (in any helicity configuration) may be decomposed as a power series in $N_c$, the number of colours~\cite{Dixon:2011xs,Dixon:1996wi,Dixon:2013uaa}:
\begin{align}\label{eq:colourdecomp}
\cA^{(L)}_n
&=g^{n-2+2L}N_c^L\sum_{\sigma\in S_n/Z_n}\tr(T^{a_{\sigma(1)}}T^{a_{\sigma(2)}}\cdots T^{a_{\sigma(n)}})
A^{(L)}(\sigma(1),\sigma(2),\cdots,\sigma(n))\nnl
&\qquad+\cO(N_c^{L-1}).
\end{align}
$S_n$ is the $n$-object permutation group and $Z_n$ is the subset of cyclic permutations, so the set $S_n/Z_n$ gives all cyclically-distinct orderings; $g$ is the strong coupling.  The objects $A^{(L)}$, coefficients of the leading single-trace terms, are known as colour-ordered (or planar) amplitudes.  The subleading terms, which appear only at loop level, carry a more complex multi-trace colour structure; their precise form will not concern us.  When studying subleading-colour terms at higher-loop order we will use alternative colour decompositions.

To achieve a trace-based decomposition from, say, an ordinary Feynman-diagram-based decomposition one should re-express all colour factors in terms of fundamental generators $T^a$ using $\tf^{abc}=\tr([T^a,T^b]T^c)$.  The Fierz identity~(\ref{eq:fierz}) should then be used to assemble these into the ordered traces given above.  Alternatively, one can write down the planar amplitudes directly using a set of colour-ordered Feynman rules~\cite{Dixon:2011xs,Dixon:2013uaa}.  For example, in Lorenz-Feynman gauge the three- and four-gluon vertices are~\cite{Dixon:1996wi}
\begin{subequations}\label{eq:orderedrules}
\begin{align}
V_{ggg}^{\mu_1\mu_2\mu_3}&=
\frac{i}{\sqrt{2}}\left(\eta^{\mu_1\mu_2}(p_1-p_2)^{\mu_3}+\eta^{\mu_2\mu_3}(p_2-p_3)^{\mu_1}+\eta^{\mu_3\mu_1}(p_3-p_1)^{\mu_2}\right),\\
V_{gggg}^{\mu_1\mu_2\mu_3\mu_4}&=
i\eta^{\mu_1\mu_3}\eta^{\mu_2\mu_4}-\frac{i}{2}(\eta^{\mu_1\mu_2}\eta^{\mu_3\mu_4}+\eta^{\mu_4\mu_1}\eta^{\mu_2\mu_3}).\label{eq:4ptvertex}
\end{align}
\end{subequations}
In this case one should only include planar diagrams for which the ordering of external legs matches that of the $T^a$ matrices in the corresponding trace.

We specialise to the all-plus sector by demanding that all external states have positive helicity.  These amplitudes satisfy
\begin{subequations}\label{eq:colourorderedsymmetries}
\begin{align}
A^{(L)}(1^+,2^+,\ldots,n^+)
&=A^{(L)}(2^+,3^+,\ldots,n^+,1^+)\label{eq:cyclicsymmetry}\\
&=(-1)^nA^{(L)}(n^+,\ldots,2^+,1^+)\label{eq:reflectionsymmetry}
\end{align}
\end{subequations}
at any loop order $L$.  The first of these follows from cyclic symmetry of the traces; the second can be deduced from antisymmetry of the Feynman vertices~(\ref{eq:orderedrules}).  This makes all-plus amplitudes much simpler than alternative helicity configurations: having different helicities present would break these symmetries.

\section{Tree-level amplitudes}\label{sec:treereview}

All-plus tree amplitudes vanish for any number of external legs.  We can see this by considering the colour-ordered amplitudes: write them in terms of colour-ordered diagrams as explained above.  Schematically, such a decomposition always takes the form~\cite{Elvang:2013cua}
\begin{align}
A^{(0)}(1^+,2^+,\ldots,n^+)\sim
\sum_{\text{cubic diagrams }\Gamma_i}\frac{\sum\,(\prod\eps^+_i\cdot\eps^+_j)(\prod\eps^+_i\cdot p_j)(\prod p_i\cdot p_j)}{\prod_{\alpha\in\Gamma_i}\cQ_\alpha(p_j)}.
\end{align}
Contributions from four-point vertices~(\ref{eq:4ptvertex}) are eliminated by inserting additional propagators.  The numerators have dimensionality $(\text{mass})^{n-2}$; we have specialised to the all-plus sector by using only positive-helicity polarisation vectors $\eps^+_\mu$.

When choosing specific helicities of external momenta it is convenient for us to adopt the spinor-helicity notation for external momenta.\footnote{Several excellent reviews of spinor helicity now exist - see, for instance, refs.~\cite{Srednicki:2007qs,Elvang:2013cua}.}  The positive- and negative-helicity polarisation vectors may be written in this notation as~\cite{Dixon:1996wi}
\begin{align}
\eps^+_{i,\mu}=\frac{\left<q_i|\gamma_\mu|p_i\right]}{\sqrt{2}\braket{q_ip_i}},&&
\eps^-_{i,\mu}=-\frac{\left[q_i|\gamma_\mu|p_i\right>}{\sqrt{2}[q_ip_i]},
\end{align}
where $q_i$ are auxiliary massless vectors whose values are undetermined; their arbitrariness reflects the freedom of on-shell gauge transformations.  By choosing all $q_i$ to lie in the same direction we can ensure that $\eps^+_i\cdot\eps^+_j=0$ for all pairs of external momenta (because $\eps_i^+\cdot\eps_j^+\propto\braket{q_iq_j}$).  Therefore, the numerators may only contain objects of the form  $\eps^+_i\cdot p_j$ or $p_i\cdot p_j$.  But each term must absorb the Lorentz indices of all $n$ polarisation vectors, which necessarily raises the mass dimension of the numerator above $n-2$.

So the colour-ordered amplitudes must vanish and, by eq.~(\ref{eq:colourdecomp}), this trivially implies the vanishing of $\cA^{(0)}$ for any number of external legs.  A similar argument works for one-minus amplitudes~\cite{Elvang:2013cua} --- they too are zero at tree level.  In this case we ensure that $\eps_1^-\cdot\eps_i^+=0$ for $i\geq2$ by selecting all $q_i$ to lie in the same direction as $p_1$, where without loss of generality we have selected $p_1$ to carry negative helicity.

\section{One-loop amplitudes}\label{sec:1lreview}

Loop-level all-plus amplitudes are generally nonvanishing.  We begin by considering the unintegrated one-loop colour-ordered amplitudes.  Starting with four external legs~\cite{Bern:1995db},
\begin{subequations}\label{eq:1lbern}
\begin{align}\label{eq:1l4gbern}
A^{(1)}(1^+,2^+,3^+,4^+)=\cT(D_s-2)I^D\bigg(\usegraph{10}{boxi}\bigg)[\mu^4],
\end{align}
where $\cT$ was given in eq.~(\ref{eq:cT}): it carries the helicity scaling of the amplitude.  We have also introduced a dependence on the spin dimension $D_s$ to account for different possible dimensional regularisation schemes (FDH or tHV, see section \ref{sec:introdim} for details).  At five points the (parity-even part) can be written as~\cite{Bern:1996ja,Bern:1993qk,Mahlon:1993fe}
\begin{align}\label{eq:1l5gbern}
&A^{(1)}(1^+,2^+,3^+,4^+,5^+)=
\frac{(D_s-2)}{\braket{12}\braket{23}\braket{34}\braket{45}\braket{51}}\times\\
&\qquad\left(\trfive(1234)I^D\bigg(\usegraph{12}{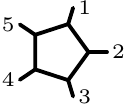}\!\bigg)[\mu^6]-
\frac{1}{2}\sum_{i=1}^5s_{i,i+1}s_{i+1,i+2}I^D\bigg(\,\usegraph{10}{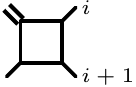}\bigg)[\mu^4]\right),\nn
\end{align}
and at six points
\begin{align}\label{eq:1l6gbern}
&A^{(1)}(1^+,2^+,3^+,4^+,5^+,6^+)=
\frac{(D_s-2)}{\braket{12}\braket{23}\braket{34}\braket{45}\braket{56}\braket{61}}\times\\
&\qquad\Bigg(\frac{1}{2}\tr(123456)I^D\bigg(\usegraph{11}{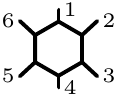}\bigg)[\mu^6]-
\frac{1}{2}\sum_{i=1}^6\trfive(i,i\!+\!1,i\!+\!2,i\!+\!3)I^D\bigg(\usegraph{13}{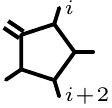}\bigg)[\mu^6]\nn\\
&\qquad\qquad-\frac{1}{4}\sum_{1\leq i_1\leq i_2\leq6}\tr(\slashed{p}_{i_1}\slashed{p}_{i_1+1,\ldots,i_2-1}\slashed{p}_{i_2}\slashed{p}_{i_2+1,\ldots,i_1-1})
I^D\bigg(\usegraph{10}{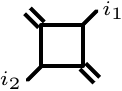}\bigg)[\mu^4]\Bigg).\nn
\end{align}
\end{subequations}
Doubled external lines indicate that multiple external momenta are clustered together.  All three of these expressions satisfy the cyclic~(\ref{eq:cyclicsymmetry}) and reflection~(\ref{eq:reflectionsymmetry}) symmetries discussed earlier.

\subsection{Analytic structure}

\begin{figure}[t]
\centering
\includegraphics[width=.5\textwidth]{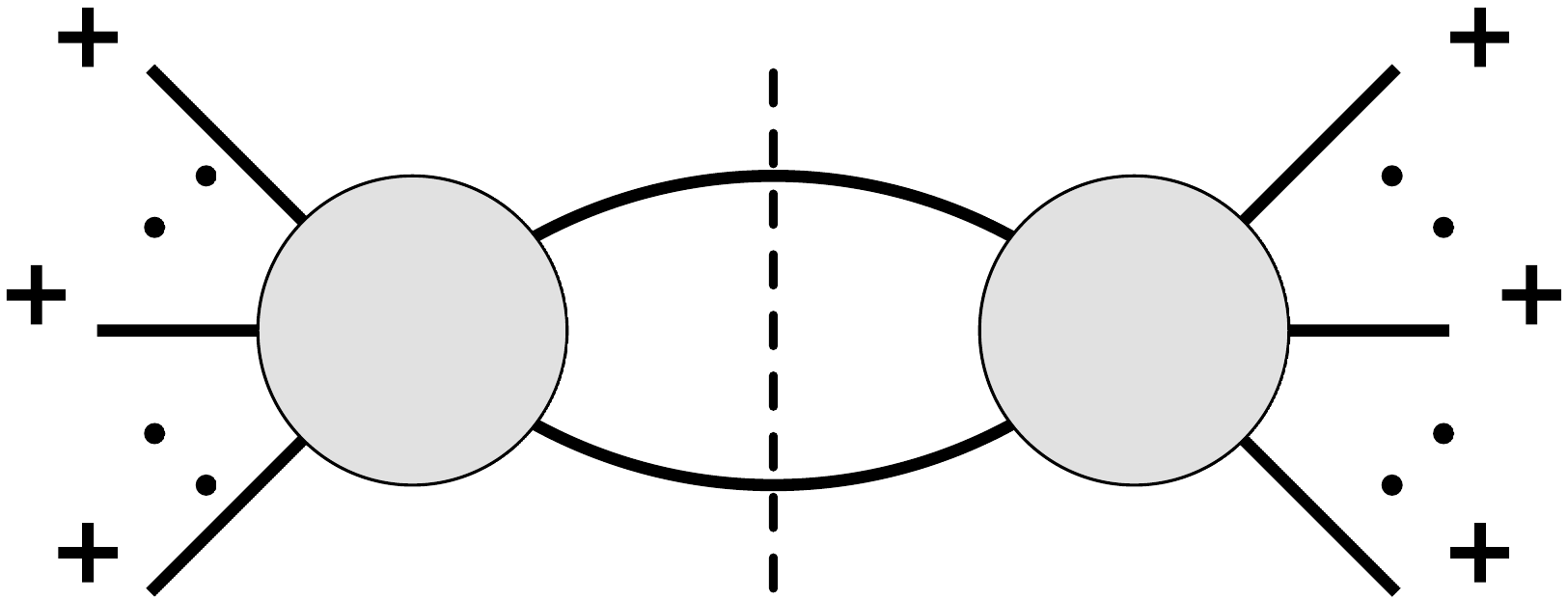}
\caption{\small A one-loop unitarity cut with all external helicities positive.  The disks represent tree-level amplitudes; no choice of helicities for the cut lines results in non-vanishing trees so this cut always vanishes in four dimensions.\label{fig:oneloopcut}}
\end{figure}

We gain further insight into these expressions by considering their origin: generalised unitarity cuts.\footnote{We will discuss generalised unitarity more carefully in section~\ref{sec:unitarity}; see also the reviews \cite{Carrasco:2011hw,Bern:2011qt,Ita:2011hi,Britto:2010xq}.}  The integrands are chosen to match the relevant one-loop cuts; these are of the schematic form displayed in Figure~\ref{fig:oneloopcut}~\cite{Bern:1994zx}.  Vanishing of the four-dimensional all-plus and one-minus tree amplitudes implies vanishing of the cuts so, naively, one would also expect the one-loop all-plus amplitudes to vanish.  However, one-loop amplitudes can also contain rational terms~\cite{vanNeerven:1983vr,Bern:1992em,Bern:1993kr}.  These are undetectable by four-dimensional unitarity; they can instead be obtained through unitarity cuts involving the non-vanishing $D$-dimensional trees~\cite{Bern:1995db,Badger:2008cm}.  This explains why the integrands all carry powers of $\mu^2$: the limit $D\to4$ involves sending $\mu^2\to0$, so the integrands vanish as expected.

The resulting one-loop integrals are conveniently calculated by shifting the loop momentum to higher dimensions.  Temporarily denoting a $D$-dimensional $n$-gon integral as $I_n^D$, the one-loop integrals are related by~\cite{Bern:1996ja,Bern:1995db}
\begin{align}\label{eq:1ldimshift}
I_n^D[\mu^{2r}]=-\eps(1-\eps)\cdots(r-1-\eps)(4\pi)^rI_n^{D+2r}[1],
\end{align}
which eliminates all powers of $\mu^2$.  Using this prescription, and a simple Feynman parametrisation, the relevant integrals are~\cite{Bern:1996ja}
\begin{align}\label{eq:1lintegrals}
I^D\bigg(\usegraph{9}{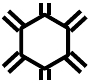}\bigg)[\mu^6]&=\cO(\eps),&&
I^D\bigg(\usegraph{11}{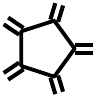}\bigg)[\mu^6]=\frac{i}{12(4\pi)^2}+\cO(\eps),\nn\\
I^D\bigg(\usegraph{10}{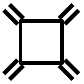}\bigg)[\mu^4]&=-\frac{i}{6(4\pi)^2}+\cO(\eps).
\end{align}
In this case the doubled external lines are used to emphasise that the integrals are valid for massive external momenta as well as massless ones.

With these integrals in hand one can check the well-known $n$-point integrated result up to six points~\cite{Bern:1993qk,Mahlon:1993si,Bern:1996ja}:
\begin{align}\label{eq:oneloopintegrated}
A^{(1)}(1^+,2^+,\ldots,n^+)=
-\frac{i}{48\pi^2}\sum_{1\leq i_1<i_2<i_3<i_4\leq n}
\frac{\tr_-(i_1i_2i_3i_4)}{\braket{12}\braket{23}\cdots\braket{n1}}+\cO(\eps),
\end{align}
which is indeed a rational function.  When using the dimensional regularisation parameter, $\eps=(4-D)/2$, there are no ultraviolet (UV) or infrared (IR) poles present.  As we shall see, both of these properties are also a consequence of the vanishing of the trees.

\subsection{Supersymmetric connection}

A remarkable property of these one-loop amplitudes is that they are related to those of the maximally-helicity violating (MHV) sector of $\cN=4$ supersymmetric Yang-Mills theory (SYM).  The amplitudes satisfy a dimension-shifting relation~\cite{Bern:1996ja},
\begin{align}\label{eq:neq4dimshift}
\left.A^{(1)}(1^+,2^+,\ldots,n^+)=
-2\eps(1-\eps)(4\pi)^2\tilde{A}^{(1),[\cN=4]}(1,2,\ldots,n)\right|_{D\to D+4},
\end{align}
where $\tilde{A}^{(1),[\cN=4]}$ is the coefficient of the supersymmetric delta function $\delta^8(Q)$ in the supersymmetric amplitude.  This delta function conserves supermomentum in the supersymmetric amplitudes and carries its helicity weight: all MHV-sector $\cN=4$ amplitudes carry it~\cite{Elvang:2009wd,Elvang:2010xn}.  Its precise form will not be important; it is enough for us to realise that it depends only on external kinematics and is totally permutation-invariant.

The dimension-shifting operation in~(\ref{eq:neq4dimshift}) is an instruction to evaluate loop integrals in $D\!+\!4$ dimensions; it has no effect on the external momenta and helicities.  However, given the relationship between one-loop integrals in different dimensions~(\ref{eq:1ldimshift}), eq.~(\ref{eq:neq4dimshift}) can be reformulated at the integrand level as a simple prescription to replace $\delta^8(Q)$ with the extra-dimensional function $(D_s-2)\mu^4$~\cite{Badger:2013gxa}.  The corresponding supersymmetric integrands at four points~\cite{Green:1982sw}, five points~\cite{Carrasco:2011mn} and six points~\cite{Bern:1996ja,Bern:2008ap} contain no diagrams smaller than a box: the so-called ``no-triangle hypothesis''~\cite{Bern:1994zx,Bern:1994cg,ArkaniHamed:2008gz}.  Clearly, this is also true for the all-plus integrands~(\ref{eq:1lbern}).

\subsection{Full-colour structure}\label{sec:ddmoneloop}

The following discussion of colour-dressed one-loop amplitudes is relevant for all external helicities (not just all-plus).  The subleading-colour terms in the trace-based colour decomposition~(\ref{eq:colourdecomp}) can be calculated as sums of permutations of the leading-colour components $A^{(1)}$~\cite{Bern:1990ux,Bern:1994zx}.  It is therefore possible to represent the full-colour amplitudes $\cA^{(1)}$ in terms of only colour-ordered amplitudes $A^{(1)}$.  Del Duca, Dixon and Maltoni (DDM) did this by expressing the colour factors as ring diagrams~\cite{DelDuca:1999ha,DelDuca:1999rs}:
\begin{align}\label{eq:1lDDM}
\mathcal{A}^{(1)}_n
=g^n\sum_{\sigma\in S_n/D_n}
c\Bigg(\!\usegraph{20}{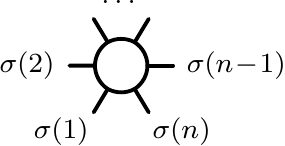}\Bigg)
A^{(1)}(\sigma(1),\sigma(2),\cdots,\sigma(n)),
\end{align}
where the sum is over $(n-1)!/2$ noncyclic reflection-inequivalent permutations; our diagrammatic notation for colour factors was introduced in section~\ref{sec:colourconventions}.  In chapter~\ref{ch:allplusbcj} we will explore a similar DDM decomposition for trees.

Using the DDM colour decomposition (\ref{eq:1lDDM}) we can, for instance, write the four-point all-plus amplitude as
\begin{align}\label{eq:1l4gddm}
\cA^{(1)}_{++++}
=g^4\cT(D_s-2)\sum_{\sigma\in S_4}\frac{1}{8}c\bigg(\usegraph{10}{boxi}\bigg)
I^D\bigg(\usegraph{10}{boxi}\bigg)[\mu^4].
\end{align}
The factor $1/8$ comes from changing the limit of the sum: it is the symmetry factor of the box.  Similar expressions are applicable for the other one-loop amplitudes.

\section{Two-loop amplitudes}\label{sec:2lreview}

Now we proceed to consider two-loop all-plus amplitudes --- the main subject of this thesis.  These amplitudes can be expressed as sums of integrals of irreducible numerators $\Delta_T$:\footnote{Requirements for numerators to be considered irreducible will be discussed in section~\ref{sec:irreduciblereview}.}
\begin{align}\label{eq:irreduciblesum}
A^{(2)}(1^+,2^+,\cdots,n^+)
=i\,\sum_TI^D\left[\Delta_T\right],
\end{align}
where the two-loop integration operator was defined in~(\ref{IntMeasure}).  The sum on $T$ runs over a complete set of two-loop topologies, many being duplicates of the same diagrams summed over different cyclic orderings (this ensures cyclic symmetry).

The move to two loops introduces nonplanar diagrams; for these diagrams there is no way of ordering the external legs without crossing the internal lines.  For instance, at five points there is the nonplanar double box, $\Delta(\includegraphics[scale=0.5,trim=0 5 0 -5]{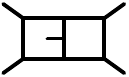})$.  This implies that there is important physical data in the subleading-colour corrections to the colour-dressed amplitudes $\cA_n^{(2)}$ that cannot be captured by the colour-ordered amplitudes $A^{(2)}$.  This is in contrast to the one-loop situation discussed earlier.

\subsection{Analytic structure}

\begin{figure}[t]
\centering
\begin{subfigure}{.4\textwidth}
  \centering
  \includegraphics[width=.8\textwidth]{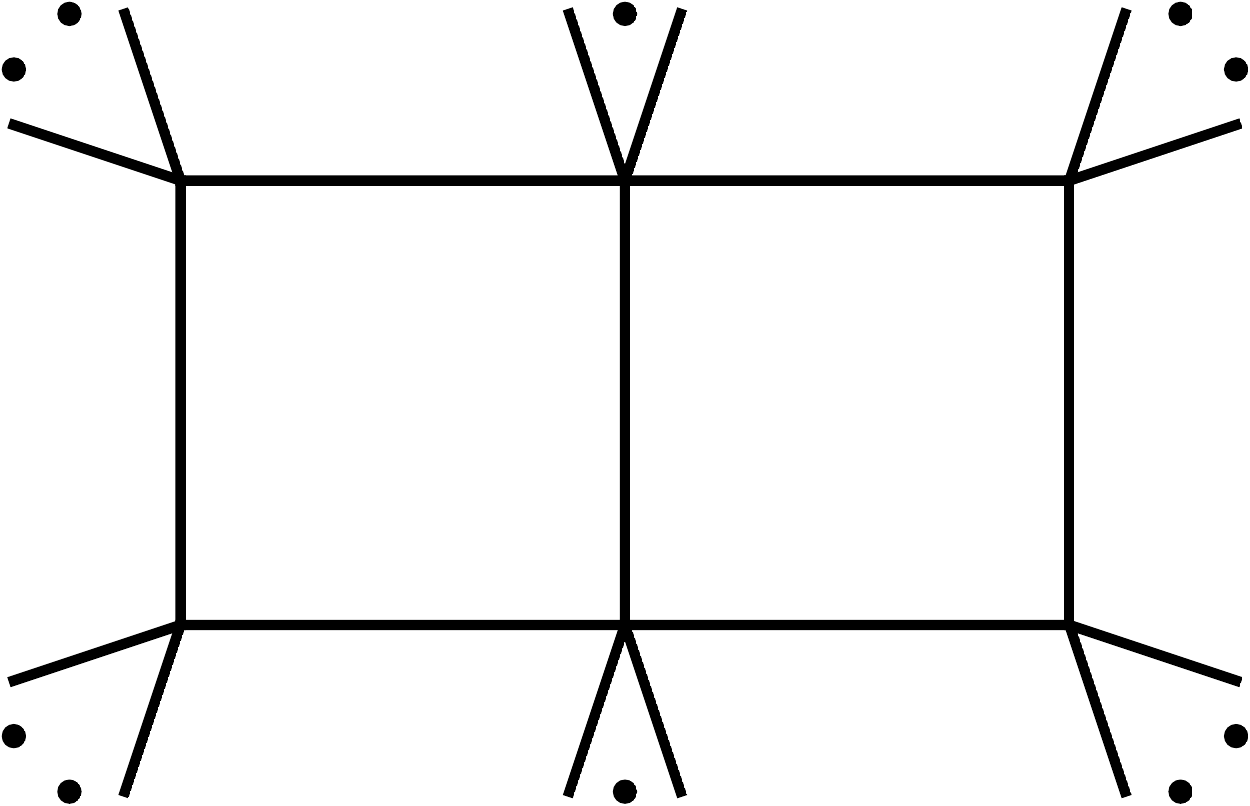}
  \caption[a]{Genuine two-loop topology}
\end{subfigure}
\begin{subfigure}{.4\textwidth}
  \centering
  \includegraphics[width=.8\textwidth]{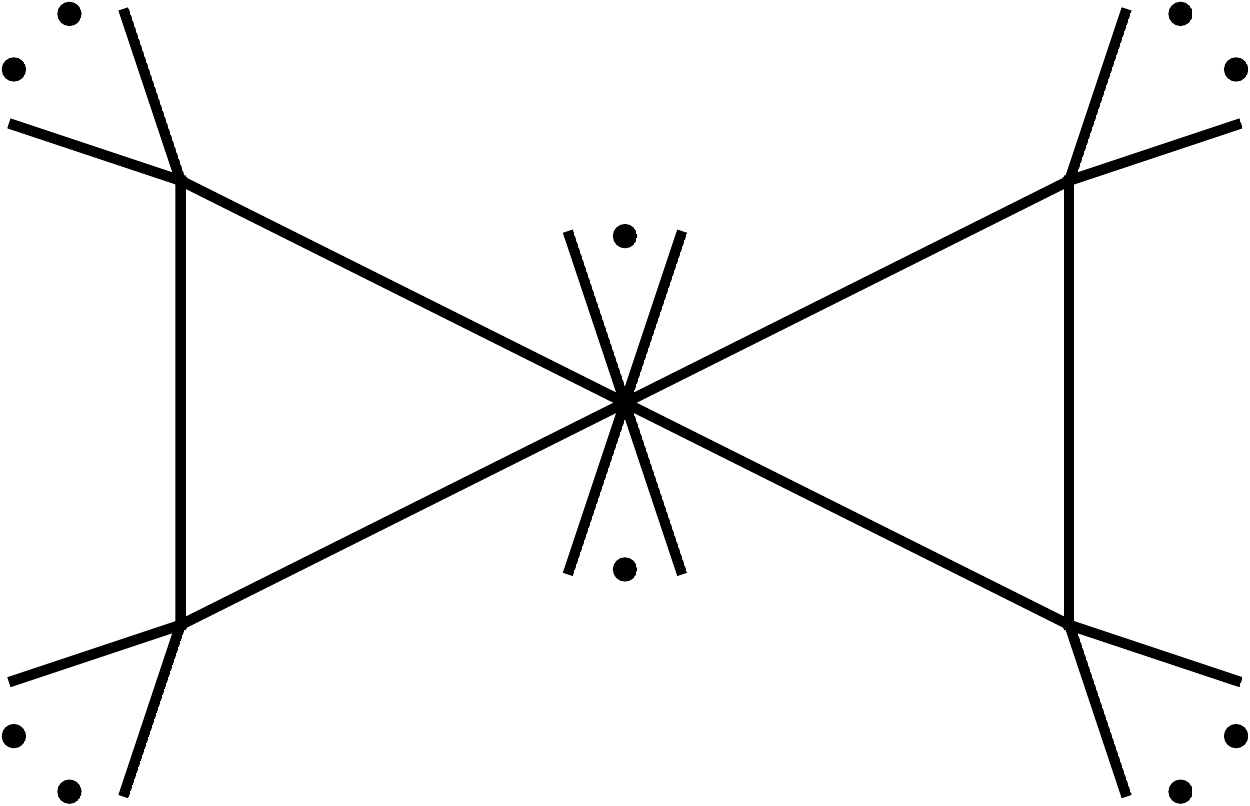}
  \caption[b]{Butterfly topology}
\end{subfigure}
\caption{\small Schematic representations of the genuine- and butterfly-type two-loop topologies.  Butterfly integrals can be evaluated as products of one-loop integrals.\label{fig:2lschematic}}
\end{figure}

The diagrams can be divided into two classes: genuine two-loop diagrams and butterfly (one-loop squared) diagrams, both of which are drawn in Figure~\ref{fig:2lschematic}.  The essential distinction is that, whereas butterfly-type integrals can be evaluated as products (or contractions) of one-loop integrals, genuine two-loop integrals cannot.  Their numerators follow a certain pattern: up to at least five external legs the numerators of all genuine two-loop topologies carry the overall $D$-dimensional prefactor~\cite{Bern:2000dn,Badger:2013gxa}
\begin{subequations}
\begin{align}\label{eq:F1def}
&F_1(\mu_1,\mu_2)\\
&\qquad
=(D_s-2)(\mu_{11}\mu_{22}+(\mu_{11}+\mu_{22})^2
+2\mu_{12}(\mu_{11}+\mu_{22}))+16(\mu_{12}^2-\mu_{11}\mu_{22}),\nn
\end{align}
while butterfly numerators are split into terms proportional to $(D_s-2)$ and $(D_s-2)^2$:
\begin{align}
F_2(\mu_1,\mu_2)
&=4(D_s-2)\mu_{12}(\mu_{11}+\mu_{22}),\label{eq:F2def}\\
F_3(\mu_1,\mu_2)
&=(D_s-2)^2\mu_{11}\mu_{22}.\label{eq:F3def}
\end{align}
\end{subequations}
In all known cases the spurious $F_2$ terms integrate to zero~\cite{Bern:2000dn,Badger:2013gxa}.

\begin{figure}[t]
\centering
\begin{subfigure}{0.25\textwidth}
  \centering
  \includegraphics[width=0.8\textwidth]{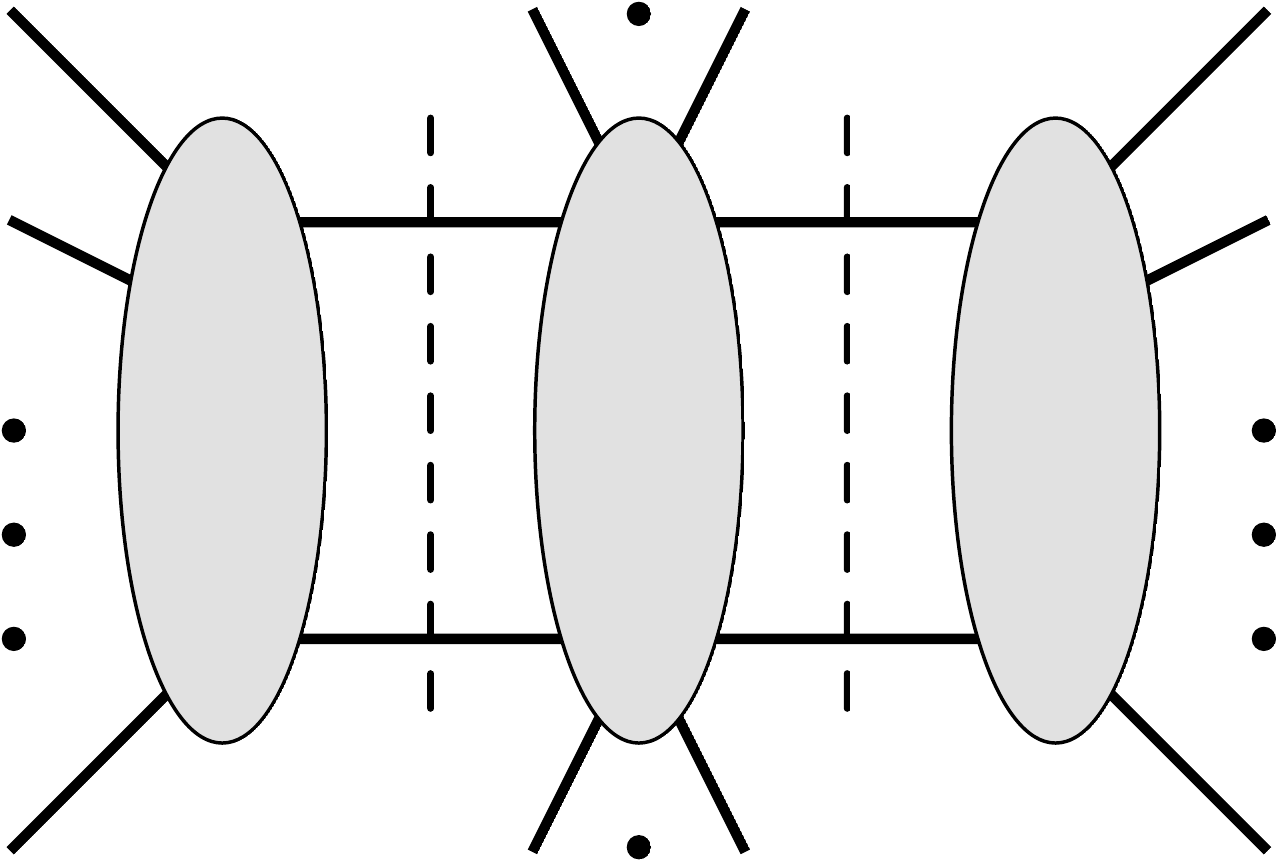}
  \caption[a]{}
\end{subfigure}
\begin{subfigure}{0.25\textwidth}
  \centering
  \includegraphics[width=0.8\textwidth]{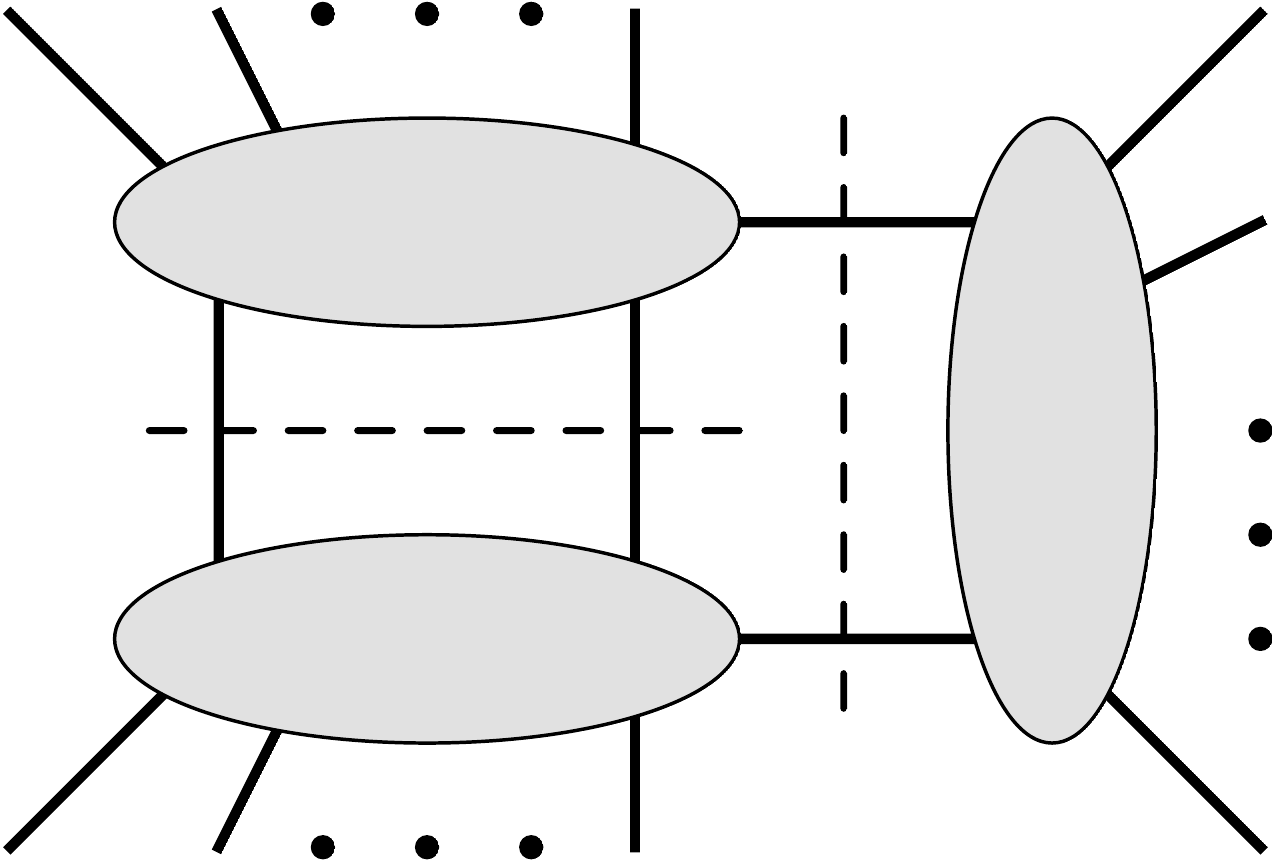}
  \caption[b]{}
\end{subfigure}
\begin{subfigure}{0.25\textwidth}
  \centering
  \includegraphics[width=0.8\textwidth]{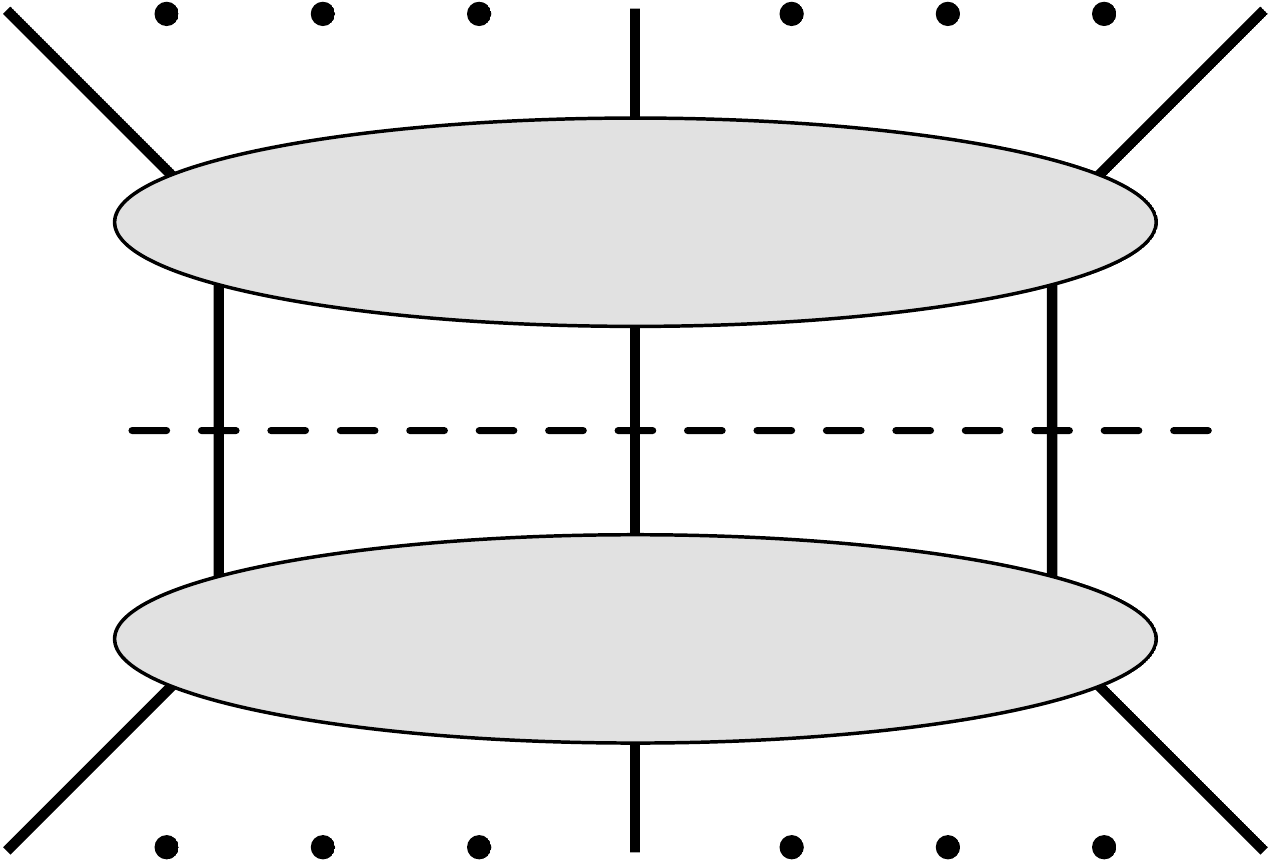}
  \caption[c]{}
\end{subfigure}
\caption{\small The three categories of spanning cuts required at two loops.  When all external helicities are positive no choice of internal helicities prevents at least one of the four-dimensional trees from vanishing.\label{fig:nptspanningcuts}}
\end{figure}

\begin{figure}[t]
\centering
\includegraphics[width=.5\textwidth]{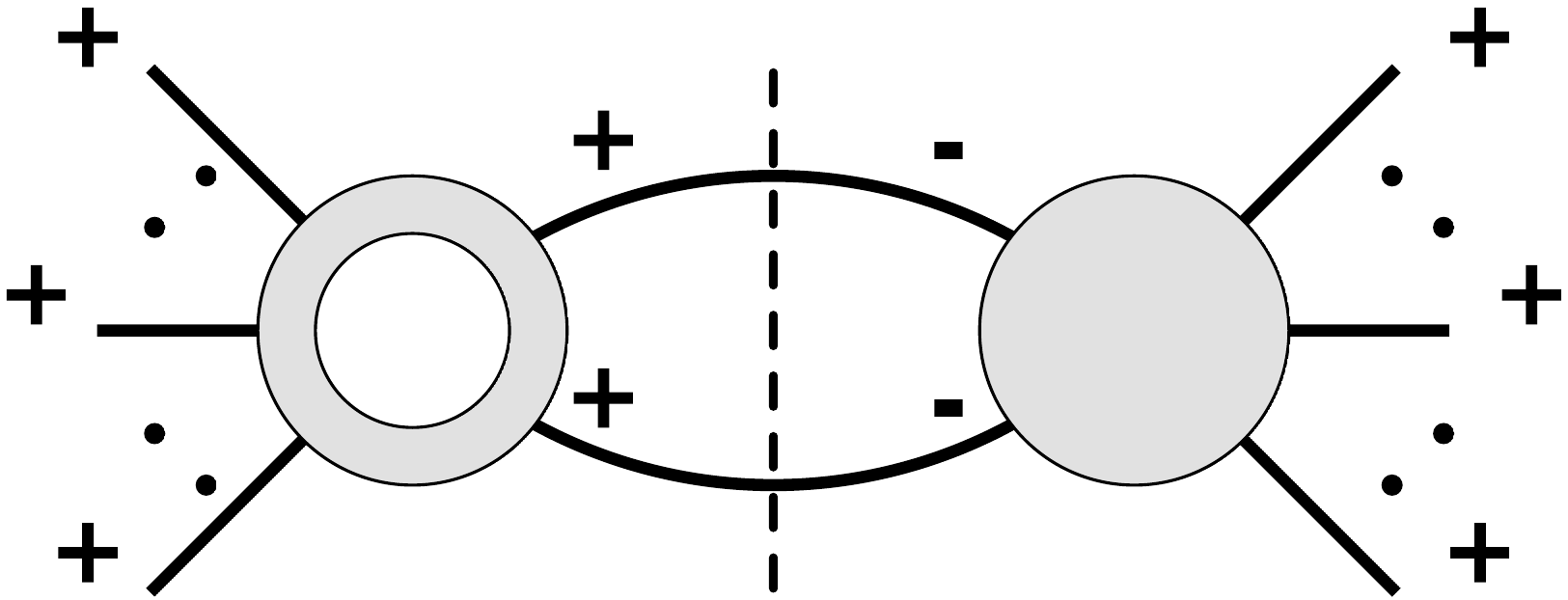}
\caption{\small A two-loop unitarity cut with all external helicities positive.  The left-hand disk is a (rational) one-loop amplitude which does not vanish in four dimensions.\label{fig:twoloopcut}}
\end{figure}

As we did at one loop, we can gain insight into this structure by examining the $n$-point spanning cuts --- all three two-loop categories are displayed in Figure~\ref{fig:nptspanningcuts}.  In general, at $L$ loops the spanning cuts form a minimal set so that any integrand chosen to match all of them is necessarily a valid integrand for the full amplitude.\footnote{For a recent review of spanning cuts in the context of generalised unitarity, see ref.~\cite{Bern:2011qt}.}  Again we see that the three spanning cuts vanish in four dimensions, so the numerators all carrying powers of $\mu_{ij}$ is to be expected.  However, unlike at one loop, this does not imply rationality of the integrated two-loop amplitudes.  To see why, consider the cut in Figure~\ref{fig:twoloopcut}: here, by only cutting into one of the two loops, we expose a cut that does not vanish in four dimensions.  Upon integration this gives a transcendental (non-rational) contribution.\footnote{Dunbar, Jehu and Perkins recently used these cuts in four dimensions to calculate complete integrated all-plus amplitudes up to six points~\cite{Dunbar:2016aux,Dunbar:2016gjb} and $n$-point polylogarithmic parts~\cite{Dunbar:2016cxp}.  The rational terms up to six points were computed separately; essentially they reduced the polylogarithmic part of the calculation to a one-loop unitarity problem.}

In this thesis we will not be interested in the transcendental parts of the integrated amplitudes.  Our aim will instead be to show how physical properties of the amplitudes can be exposed without the need for explicit integration.  We will, however, occasionally discuss integrated expressions for terms proportional to $(D_s-2)^2$: these come from butterfly-type integrals carrying $F_3$~(\ref{eq:F3def}).  The relevant integrals always become products of one-loop integrals containing $\mu^2$: by~(\ref{eq:1ldimshift}) these must be rational.

\subsection{Supersymmetric connection}

The integrand-level connection to $\cN=4$ SYM~(\ref{eq:neq4dimshift}) continues at two loops.  It only works for numerators of the genuine two-loop topologies: in this case one replaces the supersymmetric delta function $\delta^8(Q)$ with the function of extra-dimensional components $F_1$~(\ref{eq:F1def})~\cite{Badger:2013gxa}.  The butterfly topologies do not have supersymmetric counterparts; the no-triangle hypothesis mentioned earlier forbids them in many cases.

\subsection{Four-gluon amplitude}\label{sec:4g2lreview}

The planar numerators are~\cite{Bern:2000dn}
\begin{subequations}\label{eq:2L4gbern}
\begin{align}
\Delta\bigg(\usegraph{9}{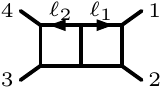}\bigg)
&=s\,F_1,
\label{eq:331bern}\\
\Delta\bigg(\usegraph{9}{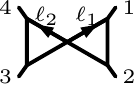}\bigg)
&=F_2+F_3\left(\frac{s+(\ell_1+\ell_2)^2}{s}\right),
\label{eq:330bern}
\end{align}
\end{subequations}
of which the first, the double box, is a genuine two-loop topology and the second, the double triangle, is a butterfly.  These numerators respect the symmetries of their graphs and adhere to the $D$-dimensional structure discussed above.  The double-box numerator equals its $\cN=4$ supersymmetric counterpart with $F_1$ replacing $\delta^8(Q)$~\cite{Bern:1997nh,Bern:1998ug}.  We have extracted overall factors of the permutation-invariant object $\mathcal{T}$~(\ref{eq:cT}), which we will do for all four-point numerators in this thesis.

The full-colour presentation of the four-point amplitude is known.  The only nonzero nonplanar numerator is the nonplanar double box~\cite{Bern:2000dn},
\begin{align}\label{eq:delta332np}
\Delta\bigg(\!\usegraph{9}{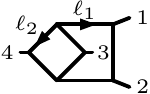}\bigg)=s\,F_1,
\end{align}
which is a genuine two-loop topology: it conforms to the same extra-dimensional structure with $F_1$.  It also matches its supersymmetric counterpart~\cite{Bern:1997nh,Bern:1998ug}.  The full-colour amplitude is
\begin{align}\label{eq:4ptamplitude}
&\mathcal{A}^{(2)}_{++++}
=ig^6\mathcal{T}\sum_{\sigma \in S_4}
I^D\Bigg[\frac{1}{4}c\bigg(\usegraph{9}{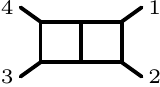}\bigg)
\Bigg(\Delta\bigg(\usegraph{9}{331}\bigg)
+\Delta\bigg(\usegraph{9}{330}\bigg)\Bigg)\nnl
&\qquad\qquad\qquad\qquad\qquad\qquad\qquad\qquad\!\!
+\frac{1}{4}c\bigg(\!\usegraph{9}{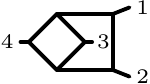}\bigg)
\Delta\bigg(\!\usegraph{9}{322}\bigg)\Bigg].
\end{align}
Beyond four points, full-colour expressions were not known prior to this work.

\subsection{Five-gluon amplitude}

An integrand presentation of the planar five-gluon, two-loop all-plus amplitude was first found by Badger, Frellesvig and Zhang (BFZ)~\cite{Badger:2013gxa}.  As it involves only six non-zero numerators we list them all here.

There are three genuine two-loop topologies: the pentabox,
\begin{subequations}\label{eq:2L5gbfz}
\begin{align}\label{eq:431bfz}
\Delta\bigg(\usegraph{9}{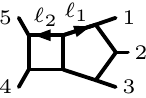}\bigg)=
-\frac{s_{12}s_{23}s_{45} F_1}{\SpDenom5\text{tr}_5}\left(\text{tr}_+(1345)
(\ell_1+p_5)^2+s_{15}s_{34}s_{45}\right),
\end{align}
the one-mass double box,
\begin{align}\label{eq:331m1bfz}
\Delta\bigg(\usegraph{9}{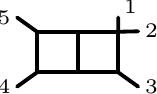}\bigg)&=
-\frac{s_{34}s_{45}^2\text{tr}_+(1235)F_1}{\SpDenom5\text{tr}_5},
\end{align}
and the five-leg double box,
\begin{align}\label{eq:3315lbfz}
\Delta\bigg(\usegraph{13}{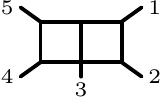}\bigg)=\frac{s_{12}s_{23}s_{34}s_{45}s_{51}F_1}{\SpDenom5\text{tr}_5},
\end{align}
where $\trfive=\trfive(1234)$.  The one-mass double box~(\ref{eq:331m1bfz}) also has a counterpart related by symmetry through the horizontal axis; the other two numerators satisfy relabelling symmetries through the horizontal and vertical axes respectively.  All three numerators are proportional to $F_1$.

The three butterfly-type numerators are the box triangle,
\begin{align}\label{eq:430bfz}
\Delta\bigg(\usegraph{13}{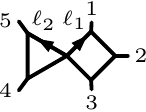}\!\!\!\:\bigg)&=
-\frac{s_{12}\text{tr}_+(1345)}{2\SpDenom5s_{13}}
(2\ell_1\cdot\omega_{123}+s_{23})\\
&\qquad\qquad\times\left(F_2 + F_3\frac{(\ell_1+\ell_2)^2+s_{45}}{s_{45}}\right),\nn
\end{align}
the one-mass double triangle (which also has a counterpart related by symmetry),
\begin{align}\label{eq:330m1bfz}
\Delta\bigg(\usegraph{10}{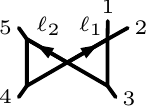}\bigg)&=
\frac{(s_{12}-s_{45})\text{tr}_+(1345)}{2\SpDenom5 s_{13}}
\left(F_2 + F_3\frac{(\ell_1+\ell_2)^2+s_{45}}{s_{45}}\right),
\end{align}
and the five-leg double triangle,
\begin{align}\label{eq:3305lbfz}
\Delta\bigg(\usegraph{13}{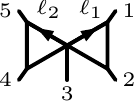}\bigg)&=
-\frac{1}{\SpDenom5}\nn\\&
\!\!\!\!\!\!\!\!\!\!\!\!\!\!\!\!\!\!\!\!\!\!\!\!\!\!\!\!\!\!\!\!\!\!\times 
\Bigg\{\frac{1}{2}\;\!
             \bigg(\text{tr}_+(1245)
                  -\frac{\text{tr}_+(1345) \text{tr}_+(1235)}{s_{13}s_{35}}
             \bigg) \nn \\ & \times
             \bigg(F_2
                  +F_3 \frac{4(\ell_1\!\cdot\!p_3)(\ell_2\!\cdot\!p_3)
                                  +(\ell_1+\ell_2)^2 (s_{12}+s_{45})
                                  + s_{12}s_{45}}
                                 {s_{12}s_{45}}
             \bigg) \nn \\ &\!\!\!\!\!\!\!\!\!\!\!\!\!\!\!\!\!\!\!\!\!\!\!\! 
           + F_3 \bigg[\,(\ell_1+\ell_2)^2 s_{15}\\ &
           \!\!\!\!\!\!\!\!\!\!\!\!\!\!\!\!\!\!\!\!\!\!\!\! 
   +\text{tr}_+(1235) \bigg( \frac{(\ell_1+\ell_2)^2}{2s_{35}}
                            -\frac{\ell_1\!\cdot\!p_3}{s_{12}}
                      \Big(1 + \frac{2(\ell_2\!\cdot\!\omega_{543})}{s_{35}}
                              + \frac{s_{12}-s_{45}}{s_{35}s_{45}}
                                (\ell_2-p_5)^2
                      \Big)
                      \bigg) \nn \\ &\!\!\!\!\!\!\!\!\!\!\!\!\!\!\!\!\!\!\!\!\!\!\!\! 
   +\text{tr}_+(1345) \bigg( \frac{(\ell_1+\ell_2)^2}{2s_{13}}
                            -\frac{\ell_2\!\cdot\!p_3}{s_{45}}
                      \Big(1 + \frac{2(\ell_1\!\cdot\!\omega_{123})}{s_{13}}
                              + \frac{s_{45}-s_{12}}{s_{12}s_{13}}
                                (\ell_1-p_1)^2
                      \Big)
                      \bigg)
                      \bigg]
      \Bigg\}.\nn
\end{align}
\end{subequations}
Again, all butterfly-type numerators carry terms proportional to $F_2$ and $F_3$.

The above representation is not unique: exchanging terms between the diagrams is allowed provided the overall integrand~(\ref{eq:irreduciblesum}) remains invariant.  For example, in the five-leg double triangle (\ref{eq:3305lbfz}) contributions from two triangle-bubble graphs have been absorbed by adding terms proportional to the missing propagators $(\ell_1-p_1)^2$ and $(\ell_2-p_5)^2$.  The numerators above were chosen to make use of the spurious directions $\omega_{ijk}$: this is convenient as numerator terms carrying, for instance, $\ell_1\cdot\omega_{123}$ are parity-odd, so integrate to zero.  These terms were omitted in the one-loop amplitudes~(\ref{eq:1lbern}).

However, the numerators given above have some disadvantages.  They contain unphysical poles in $s_{13}$, $s_{35}$ and $\trfive$; in the integrated version of this amplitude, found by Gehrmann, Henn and Lo Presti~\cite{Gehrmann:2015bfy} and based on BFZ's results, these singularities cancel away.  Besides the unsatisfactory nature of this from a aesthetic standpoint, it makes the $\cN=4$ correspondence harder to spot (though it is still valid).  The part of the all-plus integrand formed by genuine two-loop diagrams only is indeed related by $\delta^8(Q)\leftrightarrow F_1$ to the corresponding $\cN=4$ SYM integrand~\cite{Carrasco:2011mn}, but not at the level of individual numerators.  In the next chapter we will find an alternative integrand presentation that makes the cancellation of unphysical poles manifest before integration; this will help us to expose the $\cN=4$ connection.

\section{Infrared divergences}\label{sec:irreview}

A recurring feature of two-loop all-plus amplitudes that we will explore in this thesis is their infrared (IR) structure.  Besides offering a useful check of our results, the IR structure is an important physical property of the amplitudes.  The structure is universal~\cite{Catani:1998bh,Catani:2000ef}:
\begin{subequations}\label{eq:catani}
\begin{align}
\cA^{(1)}_n&=I^{(1)}\circ\cA^{(0)}_n+\cO(\eps^0),
\label{eq:1lcatani}\\
\cA^{(2)}_n&=I^{(1)}\circ\cA^{(1)}_n+I^{(2)}\circ\cA^{(0)}_n+\cO(\eps^0).
\label{eq:2lcatani}
\end{align}
\end{subequations}
$I^{(i)}\circ$ are operators acting on the colour structure of their arguments; we will explain this in more detail below, including a demonstration at four points.

Vanishing of the all-plus tree amplitudes implies IR finiteness of the one-loop  amplitudes: this is indeed the case in eq.~(\ref{eq:1lallplus}).\footnote{By a similar argument there are also no UV divergences --- these too are proportional to the finite parts of the trees~\cite{Catani:1998bh,Catani:2000ef}.}  IR divergences of the two-loop amplitudes are related to the one-loop amplitudes through the colour operator $I^{(1)}$, given by\footnote{The expression given differs from the original operator in ref.~\cite{Catani:1998bh} because we are working with UV-unrenormalised amplitudes.}
\begin{align}
I^{(1)}=g^2\sum_{i\neq j}\frac{c_\Gamma}{\eps^2}\left(\frac{\mu_R^2}{-s_{ij}}\right)^\eps\mathbf{T}_i\cdot\mathbf{T}_j.
\end{align}
From now on we will always set $\mu_R^2$, the dimensional regularisation scale, to one (we can recover it using dimensional analysis).

The colour charge $\mathbf{T}_i=\{T^a_i\}$ is a vector with respect to the colour index $a$ and an SU($N_c$) matrix with respect to the colour indices of the outgoing parton $i$.  In our case, considering only external gluons, $T^a_{bc}=if^{bac}$.  To understand the action of the operator $\mathbf{T}_i\cdot\mathbf{T}_j\,\circ$ we take a couple of examples~\cite{Bern:2000dn}:
\begin{subequations}\label{eq:colouropexamples}
\begin{align}
(\mathbf{T}_1\cdot\mathbf{T}_2)\circ c\bigg(\usegraph{10}{boxi}\bigg)
&=-\frac{1}{2}c\bigg(\usegraph{9}{331i}\bigg),\\
(\mathbf{T}_1\cdot\mathbf{T}_2)\circ c\bigg(\usegraph{10}{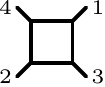}\bigg)
&=-\frac{1}{2}c\bigg(\!\usegraph{9}{322i}\bigg).
\end{align}
\end{subequations}
In each case a t-channel bridge is joined to the two external labels specified by $i$ and $j$ (factors of $1/2$ come from switching normalisations between $f^{abc}$ and $\tf^{abc}$).  The effect is to convert $L$-loop colour structures into $(L+1)$-loop colour structures.

We also need a version of the two-loop universal IR decomposition~(\ref{eq:2lcatani}) applicable to colour-ordered amplitudes.  This is found by substituting in the trace-based colour decomposition~(\ref{eq:colourdecomp}) on both sides and isolating the leading terms in $N_c$.  Structure constants are eliminated using $\tf^{abc}=\tr([T^a,T^b]T^c)$; summed fundamental-representation colour matrices $T^a$ are eliminated using the Fierz rearrangement (\ref{eq:fierz}).  We are left with
\begin{align}\label{eq:orderedcatani}
A^{(2)}(1^+,2^+,\ldots,n^+)=
-\sum_{i=1}^n\frac{c_\Gamma}{\eps^2}\left(\frac{1}{-s_{i,i+1}}\right)^\eps
A^{(1)}(1^+,2^+,\ldots,n^+)+\cO(\eps^0).
\end{align}
At five points BFZ were able to check this numerically~\cite{Badger:2013gxa}.  The relevant integrals on both sides were evaluated using the public sector-decomposition tools \textsc{FIESTA}~\cite{Smirnov:2013eza,Smirnov:2015mct} and \textsc{SecDec}~\cite{Borowka:2015mxa,Borowka:2017idc}.\footnote{The master integrals for planar $2\to3$ scattering are now available in analytic form~\cite{Gehrmann:2015bfy,Papadopoulos:2015jft}.}

\subsection{Four-point example}\label{sec:4ptirexample}

We will now illustrate the above points by explicitly checking the four-gluon two-loop IR structure.  Our calculation follows that of Bern, Dixon and Kosower~\cite{Bern:2000dn}: they provide the analytic two-loop divergences
\begin{subequations}
\begin{align}\label{eq:doubleboxirreview}
I^D\bigg(\usegraph{10}{331}\bigg)[F_1]&=
-2i(D_s-2)\frac{c_\Gamma}{\eps^2}(-s)^{-1-\eps}I^D\bigg(\usegraph{10}{boxi}\bigg)[\mu^4]
+\cO(\epsilon^0),\\
I^D\bigg(\usegraph{10}{322}\bigg)[F_1]&=
-i(D_s-2)\frac{c_\Gamma}{\eps^2}(-s)^{-1-\eps}I^D\bigg(\usegraph{10}{box1324i}\bigg)[\mu^4]
+\cO(\epsilon^0).
\end{align}
\end{subequations}
The rational double-triangle integral does not contribute to the IR structure.

First we substitute the two-loop divergences into the full-colour amplitude~(\ref{eq:4ptamplitude}):
\begin{subequations}\label{eq:4ptirchecking}
\begin{align}
&\cA^{(2)}_{++++}
=-g^6(D_s-2)\frac{c_\Gamma}{\eps^2}\cT\sum_{\sigma\in S_4}\Bigg\{
\frac{1}{4}c\bigg(\usegraph{9}{331i}\bigg)\times
2(-s)^{-\eps}I^D\bigg(\usegraph{10}{boxi}\bigg)[\mu^4]\nn\\
&\qquad\qquad\qquad
+\frac{1}{4}c\bigg(\!\usegraph{9}{322i}\bigg)\times
(-s)^{-\eps}I^D\bigg(\usegraph{10}{box1324i}\bigg)[\mu^4]\Bigg\}+\cO(\eps^0).
\end{align}
Then we insert the colour operators $\mathbf{T}_i\cdot\mathbf{T}_j$ using eqs.~(\ref{eq:colouropexamples}):
\begin{align}
&\cA^{(2)}_{++++}
=g^6(D_s-2)\frac{c_\Gamma}{\eps^2}\cT\sum_{\sigma\in S_4}
(-s)^{-\eps}\mathbf{T}_1\cdot\mathbf{T}_2\,\circ\Bigg\{
c\bigg(\usegraph{10}{boxi}\bigg)I^D\bigg(\usegraph{10}{boxi}\bigg)[\mu^4]\nnl
&\qquad\qquad\qquad\qquad\qquad\qquad
+\frac{1}{2}c\bigg(\usegraph{10}{box1324i}\bigg)I^D\bigg(\usegraph{10}{box1324i}\bigg)[\mu^4]\Bigg\}
+\cO(\eps^0).
\end{align}
Finally we use the sum on permutations to collect terms:
\begin{align}
&\cA^{(2)}_{++++}\\
&=g^6(D_s-2)\frac{c_\Gamma}{\eps^2}\cT\sum_{\sigma\in S_4}
\left(\sum_{i\neq j}\left(-s_{ij}\right)^{-\eps}\mathbf{T}_i\cdot\mathbf{T}_j\right)\circ
\Bigg\{\frac{1}{8}c\bigg(\usegraph{10}{boxi}\bigg)I^D\bigg(\usegraph{10}{boxi}\bigg)[\mu^4]\Bigg\}\nn\\
&=I^{(1)}\circ\cA^{(1)}_{++++}+\cO(\eps^0).\nn
\end{align}
\end{subequations}
The term in round brackets is permutation-invariant; we are left with the term in curly brackets which we identify as the one-loop amplitude in the DDM colour decomposition~(\ref{eq:1l4gddm}).

\section{Discussion}

In this review we have discussed various physical properties of all-plus amplitudes.  Some of these are common to all loop orders: for instance, the colour-ordered amplitudes, satisfying cyclic and reflection symmetries~(\ref{eq:colourorderedsymmetries}), can always be isolated using a trace-based colour decomposition~(\ref{eq:colourdecomp}).  One might suggest that the integrand-level connection to $\cN=4$ SYM is also an all-loop property, though the structure of a three-loop equivalent of $F_1(\mu_1,\mu_2)$ is an open question.  Most importantly though, vanishing of the tree-level amplitudes strongly influences the loop-level structure.

This explains the one-loop amplitudes being finite, rational functions of the external kinematics~(\ref{eq:oneloopintegrated}).  At two loops we saw similar structure at play: while the amplitudes are no longer rational or finite, we showed how their IR structure is nevertheless made simpler by vanishing of the trees.  We gave a full-colour integrand-level presentation of the four-gluon all-plus amplitude~(\ref{eq:4ptamplitude}) and checked its IR decomposition analytically~(\ref{eq:4ptirchecking}).  It does not contain unphysical poles, and it trivially manifests the supersymmetric connection.

These properties are obscured in BFZ's five-point two-loop amplitude by the freedom to write its integrand in different ways.  Its numerators contain unphysical poles that disappear in the integrated result~\cite{Gehrmann:2015bfy}.  This prevents the supersymmetric connection from holding diagram-by-diagram.  While the leading-colour IR decomposition~(\ref{eq:orderedcatani}) has been checked numerically, an analytic check like the one at four points would be preferable.  A full-colour amplitude is also lacking.

The freedom to write loop-level numerators in different ways will be important as we move forwards.  In the next chapter we will see how a local-integrand presentation enables us to eliminate unphysical poles from the five- and six-point planar integrands, while also making the IR structure readliy apparent.  The presentation will rely heavily on the connection to $\cN=4$ SYM.

\chapter{Local-integrand representations}\label{ch:localintegrands}

\section{Introduction}\label{sec:intro}

In the last chapter we saw that Badger, Frellesvig and Zhang's (BFZ's) planar five-point two-loop integrand~\cite{Badger:2013gxa} has spurious singularities in $s_{13}$, $s_{35}$ and $\trfive$ --- these do not correspond to physical poles.  Gehrmann, Henn and Lo Presti's integrated result~\cite{Gehrmann:2015bfy} is singularity-free, so is there a better way of writing the integrand that avoids these singularities?

Expressing amplitudes in a form manifestly free of spurious singularities can lead to remarkably efficient evaluations.  This is both because numerical stability is improved but also because the analytic formulae are more highly constrained, and thus likely to be more compact.  The bases of integral functions obtained through the multi-loop integrand reduction procedure are not unique and often hide the properties of locality and universal infrared behaviour --- we often say that functions are local to indicate that they avoid spurious singularities.  If one could select a basis of master integrals with these properties manifest then the reward could be considerably more compact amplitude expressions.

In planar $\cN=4$ SYM this problem has already been solved.  The all-loop integrand for scattering amplitudes in the planar sector of $\cN=4$ \cite{ArkaniHamed:2010kv,ArkaniHamed:2010gh} generalises the BCFW recursion relations for tree amplitudes \cite{Britto:2004ap,Britto:2005fq} to loop level, building amplitudes out of chiral integrals with unit leading singularities.  Amongst many other important properties, the presentation is local and makes use of infrared (IR) finite integrals. This technique has been applied to a variety of explicit cases, most recently to all-multiplicity two-loop amplitudes in planar $\cN=4$~\cite{Bourjaily:2015jna}. One naturally questions whether such an approach could be applied in pure Yang-Mills theory where it is convenient to work in $D\neq4$.

In this chapter we investigate the possibility of obtaining compact two-loop planar QCD amplitudes free of spurious singularities using integrand reduction and generalised unitarity cuts in $D$ dimensions.  Rather than constructing $D$-dimensional local integrands from first principles, we use the all-plus sector's connection with $\cN=4$ to directly recycle the supersymmetric expressions into all-plus ones.  The result will be a partially-determined basis of master integrals onto which we can fit cut solutions.  In addition to the five-point case reviewed in the last chapter we will also consider the planar six-gluon all-plus amplitude at two loops, an integrated expression for which has been written down by Dunbar and Perkins \cite{Dunbar:2016aux,Dunbar:2016cxp,Dunbar:2016gjb}.  The resulting amplitude expressions are free of unphysical poles before integration and have a remarkably simple infrared structure.

\section{One-loop local integrands}\label{sec:oneloop}

Here we discuss some one-loop applications of $D$-dimensional local integrands.  As the relevant integrals and amplitudes are already known we can recycle results from the literature; in the next section we will review $D$-dimensional unitarity which will enable us to study two-loop examples.

\subsection{The box integral in \texorpdfstring{$D=4-2\epsilon$}{D=4-2e}}

To motivate our discussion of local integrands we begin by considering the scalar box integral:
\begin{align}\label{eq:boxintegral}
I^D\bigg(\usegraph{10}{box}\bigg)
=i\,\frac{c_\Gamma}{st}\left(\frac{2}{\epsilon^2}\left((-s)^{-\epsilon}+(-t)^{-\epsilon}\right)
-\ln^2\left(\frac{s}{t}\right)-\pi^2\right)+\cO(\epsilon).
\end{align}
The $\epsilon$ pole structure is entirely due to IR divergences from soft regions, such as when $\ell\to0$ or $\ell-p_1\to0$, and collinear regions, such as when $\ell$ approaches collinearity with $p_1$.  The possibility of simultaneous soft and collinear divergences gives leading poles at $\cO(\epsilon^{-2})$.

The box integral can be rendered finite by introducing a local numerator\footnote{The ``wavy line'' notation was first introduced by Arkani-Hamed et al. to denote local integrands in maximally supersymmetric $\mathcal{N}=4$ \cite{ArkaniHamed:2010kv,ArkaniHamed:2010gh}.  This connection will be explored in the next section.}
\begin{align}\label{eq:localboxintegral}
I^D\bigg(\usegraph{10}{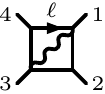}\bigg)\equiv
I^D\bigg(\usegraph{10}{box}\bigg)\left[\trp(1(\ell-p_1)(\ell-p_{12})3)\right].
\end{align}
This numerator vanishes in all of the soft and collinear regions.\footnote{$D$-dimensional Dirac traces are formally introduced in appendix~\ref{app:traces}.}  To see explicitly that the new integral is finite we evaluate the Dirac trace
\begin{align}
&\trp(1(\ell-p_1)(\ell-p_{12})3)
=\frac{1}{2}\tr(1(\ell-p_1)(\ell-p_{12})3)+\frac{1}{2}\trfive(1(\ell-p_1)(\ell-p_{12})3)
\nonumber\\&\qquad
=\frac{st}{2}-\frac{t}{2}\ell^2-\frac{s}{2}(\ell-p_1)^2-\frac{t}{2}(\ell-p_{12})^2-\frac{s}{2}(\ell+p_4)^2
-\frac{1}{2}\trfive(123\ell).
\end{align}
The spurious $\trfive$ term integrates to zero.  When the propagators are cancelled against their counterparts in the denominator the new box integral becomes a linear combination of scalar box and triangle integrals
\begin{align}
&I^D\bigg(\usegraph{10}{boxs}\bigg)
\nn\\&\qquad
=\frac{st}{2}I^D\bigg(\usegraph{10}{box}\bigg)\!
-\!\frac{s}{2}I^D\bigg(\!\usegraph{9}{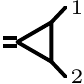}\bigg)\!
-\!\frac{t}{2}I^D\bigg(\!\usegraph{9}{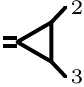}\bigg)\!
-\!\frac{s}{2}I^D\bigg(\!\usegraph{9}{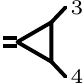}\bigg)\!
-\!\frac{t}{2}I^D\bigg(\!\usegraph{9}{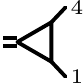}\bigg)
\nn\\&\qquad
=-i\frac{c_\Gamma}{2}\left(\ln^2\left(\frac{s}{t}\right)+\pi^2\right)+\cO(\epsilon).
\end{align}
Here we have used the scalar triangle integral
\begin{align}\label{eq:triintegral}
I^D\bigg(\!\usegraph{9}{tri12}\bigg)=-i\frac{c_\Gamma}{\epsilon^2}(-s)^{-1-\epsilon}.
\end{align}
The regulated box integral is also known to satisfy a dimension-shifting formula~\cite{Bern:1992em,Bern:1993kr}:
\begin{align}
I^D\bigg(\usegraph{10}{boxs}\bigg)
=(-1+2\epsilon)u(4\pi)^2I^{D+2}\bigg(\usegraph{10}{box}\bigg),
\end{align}
where $u=(p_1+p_3)^2$.  IR finiteness follows trivially as the box integral is finite in six dimensions.

\subsection{One-loop all-plus amplitudes}\label{sec:oneloopamplitudes}

To continue our motivation of local integrands we now rewrite the planar five- and six-gluon one-loop all-plus amplitudes~(\ref{eq:1lbern}) into such a representation.  Our starting point is the local all-loop integrand for scattering amplitudes in the planar MHV sector of $\cN=4$ SYM \cite{ArkaniHamed:2010kv,ArkaniHamed:2010gh}.  These supersymmetric expressions are by now quite familiar and, amongst many other interesting properties, they are known to exhibit simple IR behaviour.  To use them in the context of all-plus Yang-Mills theory we will use the dimension shift reviewed in section~\ref{sec:1lreview}: this amounts to replacing the supersymmetric delta function $\delta^8(Q)$ with $(D_s-2)\mu^4$ in the all-plus integrand.

The momentum-twistor formalism used to write the $\cN=4$ expressions seemingly ties them to four dimensions so we begin by translating to a manifestly $D$-dimensional language.  Full details of the procedure up to two loops are given in appendix~\ref{sec:N4localintegrands}; for now we merely notice that, at one loop, pentagon integrals always operate on twistor brackets involving loop momenta.  These twistor brackets are, up to a helicity-dependent scaling, equivalent to Dirac traces with a positive projector:
\begin{align}\label{eq:squiggledef}
\usegraph{13}{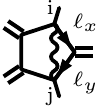}\sim\braket{AB|(i-1,i,i+1)\cap(j-1,j,j+1)}\sim\trp(i\ell_x\ell_yj).
\end{align}
The only other one-loop integrals used in the four-dimensional local integrands are scalar box integrals.

With this correspondence in mind we define $D$-dimensional regulated pentagon integrals as
\begin{align}\label{eq:regulatedpentagon}
I^D\bigg(\!\usegraph{13}{pentschemes}\!\bigg)\left[\mathcal{P}(p_i,\ell,\mu^2)\right]
\equiv I^D\bigg(\!\usegraph{13}{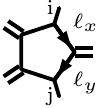}\!\bigg)
\left[\trp(i\ell_x\ell_yj)\mathcal{P}(p_i,\ell,\mu^2)\right],
\end{align}
which is consistent with the definition we made for the regulated box in eq.~(\ref{eq:localboxintegral}).
This ``wavy line'' notation is similar to that used by Arkani-Hamed et al.: it has the same property of controlling IR divergences.

The five-point one-loop all-plus amplitude~(\ref{eq:1l5gbern}) can now be re-expressed as
\begin{subequations}\label{eq:1lallplus}
\begin{align}
&A^{(1)}(1^+,2^+,3^+,4^+,5^+)=
\frac{(D_s-2)}{\braket{12}\braket{23}\braket{34}\braket{45}\braket{51}}
\times\label{eq:5pt1lallplus}\\
&
\Bigg(\frac{\trp(1345)}{s_{13}}I^D\bigg(\!\usegraph{12}{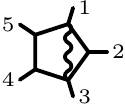}\!\bigg)[\mu^4]
-s_{23}s_{34}I^D\bigg(\usegraph{10}{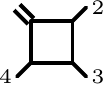}\bigg)[\mu^4]
-s_{12}s_{15}I^D\bigg(\usegraph{10}{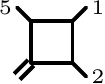}\bigg)[\mu^4]\Bigg),\nn
\end{align}
and the six-point amplitude as
\begin{align}
&A^{(1)}(1^+,2^+,3^+,4^+,5^+,6^+)=
\frac{(D_s-2)}{\braket{12}\braket{23}\braket{34}\braket{45}\braket{56}\braket{61}}
\times\nonumber\\
&\Bigg(\!
-\trp(123456)I^D\bigg(\usegraph{12}{hex}\bigg)[\mu^6]
+\frac{\trp(1456)}{s_{14}}I^D\bigg(\!\usegraph{12}{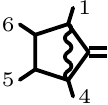}\bigg)[\mu^4]\label{eq:6pt1lallplus}\\
&\qquad
+\frac{\trp(13(4\!+\!5)6)}{s_{13}}I^D\bigg(\!\usegraph{12}{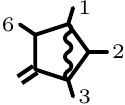}\!\bigg)[\mu^4]
+\frac{\trp(245(6\!+\!1))}{s_{24}}I^D\bigg(\!\usegraph{12}{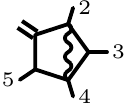}\!\bigg)[\mu^4]\nonumber\\
&\qquad
-s_{12}s_{61}I^D\bigg(\usegraph{10}{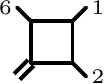}\bigg)[\mu^4]
-s_{23}s_{345}I^D\bigg(\,\usegraph{10}{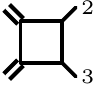}\bigg)[\mu^4]
-s_{34}s_{45}I^D\bigg(\usegraph{10}{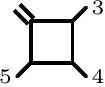}\bigg)[\mu^4]\Bigg),\nn
\end{align}
\end{subequations}
both of which include parity-odd terms.  The hexagon integral, being a $D$-dimensional contribution, is not present in the four-dimensional $\cN=4$ local integrand presentation.  We obtained its value from the $D$-dimensional presentation of the (parity-even) part of the same $\cN=4$ amplitude given in ref.~\cite{Bern:2008ap}.  Upon integration, we have checked that the above expressions agree with eq.~(\ref{eq:oneloopintegrated}).  The main difference here is that, whereas the representations given in eqs.~(\ref{eq:1lbern}) made use of $\mu^6$ pentagons, we instead use Dirac traces --- this keeps power counting of loop momentum within the expectations of pure Yang-Mills theory.  Besides the hexagon and box integrals~(\ref{eq:1lintegrals}) we also need
\begin{align}
I^D\bigg(\usegraph{13}{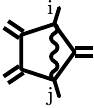}\bigg)[\mu^4]=-\frac{i}{6(4\pi)^2}s_{ij}+\cO(\eps).
\end{align}
Spurious poles in $s_{ij}$ associated with pentagons all cancel.

One can also infer this cancellation of spurious poles, and therefore the locality of the amplitudes, to all orders in $\epsilon$ by considering the unintegrated expressions.  For instance, in the pentagon integral from eq.~(\ref{eq:5pt1lallplus}) the identity (\ref{eq:trstacking}) gives
\begin{align}
\trp(1(\ell-p_1)(\ell-p_{12})345)=\frac{\trp(1345)}{s_{13}}\trp(1(\ell-p_1)(\ell-p_{12})3).
\end{align}
Similar relationships are applicable to the six-point pentagon integrals.  Therefore the integrands are all local despite there being unphysical kinematic variables in the denominators.

\section{Review: \texorpdfstring{$D$}{D}-dimensional unitarity}\label{sec:unitarity}

So far we have not reviewed the methods used to find two-loop all-plus integrands; in the previous chapter we merely discussed their physical properties at four and five points.  As our intention is to find new two-loop results we will now outline the relevant techniques; these are applicable to both this and the next chapter.

In this thesis we use a multi-loop integrand-reduction algorithm \cite{Mastrolia:2011pr,Zhang:2012ce,Mastrolia:2012an,Badger:2013gxa,Badger:2013sta,Zhang:2016kfo} which uses multivariate polynomial division to find integrand representations of two-loop amplitudes.  The algorithm was successfully used by BFZ to derive their planar five-gluon two-loop all-plus integrand~(\ref{eq:2L5gbfz}).  This section is intended as a brief overview of the approach; we encourage the reader to refer to the literature for more detailed information.

\subsection{Integrand reduction}

A generic contribution to a planar loop amplitude may be expressed, by means of integrand reduction, as a sum of irreducible integrands:
  \begin{align}\label{eq:intreductionbasics}
    \frac{\mathcal{P}(p_i,\ell_i,\mu_i)}
{\prod_{\alpha} \cQ_{\alpha}(p_i, \ell_i)} = \sum_{T} \frac{\Delta_{T}(p_i,\ell_i)}{\prod_{\alpha\in
T} \cQ_{\alpha}(p_i,\ell_i)},
  \end{align}
where the sum $T$ runs over all the subtopologies of the parent topology.  To achieve this, one expands the numerator $\mathcal{P}$ into a (possibly vanishing) irreducible component, $\Delta$, and terms multiplying the propagators $\cQ_\alpha$ of topology $T$:
\begin{align}
\mathcal{P}(p_i,\ell_i,\mu_i)
=\Delta(p_i,\ell_i)+
\sum_T\kappa_T(p_i)\left(\prod_{\alpha\not\in T}\cQ_{\alpha}(p_i,\ell_i)\right).
\end{align}
The procedure is not unique, and the notion of irreducibility will be more carefully explained below.  The terms proportional to propagators may then be cancelled against those same terms in the denominator.  By proceeding in this way through all subtopologies, one arrives at the decomposition (\ref{eq:intreductionbasics}).  Of course, some of the irreducible numerators $\Delta_T$ obtained in this way may simply be zero.

This prescription could of course be applied to a colour-ordered Feynman-diagram expansion of the integrand.  However, it is often more convenient to determine irreducible numerators $\Delta_T$ directly from generalised unitarity cuts.  One evaluates the integrand on values of the loop momenta satisfying the multiple-cut conditions $\{\cQ_\alpha = 0, \alpha \in T\}$.  Explicitly, (\ref{eq:intreductionbasics}) is rearranged into
  \begin{align}
    \Delta_{T}(p_i,\ell_i) =     \left(\frac{\mathcal{P}(p_i,\ell_i,\mu_i)}
{\prod_{\beta \not \in T} \cQ_{\beta}(p_i,\ell_i)} - \sum_{T' \supset T}
\frac{\Delta_{T'}(p_i,\ell_i)}{\prod_{\beta\in T' \setminus T} \cQ_{\beta}(p_i,\ell_i)}\right),
\quad {\textrm {if } \alpha \in T, \cQ_\alpha = 0}.
  \end{align}

Making the distinction between the cut associated with a graph $T$ and the irreducible numerator associated with the same graph is crucial for understanding this construction. The irreducible numerator contains only that information which is required on the cut associated with $T$, and which is not captured by irreducible numerators of graphs $T'$ that are ``larger'' than $T$, in the sense that the propagators contained in $T'$ are a proper superset of the propagators contained in $T$. In other words, by applying the cuts in a top-down approach we can isolate each topology systematically subtracting the higher-point singularities.

With the propagators on shell, the cut integrand $\mathcal{P}$ factorises into a product of tree-level amplitudes. These tree-level amplitudes must be evaluated in $\mathcal{D}>4$ in order to extract the $\mu_{ij}$ terms.  At two loops the minimum embedding dimension is six so we make use of the six-dimensional spinor-helicity formalism~\cite{Cheung:2009dc,Bern:2010qa,Davies:2011vt}.  The $\cD=6$ dimensional cuts can be dimensionally reduced to the tHV or FDH schemes by considering additional scalar loops~\cite{Bern:2002zk}. We include $D_s-\cD$ contributions with a single scalar loop and $(D_s-\cD)^2$ contributions with two scalar loops:\footnote{The analogous one-loop statement is that $\Delta_T=\Delta_T^{(g,\cD)}+(D_s-\cD)\Delta_T^{(s,\cD)}$, where the minimum embedding dimension is $\cD=5$.}
\begin{align}
  \Delta_T = \Delta_T^{(g,\cD)} + (D_s-\cD)\, \Delta_T^{(s,\cD)} + (D_s-\cD)^2\, \Delta_T^{(s^2,\cD)}.
\end{align}
The full reduction procedure starts from the top-level topology and recursively proceeds to lower topologies. At each step, the previously-computed irreducible numerators are used to remove poles appearing in the cut numerator.

The external kinematics can be conveniently parameterised in terms of momentum-twistor
variables~\cite{Hodges:2009hk}. In our implementation the six-dimensional spinors are evaluated directly
in terms of the explicit parameterisation given in ref.~\cite{Badger:2016uuq}. In the case of the six-gluon tree-level
amplitudes appearing in this work, we obtained the results through BCFW recursion relations.\footnote{We are grateful to Christian Br\o{}nnum-Hansen for providing his Mathematica code for the evaluation of 6D amplitudes via BCFW recursion.}  This approach is particularly convenient since it can be applied both numerically or analytically.  We have made use of the ability to evaluate using rational numerics to reconstruct the full analytic form of the cuts in cases where factorisation of intermediate polynomials became computationally expensive.

\subsection{Irreducible numerators}\label{sec:irreduciblereview}

For a kinematic numerator $\Delta_{T}(p_i,\ell_i)$ of some diagram topology $T$ to be considered irreducible, it must be expressible polynomially in terms of a set of scalar products $(x_1,x_2,\ldots,x_n)$ of loop momenta known as irreducible scalar products (ISPs):
\begin{align} \label{eq:deltaparametrization}
\Delta_{T}(p_i,\ell_i)
=\sum_{i_1,\ldots,i_n}c_{i_1i_2\cdots i_n}x_1^{i_1}x_2^{i_2}\cdots x_n^{i_n},
\end{align}
where the upper limits in this sum remain to be determined.  A set of ISPs may be chosen for each topology individually: their defining property is that it is not possible to express them polynomially in terms of the propagators of that topology.  The only other requirements are that (i) they should be linearly independent and (ii) there should be enough of them to capture all possible loop-momentum dependence.

The ISPs will, however, satisfy quadratic relationships between themselves and the propagators.  These relationships can be derived using  $5\times5$ Gram matrices: the determinants of these matrices vanish when acting on strictly four-dimensional vector arguments.  One should insert various combinations of external momenta $p_i$, orthogonal directions $\omega_{ijk}$ and four-dimensional components of loop momenta $\bar{\ell_i}$ to obtain all independent relationships.

The monomial coefficients of $c_{i_1i_2\cdots i_n}$ should be linearly independent, so these quadratic relationships between ISPs must be taken into account when determining the limits of the sum in eq.~(\ref{eq:deltaparametrization}).  A useful technique for determining a valid set of limits is polynomial division with respect to a Gr\"{o}bner basis of the quadratic relations.  While we will not review this method here, a good introduction may be found in ref.~\cite{Zhang:2016kfo}; in ref.~\cite{Zhang:2012ce} the strategy was developed into the public code \textsc{BasisDet}.

As explained above, generalised unitarity allows us to evaluate irreducible numerators on a set of generalised unitarity cuts, in which case all the propagators of the topology in question are set to zero: $\cQ_\alpha(p_i,\ell_i)=0$ for $\alpha\in T$.  In this case, the remaining degrees of freedom can be parametrised through a set of parameters $\tau_i$.  On the cut~\cite{Badger:2012dp,Badger:2013gxa}
\begin{align}\label{eq:tauparametrisation}
\left.\Delta_{T}(p_i,\ell_i)\right|_\text{cut}
=\sum_{j_1,\ldots,j_n}d_{j_1j_2\cdots j_n}\tau_1^{j_1}\tau_2^{j_2}\cdots\tau_n^{j_n}.
\end{align}
Inserting the constrained loop momenta into eq.~(\ref{eq:deltaparametrization}) then allows us to set up a linear relationship between the coefficients,
\begin{align}\label{eq:coeffmatrix}
\mathbf{d}=M\,\mathbf{c},
\end{align}
where $\mathbf{c}$ and $\mathbf{d}$ are vectors of the coefficients given in eqs.~(\ref{eq:deltaparametrization}) and (\ref{eq:tauparametrisation}) respectively.  As shown in \cite{Badger:2013gxa}, there is only ever a single branch to the set of solutions to the on-shell equation in $D$ dimensions, which simplifies the inversion of the system in \eqn{eq:coeffmatrix} to find the coefficients $c_{i_1i_2\cdots i_n}$.

To illustrate these points, consider the one-loop box~(\ref{eq:boxintegral}) for some arbitrary external helicity configuration.  Scalar products of the form $\ell^2$ and $\ell\cdot p_i$ are not ISPs as we can write them as linear combinations of propagators.  For instance, $\ell\cdot p_1=\frac{1}{2}\ell^2-\frac{1}{2}(\ell-p_1)^2$.  We refer to these objects as reducible scalar products (RSPs).  The only other possible dependence on loop momentum is through $\mu^2$ and $\ell\cdot\omega_{123}$, where $\omega_{123}$ is the spurious direction orthogonal to the external momenta~(\ref{omega}).  They are linearly independent, and therefore valid ISPs.

These two ISPs do, however, satisfy a quadratic relationship:
\begin{align}\label{eq:onelooplinearcombo}
(\ell\cdot\omega_{123})^2\!-\!\mu^2=
\text{linear combination of $\ell^2$, $(\ell\!-\!p_1)^2$,
$(\ell\!-\!p_{12})^2$, $(\ell\!+\!p_4)^2$}
\end{align}
because only five degrees of freedom exist in the $D$-dimensional loop momentum $\ell$: the four propagators and two ISPs must therefore be related.  The precise form of eq.~(\ref{eq:onelooplinearcombo}) follows from the vanishing of a Gram determinant:
\begin{align}
\text{det}\,G\left(
\begin{matrix}
p_1 & p_2 & p_3 & \omega_{123} & \bar{\ell}\,\,\\
p_1 & p_2 & p_3 & \omega_{123} & \bar{\ell}\,\,
\end{matrix}
\right)=0,
\end{align}
where the four-dimensional part of the loop momentum $\bar{\ell}$ is eliminated using $\bar{\ell}^2=\ell^2+\mu^2$ and $\bar{\ell}\cdot v=\ell\cdot v$ ($v$ is any four-dimensional vector).  So we can write
\begin{align}\label{eq:boxdecompexample}
\Delta\bigg(\usegraph{10}{box}\bigg)=
c_{00}+c_{01}\ell\cdot\omega_{123}+c_{10}\mu^2+c_{11}\mu^2\ell\cdot\omega_{123}+c_{20}\mu^4
\end{align}
if we assume that loop-momentum power counting is capped at four powers.  All monomials are linearly independent.

It remains to determine the coefficients $c_{ij}$.  So now let us suppose that we know the box numerator on its maximal cut.  We parametrise loop-momentum dependence on a single branch of solutions to the multiple cut conditions:
\begin{align}
\bar{\ell}^\mu=
\tau\frac{\braket{12}}{\braket{24}}\frac{\left<4|\gamma^\mu|1\right]}{2}-
(1+\tau)\frac{[12]}{[24]}\frac{\left<1|\gamma^\mu|4\right]}{2}, &&
\mu^2=\frac{st}{u}\tau(1+\tau),
\end{align}
where the four cut conditions $\ell^2=(\ell-p_1)^2=(\ell-p_{12})^2=(\ell+p_4)^2=0$ have left one degree of freedom, $\tau$.  Notice that taking the limit $D\to4$, which involves sending $\mu^2\to0$, breaks this solution into two branches: $\tau=0$ and $\tau=-1$.  We find that $\ell\cdot\omega_{123}=\frac{1}{2}t(1+2\tau)$, so (\ref{eq:boxdecompexample}) can be decomposed on this cut into
\begin{align}
\left.\Delta\bigg(\usegraph{10}{box}\bigg)\right|_\text{cut}=
\sum_{i=0}^4d_i\tau^i
\end{align}
where, for instance, $d_0=c_{00}+\frac{1}{2}tc_{01}$.  Having determined this expression from on-shell data, we can invert to find the coefficients $c_{ij}$.

This simple example exposes a drawback of using irreducible numerators: they are still to some extent arbitrary.  Different choices of ISPs give different, but equally valid, sets of $\Delta$s.  In this thesis we explore how specific choices of ISPs, satisfying additional properties to the ones described above, allow us to expose physical properties of the amplitudes.

\subsection{Local integrands as ISPs}

In this chapter we will use the local integrand structures introduced in the previous section as a partial basis of ISPs.  When deciding what specifically to use for a given topology we follow a set of guidelines:
\begin{itemize}
\item the infrared pole structure of the amplitude should follow from its integrand
  representation,
\item the integrand should not contain spurious singularities with
  respect to the external invariants,
\item an $n$-point integrand should match the $(n-1)$-point result
  when taking soft limits of the external particles.
\end{itemize}
We will show how the infrared poles can be extracted from the integrand in section \ref{sec:softlimits}. As we shall see, this leads to an integrand form for the all-plus amplitudes presented here with a significantly lower number of non-vanishing terms compared to the ones presented in chapter~\ref{ch:review}.

\section{Two-loop local integrands}\label{sec:twoloops}

In this section we present the leading-colour parts of the two-loop all-plus five- and six-gluon scattering amplitudes.  While the five-gluon numerators were already written down in eqs.~(\ref{eq:2L5gbfz}), here we show how a local-integrand presentation allows us to write this result more compactly and eliminate unphysical poles before integration.  The same functions of extra-dimensional components $F_1$, $F_2$ and $F_3$, given in eqs.~(\ref{eq:F1def}), (\ref{eq:F2def}) and (\ref{eq:F3def}) respectively, are still useful for writing these results.

\subsection{The five-gluon integrand}

The five-point numerators are
\begin{subequations}\label{eq:2L5g}
\begin{align}
\Delta\bigg(\usegraph{9}{431}\bigg)
&=\frac{s_{45}\trp(1(\ell_1-p_1)(\ell_1-p_{12})345)F_1}{\SpDenom5},
\label{eq:431}\\
\Delta\bigg(\usegraph{13}{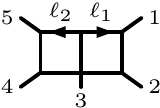}\bigg)
&=-\frac{s_{12}s_{45}s_{15}F_1}{\SpDenom5},\label{eq:3315L}\\
\Delta\bigg(\usegraph{13}{430}\!\bigg)
&=\frac{\trp(1(\ell_1-p_1)(\ell_1-p_{12})345)}{\SpDenom5}
\left(F_2+F_3\frac{s_{45}+(\ell_1+\ell_2)^2}{s_{45}}\right),
\label{eq:430}\\
\Delta\bigg(\usegraph{13}{3305L}\bigg)
&=\frac{1}{\SpDenom5}\label{eq:3305L}\\
&\qquad\times\Bigg(\trp(1245)(F_2+F_3)+\frac{F_3}{s_{12}s_{45}}
\bigg(\trp(123\ell_1\ell_2345)
\nonumber\\&\qquad\qquad
+(s_{12}s_{45}s_{15}+(s_{12}+s_{45})\trp(1245))(\ell_1+\ell_2)^2\bigg)\Bigg),\nn
\end{align}
\end{subequations}
of which the first two are genuine two-loop topologies.  Two of the numerators given in section~\ref{sec:2lreview} are now zero: these are the one-mass double box and one-mass double triangle, given in eqs.~(\ref{eq:331m1bfz}) and (\ref{eq:330m1bfz}) respectively.  Their contributions have been absorbed into other topologies.

In this new presentation, both genuine two-loop topologies match their counterparts from the $\cN=4$ all-loop integrand.  The procedure for translating these supersymmetric expressions to the $D$-dimensional language used above is the same as that outlined in section \ref{sec:2lreview} and elaborated on in appendix \ref{sec:N4localintegrands}; we see that the supersymmetric delta function $\delta^8(Q)$ is indeed now replaced with the $D$-dimensional prefactor $F_1$.  However, we also notice that, even when there are no supersymmetric counterparts to the diagrams, the same local integrands involving Dirac $\trp$ objects continue to be a useful means of expressing loop-momentum dependence.

When forming integrated expressions the rearrangement of Dirac $\trp$ objects demonstrated in eq.~(\ref{eq:trstacking}) allows the use of regulated pentabox and box-triangle integrals:
\begin{subequations}
\begin{align}
&I^D\left[\Delta\bigg(\usegraph{9}{431}\bigg)\right]
=\frac{s_{45}\trp(1345)}{\SpDenom5s_{13}}I^D\bigg(\usegraph{9}{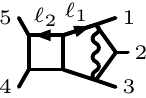}\bigg)[F_1],\\
&I^D\left[\Delta\bigg(\usegraph{13}{430}\!\bigg)\right]
=\frac{1}{\SpDenom5}\\&\qquad
\times\Bigg(\frac{\trp(1345)}{s_{13}}I^D\bigg(\usegraph{13}{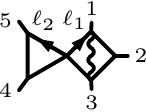}\!\bigg)[F_3]
+\frac{\trp(1345)}{s_{13}s_{45}}I^D\bigg(\usegraph{13}{430s}\!\bigg)[F_3(\ell_1+\ell_2)^2]\Bigg).\nn
\end{align}
\end{subequations}
The two-loop ``wavy line'' notation used here follows precisely the same one-loop conventions introduced in eqs.~(\ref{eq:localboxintegral}) and (\ref{eq:regulatedpentagon}).

As a final remark on this local representation we notice that the soft limits of the irreducible numerators match directly onto the four-point numerators.  For example,
\begin{subequations}\label{eq:5ptsoftlimits}
\begin{align}
\Delta\bigg(\usegraph{13}{430}\!\bigg)
&\xrightarrow{p_2\to 0}\Delta\bigg(\usegraph{10}{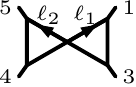}\!\bigg),\\
\Delta\bigg(\usegraph{13}{3305L}\!\bigg)
&\xrightarrow{p_3\to 0}\Delta\bigg(\usegraph{10}{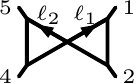}\!\bigg),
\end{align}
\end{subequations}
which can be checked by explicit evaluation and comparison with~(\ref{eq:330bern}).

\subsection{The six-gluon integrand}

With six gluons scattering there are twelve nonzero topologies.  The six genuine two-loop topologies are
\begin{subequations}\label{eq:2L6ga}
\begin{align}
\Delta\bigg(\!\usegraph{10}{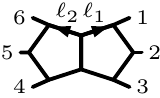}\!\bigg)
&=\frac{s_{123}\trp(1(\ell_1-p_1)(\ell_1-p_{12})34(\ell_2-p_{56})(\ell_2-p_6)6)F_1}
{\braket{12}\braket{23}\braket{34}\braket{45}\braket{56}\braket{61}},
\label{eq:441}\\
\Delta\bigg(\usegraph{10}{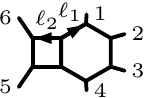}\!\bigg)
&=-\frac{s_{56}\trp(123456)\mu_{11}F_1}
{\braket{12}\braket{23}\braket{34}\braket{45}\braket{56}\braket{61}},\label{eq:531}\\
\Delta\bigg(\usegraph{10}{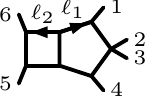}\!\bigg)
&=\frac{s_{56}\trp(1(\ell_1-p_1)(\ell_1-p_{123})456)F_1}
{\braket{12}\braket{23}\braket{34}\braket{45}\braket{56}\braket{61}},\label{eq:431M2}\\
\Delta\bigg(\usegraph{10}{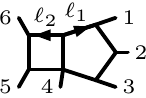}\!\bigg)
&=\frac{s_{56}\trp(1(\ell_1-p_1)(\ell_1-p_{12})3(4\!+\!5)6)F_1}
{\braket{12}\braket{23}\braket{34}\braket{45}\braket{56}\braket{61}},\label{eq:4316L}\\
\Delta\bigg(\usegraph{10}{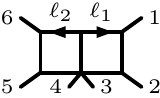}\bigg)&=-\frac{s_{12}s_{56}s_{61}F_1}
{\braket{12}\braket{23}\braket{34}\braket{45}\braket{56}\braket{61}},\label{eq:3315LM}\\
\Delta\bigg(\usegraph{13}{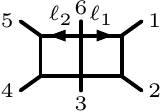}\bigg)&=-\frac{s_{12}s_{45}s_{234}F_1}
{\braket{12}\braket{23}\braket{34}\braket{45}\braket{56}\braket{61}}.\label{eq:3316L}
\end{align}
\end{subequations}
With the exception of the hexagon box (\ref{eq:531}) these agree with their supersymmetric counterparts.  The hexagon box is a manifestly $D$-dimensional contribution; to obtain its value we again referred to the $D$-dimensional presentation of the (parity-even) part of the supersymmetric amplitude given in ref.~\cite{Bern:2008ap}.  The six-leg pentabox (\ref{eq:4316L}) has a counterpart related by symmetry through the horizontal axis, an expression for which can determined by relabelling the one given above.

The one-loop-squared numerators are
\begingroup
\allowdisplaybreaks
\begin{subequations}
\begin{align}
\Delta\bigg(\!\usegraph{9}{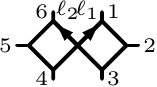}\!\bigg)
&=\frac{\trp(1(\ell_1-p_1)(\ell_1-p_{12})34(\ell_2-p_{56})(\ell_2-p_6)6)}
{\braket{12}\braket{23}\braket{34}\braket{45}\braket{56}\braket{61}}\times\nonumber\\
&\qquad
\left(F_2+F_3\frac{s_{123}+(\ell_1+\ell_2)^2}{s_{123}}\right),
\label{eq:440}\\
\Delta\bigg(\usegraph{10}{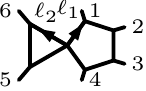}\!\bigg)
&=-\frac{\trp(123456)\mu_{11}}
{\braket{12}\braket{23}\braket{34}\braket{45}\braket{56}\braket{61}}
\left(F_2+F_3\frac{s_{56}+(\ell_1+\ell_2)^2}{s_{56}}\right),
\\
\Delta\bigg(\usegraph{10}{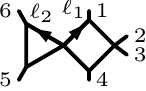}\!\bigg)
&=\frac{\trp(1(\ell_1-p_1)(\ell_1-p_{123})456)}
{\braket{12}\braket{23}\braket{34}\braket{45}\braket{56}\braket{61}}
\left(F_2+F_3\frac{s_{56}+(\ell_1+\ell_2)^2}{s_{56}}\right),
\\ 
\Delta\bigg(\usegraph{10}{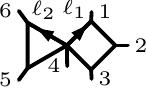}\!\bigg)
&=\frac{\trp(1(\ell_1-p_1)(\ell_1-p_{12})3)}
{\braket{12}\braket{23}\braket{34}\braket{45}\braket{56}\braket{61}s_{13}}
\bigg(\text{tr}_+(1356)(F_2+F_3)\nn\\
&\qquad+\frac{F_3}{s_{123}s_{56}}\bigg(\trp(134\ell_1\ell_2456)\label{eq:4306L}\\
&\qquad\qquad\qquad\qquad
+\big(s_{123}\trp(1356)-s_{56}\trp(1346)\big)(\ell_1+\ell_2)^2\bigg)\bigg),
\nn\\
\Delta\bigg(\usegraph{10}{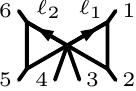}\!\bigg)
&=\frac{1}{\braket{12}\braket{23}\braket{34}\braket{45}\braket{56}\braket{61}}\bigg(
\trp(1256)(F_2+F_3)+\frac{F_3}{s_{12}s_{123}s_{56}}\times\nn\\
&\qquad\bigg(-s_{12}\trp(134\ell_1\ell_2456)-s_{56}\trp(123\ell_1\ell_2346)\nn\\
&\qquad\qquad+s_{123}\trp(12(3\!+\!4)\ell_1\ell_2(3\!+\!4)56)
+s_{34}\mu_{12}\trp(123456)\nonumber\\
&\qquad\qquad
+\left(s_{34}+s_{123}\frac{\trp(2345)}{s_{23}s_{45}}\right)\trp(123\ell_1\ell_2456)
\nonumber\\
&\qquad\qquad
+\Big(s_{12}s_{56}\trp(1346)-s_{123}s_{34}\trp(1256)\\
&\qquad\qquad\,\,\,\,\,\,\,
+s_{123}(s_{12}s_{56}s_{16}+(s_{12}+s_{56})\trp(1256))\Big)(\ell_1+\ell_2)^2\bigg)\bigg),
\nn\\
\Delta\bigg(\usegraph{13}{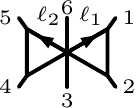}\!\bigg)
&=\frac{1}{\braket{12}\braket{23}\braket{34}\braket{45}\braket{56}\braket{61}}\bigg(
\trp(1245)(F_2+F_3)+\frac{F_3}{s_{12}s_{45}s_{123}s_{345}}\times\nn\\
&\qquad\bigg(
f_0(123456;\ell_1-p_1,\ell_2-p_5)+f_1(123456;\ell_1-p_1)\nn\\
&\qquad-f_1(216543;\ell_1-p_1)-f_1(456123; \ell_2-p_5)\nn\\
&\qquad+f_1(543216; \ell_2-p_5)+f_2(123456; \ell_1-p_1,\ell_2-p_5)\nn\\
&\qquad+f_2(456123;\ell_2-p_5,\ell_1-p_1)+f_3(123456;\ell_1-p_1,\ell_2-p_5)\,s_{123}\nn\\
&\qquad+f_3(456123; \ell_1-p_1,  \ell_2-p_5)\, s_{345}
+(\ell_1-p_1)^2\,f_4(123456)\nn\\
&\qquad+ (\ell_2-p_5)^2\, f_4(456123)\label{eq:3306L}
 \bigg)\bigg).
\end{align}
\end{subequations}
\endgroup
The $s_{13}$ pole in the six-leg box triangle (\ref{eq:4306L}) can be removed using trace identities --- we write the expression this way for compactness.  The graph has a counterpart related by symmetry through the horizontal axis.  The expression for the six-leg double triangle (\ref{eq:3306L}) is somewhat less compact: the functions $f_i$ can be expressed as
\begin{align}
 f_0(123456;\ell_1,\ell_2) ={}&-s_{123} s_{345} \trp(1245)^2
+ 2 (\ell_1\cdot \ell_2) \Big(s_{12} s_{123} s_{345} \trp(1245)\nn\\&+s_{45} s_{123} s_{345} \trp(1245)+s_{12} s_{45} s_{345} \trp(1346)\nn\\&-s_{345} \trp(1236) \trp(4563)-s_{123} \trp(1236) \trp(4563)\nn\\&+s_{12} s_{45} s_{123} \trp(5326)+s_{12} s_{45} s_{123} s_{234} s_{345}\Big)
\nonumber \\
&-\mu_{12} (s_{123}+s_{345}) \trp(1236) \trp(4563),\\
f_1(123456,\ell) = {}&-s_{123} (s_{12} s_{45} s_{56} \trp(\ell\, 234)+s_{12} s_{34} \trp(\ell\, 23654)\nn\\&+s_{12} s_{345} \trp(\ell\, 24654)+s_{345} \trp(\ell\, 2451245)),
\\
f_2(123456;\ell_1,\ell_2)={}&s_{123} s_{345} \trp(123\, \ell_1\, \ell_2\, 345)-s_{45} s_{345} \trp(123\, \ell_1\, \ell_2\, 346)\nn\\&-s_{12} s_{123} \trp(623\, \ell_1\, \ell_2\, 345),
\\
f_3(123456;\ell_1,\ell_2)={}&\trp(123\, \ell_1\, \ell_2\, 65436),
\\
f_4(123456)={}&s_{123} (s_{45} s_{345} \trp(1245)-s_{12} s_{45} \trp(2653)+s_{12} s_{345} \trp(4563)\nn\\&+s_{12} s_{45} \trp(5326)).
\end{align}
This integrand obeys the soft limits on $p_3$ and $p_6$:
\begin{subequations}
\begin{align}
  &\Delta\bigg(\usegraph{13}{3306L}\!\bigg)\xrightarrow{p_3\to 0}
  \Delta\bigg(\usegraph{9}{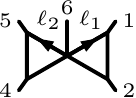}\!\bigg), \\&
  \Delta\bigg(\usegraph{13}{3306L}\!\bigg)\xrightarrow{p_6\to 0}
  \Delta\bigg(\usegraph{13}{3305L}\!\bigg),
\end{align}
\end{subequations}
which can be checked using spinor algebra.

\section{Infrared pole structure}\label{sec:ir}

As discussed in section~\ref{sec:irreview}, since the all-plus tree amplitudes are zero the universal IR structure should be that of a one-loop amplitude:
\begin{align}\label{eq:ircatani}
&A^{(2)}(1^+,2^+,\cdots,n^+)\\
&\qquad
=i\,\sum_{i=1}^ns_{i,i+1}I^D\bigg(\!\usegraph{9}{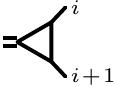}\bigg)
A^{(1)}(1^+,2^+,\cdots,n^+)+F^{(2)}(1^+,2^+,\cdots,n^+),\nn
\end{align}
where $F^{(2)}$ is finite in the limit $\eps\to0$.  This is precisely the same as eq.~(\ref{eq:orderedcatani}) --- we have simply written it in a more suggestive form using the triangle integral~(\ref{eq:triintegral}).

Reproducing this behaviour requires us to find the IR divergences up to $\cO(\epsilon^{-1})$ in all of our two-loop integrals, like the decomposition we gave for the double box in eq.~(\ref{eq:doubleboxirreview}).  As we shall see, the two-loop integrals can only contain a single soft divergence, rather than a double-soft divergence.  This enables us to break them into sums of regions with soft singularities and evaluate them in the various limits.  The integrals then factorise into products of one-loop integrals, which it is unnecessary for us to evaluate explicitly.  Local integrands are a crucial element: using them enables us to get the subleading $1/\epsilon$ poles as well as the leading $1/\epsilon^2$.

\subsection{Soft limits of two-loop integrals \label{sec:softlimits}}

All IR-divergent integrals in our two-loop amplitudes come from topologies containing
\begin{align}\label{eq:F1}
F_1=(D_s-2)(\mu_{11}\mu_{22}+(\mu_{11}+\mu_{22})^2
+2\mu_{12}(\mu_{11}+\mu_{22}))+16(\mu_{12}^2-\mu_{11}\mu_{22}).
\end{align}
As the external momenta $p_i$ live in four dimensions, going into any soft region requires taking the $(-2\epsilon)$-dimensional part of one of the loop momenta $\ell^{[-2\epsilon]}_i\to0$.  In this limit
\begin{align}
F_1\xrightarrow{\ell^{[-2\epsilon]}_1\to0}(D_s-2)\mu_{22}^2, &&
F_1\xrightarrow{\ell^{[-2\epsilon]}_2\to0}(D_s-2)\mu_{11}^2.
\end{align}
Collinear limits also require $\ell^{[-2\epsilon]}_i\to0$.  The $F_1$ numerator therefore prevents any divergences beyond $\cO(\epsilon^{-2})$ as only one of the loop momenta can enter a soft or collinear region at a time without $F_1$ vanishing.

Again taking the example of the double box, we find two soft regions by taking the limit of either loop.  We find a soft singularity whenever we have two adjacent massless legs in one of the loop integrations.  In each case, we factorise into an IR-divergent triangle and a dimension-shifted box:
\begin{subequations}
\begin{align}
I^D\bigg(\usegraph{10}{331}\bigg)[F_1]
\xrightarrow{\ell_1\to p_1}(D_s-2)I^D\bigg(\usegraph{10}{boxi}\bigg)[\mu^4]
I^D\bigg(\!\usegraph{9}{tri12}\bigg),\\
I^D\bigg(\usegraph{10}{331}\bigg)[F_1]
\xrightarrow{\ell_2\to p_4}(D_s-2)I^D\bigg(\usegraph{9}{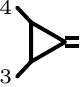}\!\bigg)
I^D\bigg(\usegraph{10}{boxi}\bigg)[\mu^4].
\end{align}
\end{subequations}
Using the triangle integral given in eq.~(\ref{eq:triintegral}) and summing these two regions, we see that this reproduces the known result given in eq.~(\ref{eq:doubleboxirreview}), including the subleading $\eps^{-1}$ terms.

Now proceeding to five external legs, we notice that the introduction of local integrands renders many seemingly soft regions finite.  Taking the regulated pentabox integral as an example, we notice that the integral has only one soft region: $\ell_2\to p_5$.  In this limit,
\begin{align}\label{eq:431ir}
I^D\bigg(\usegraph{10}{431s}\bigg)[F_1]
\xrightarrow{\ell_2\to p_5}(D_s-2)I^D\bigg(\usegraph{9}{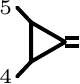}\!\bigg)
I^D\bigg(\!\usegraph{12}{pents}\!\bigg)[\mu^4].
\end{align}
The other two supposedly soft limits $\ell_1\to p_1$ and $\ell_1\to p_{12}$ are finite as the numerator $\trp(1(\ell_1-p_1)(\ell_1-p_{12})3)$ vanishes in these regions - this is the same phenomenon as we saw in the regulated box integral (\ref{eq:localboxintegral}).

Having numerically evaluated the regulated pentabox integral using FIESTA's numerical sector decomposition algorithm \cite{Smirnov:2013eza,Smirnov:2015mct}, we find that the decomposition (\ref{eq:431ir}) correctly predicts the IR structure at $\cO(\epsilon^{-1})$ as well as the leading $1/\epsilon^2$ pole.  The decomposition must therefore be accounting for collinear as well as soft singularities.  This property does not hold true for a scalar pentabox integral with its additional soft regions $\ell_1\to p_1$ and $\ell_1\to p_{12}$ - in that case the same technique correctly predicts only the $1/\epsilon^2$ pole.

The only other IR-divergent five-point integral is the five-leg double box.  This integral has two soft regions: $\ell_1\to p_1$ and $\ell_2\to p_5$.  The same procedure reveals that
\begin{align}\label{eq:3315Lir}
&I^D\bigg(\usegraph{13}{3315L}\bigg)[F_1]=(D_s-2)\times\nonumber\\
&\qquad\left(
I^D\bigg(\usegraph{9}{tri45}\!\bigg)I^D\bigg(\usegraph{10}{box34}\bigg)[\mu^4]+
I^D\bigg(\usegraph{10}{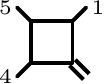}\bigg)[\mu^4]I^D\bigg(\!\usegraph{9}{tri12}\bigg)
\right)+\cO(\epsilon^0).
\end{align}

At six points, the genuine two-loop integrals are
\begin{subequations}\label{eq:6ptir}
\begin{align}
I^D\bigg(\!\usegraph{10}{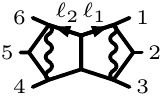}\!\bigg)[F_1]
&=\cO(\epsilon^0),\label{eq:441ir}\\
I^D\bigg(\usegraph{10}{531}\!\bigg)[\mu_{11}F_1]
&=(D_s-2)I^D\bigg(\usegraph{9}{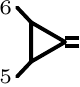}\!\bigg)I^D\bigg(\usegraph{12}{hex}\bigg)[\mu^6]
+\cO(\epsilon^0),\label{eq:531ir}\\
I^D\bigg(\usegraph{10}{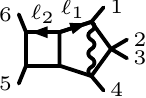}\!\bigg)[F_1]
&=(D_s-2)I^D\bigg(\usegraph{9}{tri56}\!\bigg)
I^D\bigg(\usegraph{12}{pent23s}\!\bigg)[\mu^4]+\cO(\epsilon^0),\label{eq:431M2ir}\\
I^D\bigg(\usegraph{10}{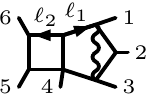}\!\bigg)[F_1]
&=(D_s-2)I^D\bigg(\usegraph{9}{tri56}\!\bigg)
I^D\bigg(\usegraph{12}{pent45s}\!\bigg)[\mu^4]+\cO(\epsilon^0),\label{eq:4316Lir}\\
I^D\bigg(\usegraph{9}{3315LM}\bigg)[F_1]
&=(D_s-2)\times\label{eq:3315LMir}\\
&\!\!\!\!\!\!\!\!\!\!\!\!\!\!\!\!\!\!\!\!\!\!\!\!\!\!\!\!\!\!\!\!\!\!\!\!\!\!\!\!\!\!\left(
I^D\bigg(\usegraph{9}{tri56}\!\bigg)I^D\bigg(\usegraph{10}{box345}\bigg)[\mu^4]
+I^D\bigg(\usegraph{10}{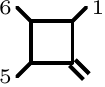}\bigg)[\mu^4]I^D\bigg(\!\usegraph{9}{tri12}\bigg)
\right)+\cO(\epsilon^0),\nn\\
I^D\bigg(\usegraph{13}{3316L}\bigg)[F_1]
&=(D_s-2)\times\label{eq:3316Lir}\\
&\!\!\!\!\!\!\!\!\!\!\!\!\!\!\!\!\!\!\!\!\!\!\!\!\!\!\!\!\!\!\!\!\!\!\!\!\!\!\!\!\!\!\left(
I^D\bigg(\usegraph{9}{tri45}\!\bigg)I^D\bigg(\,\usegraph{10}{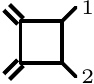}\bigg)[\mu^4]
+I^D\bigg(\usegraph{10}{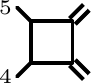}\,\bigg)[\mu^4]I^D\bigg(\!\usegraph{9}{tri12}\bigg)
\right)+\cO(\epsilon^0).\nn
\end{align}
\end{subequations}
where we have checked all topologies with 8 or fewer propagators using the sector decomposition algorithms implemented in FIESTA~\cite{Smirnov:2013eza,Smirnov:2015mct} and SecDec3.0 \cite{Borowka:2015mxa}. In the ``hexabox'' integral (\ref{eq:531ir}) the extra $\mu_{11}$ term regulates the $\ell_1$ loop.  In this sense it plays a role analogous to the $\trp$ ``wavy line'' structures used in the pentagon integrals.

\subsection{Integrand-level infrared structure}

The IR divergences in the two-loop integrals given above all arise from unregulated box subintegrals.  Therefore, as the butterfly-type integrals are all finite,
the $\epsilon$ poles all come from triangle integrals.  This observation explains our decision to rewrite the universal IR factorisation (\ref{eq:ircatani}) in terms of triangle integrals, and motivates a simple approach to verifying the universal IR structure.

In the five-point example we proceed as follows. First we take the full two-loop amplitude and substitute the IR divergences given in eqs.~(\ref{eq:431ir}) and (\ref{eq:3315Lir}), summing over cyclic permutations:
\begin{align}
&\braket{12}\braket{23}\braket{34}\braket{45}\braket{51}
A^{(2)}(1^+,2^+,3^+,4^+,5^+)\nonumber\\
&=i\,\sum_{\sigma\in Z_5}\Bigg\{
\frac{s_{45}\trp(1345)}{s_{13}}I^D\bigg(\usegraph{10}{431s}\bigg)[F_1]
-s_{12}s_{45}s_{15}I^D\bigg(\usegraph{13}{3315L}\bigg)[F_1]+\cdots\Bigg\}\nonumber\\
&=i(D_s-2)\sum_{\sigma\in Z_5}\Bigg\{
\frac{s_{45}\trp(1345)}{s_{13}}
I^D\bigg(\usegraph{9}{tri45}\!\bigg)I^D\bigg(\!\usegraph{12}{pents}\!\bigg)[\mu^4]\nn\\
&\qquad\!\!
-s_{12}s_{45}s_{15}
I^D\bigg(\!\usegraph{9}{tri12}\bigg)I^D\bigg(\usegraph{10}{box23}\bigg)[\mu^4]
-s_{12}s_{45}s_{15}
I^D\bigg(\usegraph{9}{tri45}\!\bigg)I^D\bigg(\usegraph{10}{box34}\bigg)[\mu^4]
\Bigg\}\nn\\
&\qquad+\cO(\epsilon^0).
\end{align}
Next, we exploit the sum on cyclic permutations to relabel the box integral with a massive $p_{23}$ leg:
\begin{align}
&\braket{12}\braket{23}\braket{34}\braket{45}\braket{51}
A^{(2)}(1^+,2^+,3^+,4^+,5^+)\nonumber\\
&=i(D_s-2)\sum_{\sigma\in Z_5}\Bigg\{s_{45}I^D\bigg(\usegraph{9}{tri45}\!\bigg)\Bigg(
\frac{\trp(1345)}{s_{13}}I^D\bigg(\!\usegraph{12}{pents}\!\bigg)[\mu^4]\\
&\qquad\qquad\qquad\qquad
-s_{23}s_{34}I^D\bigg(\usegraph{10}{box51}\bigg)[\mu^4]
-s_{12}s_{15}I^D\bigg(\usegraph{10}{box34}\bigg)[\mu^4]
\Bigg)\Bigg\}+\cO(\epsilon^0),\nn
\end{align}
where we have extracted an overall triangle integral inside the sum, together with $s_{45}$.  The term in brackets we recognise from eq.~(\ref{eq:5pt1lallplus}) as the planar five-gluon, one-loop all-plus amplitude (multiplied by some extra factors) --- this is of course invariant under cyclic permutations.  Rearranging, we arrive at the desired result:
\begin{align}
A^{(2)}(1^+,2^+,3^+,4^+,5^+)
=i\sum_{i=1}^5s_{i,i+1}I^D\bigg(\!\usegraph{9}{trig}\bigg)
A^{(1)}(1^+,2^+,3^+,4^+,5^+)+\cO(\eps^0),
\end{align}
which agrees with our expectation from eq.~(\ref{eq:ircatani}).

The six-gluon calculation is completely analogous.  By applying the IR singularities given in eqs.~(\ref{eq:6ptir}) to the integrated versions of the six-point numerators presented in eqs.~(\ref{eq:2L6ga}) one can reproduce the one-loop amplitude as presented in eq.~(\ref{eq:6pt1lallplus}).

\section{Rational terms}\label{sec:rational}

The integrated forms of the finite contributions to the leading-colour amplitudes are
\begin{align}\label{eq:A6}
F^{(2)}(1^+,2^+,\cdots,n^+)=(D_s-2)P_n^{(2)}+(D_s-2)^2R_n^{(2)}+\cO(\eps),
\end{align}
where $F^{(2)}$ was introduced in eq.~(\ref{eq:ircatani}).  The rational terms $R_n^{(2)}$ all come from the one-loop-squared topologies while the polylogarithmic terms $P_n^{(2)}$ come from the topologies shared with $\cN=4$.  These have been identified by Dunbar, Jehu and Perkins~\cite{Dunbar:2016cxp}.\footnote{We have not explicitly checked that the finite part our expressions match those obtained in ref.~\cite{Dunbar:2016cxp} owing to the complexity of the multi-scale two-loop integrals appearing.  The $(\cN=4) \times F_1$ property has been explicitly checked at five points and we have no reason to expect different behaviour at six points.}

The rational terms can be found using the one-loop-squared integrals listed in appendix \ref{app:2lintegrals}.  Inserting these expressions into our $D$-dimensional integrands we find that $R_5^{(2)}$ precisely matches the rational part of the full amplitude given in ref.~\cite{Gehrmann:2015bfy}.  At six points we find that $R_6^{(2)}$ can be written as\footnote{The integrands of eqs.~(\ref{eq:440}-\ref{eq:3306L}) should be combined together with the appropriate symmetry factors.  The complete expression is available in an ancillary file included in the \texttt{arXiv} submission of ref.~\cite{Badger:2016ozq}.}
\begin{align}
R_6^{(2)}=-\frac{1}{144}\frac{i(D_s-2)^2}{\braket{12}\braket{23}\braket{34}\braket{45}\braket{56}\braket{61}(4\pi)^{4-2\epsilon}}
\sum_{\sigma\in Z_6}\Big\{f_R(123456)+f_R(654321)\Big\},
\end{align}
which makes its cyclic and reversal symmetry manifest.  The function
$f_R$ can be written as a sum of contributions corresponding to
physical pole structures in the external invariants
\begin{align}
&f_R(123456)\nn\\
&=\frac{2 s_{23} s_{34} s_{45} \trp(1256)}{s_{12} s_{56} s_{123}}+
\frac{\trp(1236)}{s_{12}s_{123}}\bigg(-4s_{23}s_{34}+2\trp(1345)+6\trp(2345)\bigg)\nn\\
&\qquad+\frac{1}{s_{123}}\bigg(-12 s_{34}\trp(1236)-s_{34}\trp(1256)\nn\\
&\qquad\qquad+3(s_{12}+s_{34}+s_{56})\trp(1346)-16 s_{12} s_{16} s_{34}\bigg)\nn\\
&\qquad-\frac{\trp(1245)^2}{s_{12} s_{45}}
+\frac{\trp(1256)}{s_{12} s_{56}}\bigg(-2\trp(p_{14}345)-2\trp(1256)\bigg)\nn\\
&\qquad+\frac{1}{s_{12}}\bigg(2 (s_{16}-s_{34}+s_{45})\trp(1234)
+2(s_{23}-s_{34}+s_{45}-s_{123})\trp(1235)\nn\\
&\qquad\qquad+2(s_{23}+s_{45}-3 s_{123}) \trp(1245)\bigg)\nn\\
&\qquad-\frac{1}{4}s_{12}(59s_{23}-8 s_{34}-56 s_{45})
+\frac{1}{4} s_{123} (-4 s_{12}-4 s_{23}+39 s_{34}-40 s_{234})\nn\\
&\qquad+\frac{9}{4} \trp(1234)+\frac{35}{4} \trp(1235)+\frac{15}{4} \trp(1245).
\end{align}
We have checked that this expression satisfies all universal collinear limits and agrees with the computation of Dunbar and Perkins \cite{Dunbar:2016gjb}.

\section{Discussion}\label{sec:localintegranddiscussion}

In this chapter we have explored the use of $D$-dimensional local integrands as a means of obtaining compact analytic representations of multi-leg two-loop amplitudes. The integrands, introduced in studies of planar $\mathcal{N}=4$ supersymmetric Yang-Mills, were shown to be powerful tools in the context of dimensionally regulated amplitudes. The all-plus sector proved a useful testing ground: we found $D$-dimensional local representations of the five- and six-gluon amplitudes.

The representations benefit from highlighting certain physical properties.  The infrared structure is manifest at the integrand level and the integral coefficients are free of spurious poles.  This results in a reduction in the number of basis integrals, thereby constraining the overall analytic form of the amplitude.  The rational terms result from one-loop squared topologies and were obtained in a compact form --- at six points this is in agreement with the expressions obtained by Dunbar and Perkins using augmented BCFW recursion relations~\cite{Dunbar:2016gjb}.

Finding expressions for the one-loop-squared topologies, which depend on a six-gluon tree amplitude, proved the most
complicated part of the computation. Though it was possible to find a number of different local representations,
none were as compact as the other topologies. The extremely simple form of the integrated expression
suggests that the integrand expression could yet be improved further. An important additional
check on the integrated expression came from the known collinear limits. It may be that collinear
limits at the integrand level give additional information but would require the development of
additional technology.

General two-loop amplitudes are of course far more complicated than the all-plus amplitudes
considered here. Nevertheless, the techniques we have used should be applicable to the general case
as well. There remain many open questions however: for instance,
one would need to identify a complete basis of local integrands outside of the specific examples considered.

Finally, we note that the present study was restricted to planar amplitudes.  In the next chapter we will discuss subleading colour: we will see how the nonplanar sector can be connnected to the planar sector using colour-kinematics (BCJ) relations.  An interesting question that we shall return to later is whether the results presented here may be of use in identifying local representations of nonplanar amplitudes.
It would be interesting to see if all-plus amplitudes continue to connect with recent studies in $\cN=4$ supersymmetric Yang-Mills theory \cite{Arkani-Hamed:2014via,Arkani-Hamed:2014bca,Bern:2014kca,Bern:2015ple}.

\chapter{Colour-dressed representations}\label{ch:allplusfull}

\section{Introduction}\label{sec:introduction}

In this chapter we further develop our understanding of two-loop all-plus amplitudes by exploring the nonplanar sector.  In particular, we complete the integrand-level computation of the two-loop, five-gluon, all-plus helicity amplitude.

There are two major aspects to this work.  Firstly, in order to deal with the increase in complexity of the full-colour amplitude, we introduce a method to find compact colour decompositions that makes full use of the underlying Kleiss-Kuijf (KK) relations \cite{Kleiss:1988ne}.  This is similar to the previous treatment at one loop by Del Duca, Dixon and Maltoni (DDM), which we reviewed in section~\ref{sec:ddmoneloop}.  The method is fully generalisable to higher loops and alternative helicity configurations; indeed, it was recently described in more detail by Ochirov and Page~\cite{Ochirov:2016ewn}.

The second aspect is the determination of nonplanar cuts and corresponding off-shell irreducible numerators.  We find that all of the nonplanar cuts can be obtained from planar cuts using the tree-level BCJ relations~\cite{Bern:2008qj}.  This allows us to write on-shell expressions for the nonplanar irreducible numerators in terms of the planar numerators.  We restrict the form of our irreducible numerators to ensure that the choice of ISPs and monomials satisfy the basic symmetries required by our colour decomposition.

\section{Colour decomposition}\label{sec:colour}

An integrand-reduced colour-dressed two-loop amplitude has the form
\begin{align}
   \mathcal{A}_n^{(2)}(a_i,p_i) =
      i g^{n+2} \int \frac{\d^D\ell_1\d^D\ell_2}{(2\pi)^{2D}} \sum_T\frac{1}{S_T}
      \frac{\tilde{\Delta}_T(a_i,p_i, \ell_i)}
           {\prod_{\alpha\in T}\cQ_\alpha(p_i,\ell_i)},
  \label{IrredIntegrand}
\end{align}
where $a_i$ are the external colour indices.  Associated with each graph topology $T$ in the sum is a symmetry factor $S_T$ and a colour-dressed irreducible numerator $\tilde{\Delta}_T$; the latter are functions of the colour indices $a_i$ as well as the external and loop momenta. Each of these numerators has a colour decomposition
\begin{align}\label{eq:irrednumdecomp}
   \tilde{\Delta}_T(a_i,p_i,\ell_i) =
     \sum_\sigma c_T(a_i)\,
         \Delta_T(p_i,\ell_i),
\end{align}
where we must explicitly determine the permutation sum $\sigma$ on external momenta and the associated colour factors $c_T$.

When such a colour decomposition is chosen it picks a set of colour tensors describing the colour structure of the amplitude. At the same time, it specifies an associated set of cut diagrams which must be computed. Each of these cut diagrams is, in turn, associated with a unique irreducible numerator. Thus the colour decomposition that we pick is of central importance, because it determines the set of irreducible numerators that we need to calculate.

\subsection{Multi-peripheral colour decomposition}\label{sec:reducedexpansion}

Our algorithm is applicable to the general case of an $L$-loop Yang-Mills
amplitude. Following the generalised unitarity principle, we begin by writing
the amplitude as a sum over all colour-dressed cuts. Diagrammatically, these
cuts consist of vertices formed from colour-dressed tree amplitudes which are
joined by on-shell propagators.  As discussed in section~\ref{sec:2lreview}, at two loops the set of colour-dressed cuts can be classified as genuine two-loop topologies and one-loop squared (or butterfly) topologies, both shown in Figure~\ref{fig:2lschematic}.

The central idea is to build the loop-level colour decomposition using knowledge of the underlying tree-level amplitudes.  Instead of the standard trace-based decomposition~(\ref{eq:colourdecomp}), we find it convenient to use the DDM form~\cite{DelDuca:1999ha,DelDuca:1999rs}:
\begin{align}\label{DDM}
\cA^{(0)}_n=g^{n-2}\!\sum_{\sigma\in S_{n-2}}\!c\bigg(\,\usegraph{10}{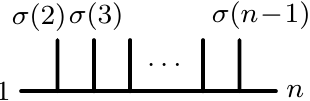}\,\bigg)\,
A^{(0)}(1,\sigma(2),\dots,\sigma(n-1),n).
\end{align}
This is the tree-level equivalent of the one-loop DDM decomposition that we introduced in section~\ref{sec:1lreview}.

The main advantage of this form of the amplitude is that
it contains $(n-2)!$ colour structures,
as compared to $(n-1)!$ in the trace-based decomposition.
This fact helps to reduce the number of generated diagrams; in particular,
an algorithm based directly on the trace decomposition of tree amplitudes
generates a larger set of diagrams, some of which are rather obscure.

Each of the colour structures in the tree decomposition is a string of group
theoretic structure constants $\tf^{abc}$. For an $n$-gluon amplitude
the decomposition is constructed by fixing the position of any two gluons
at either end of this string. The $(n-2)!$ permutations of the remaining gluons
between the ends of the string form the set of colour factors each of which is associated with a colour-ordered tree of the same ordering.
Pictorially, the colour structures look like combs.

It is straightforward to build the loop colour structure from these DDM tree
colour structures. The loop structure follows directly from the cut diagram:
one simply inserts the DDM trees at the vertices; propagator lines connecting
trees indicate that the ends of the DDM combs at either end of the propagator
have the same colour index to be summed over. Notice, however, that we
must pick two special lines in the DDM form of the tree amplitudes
(corresponding to lines $1$ and $n$ in eq.~(\ref{DDM}).)
These lines are on opposite ends of the DDM colour strings, so one can
informally think of this choice as picking two lines and ``stretching'' the
colour-ordered tree between these two ends. We make canonical choices of which
legs to pick as special, depending on the number of propagators that connect
to the three-point amplitude. These choices are:

\begin{figure}[t]
  \centering
  \includegraphics[width=0.8\textwidth]{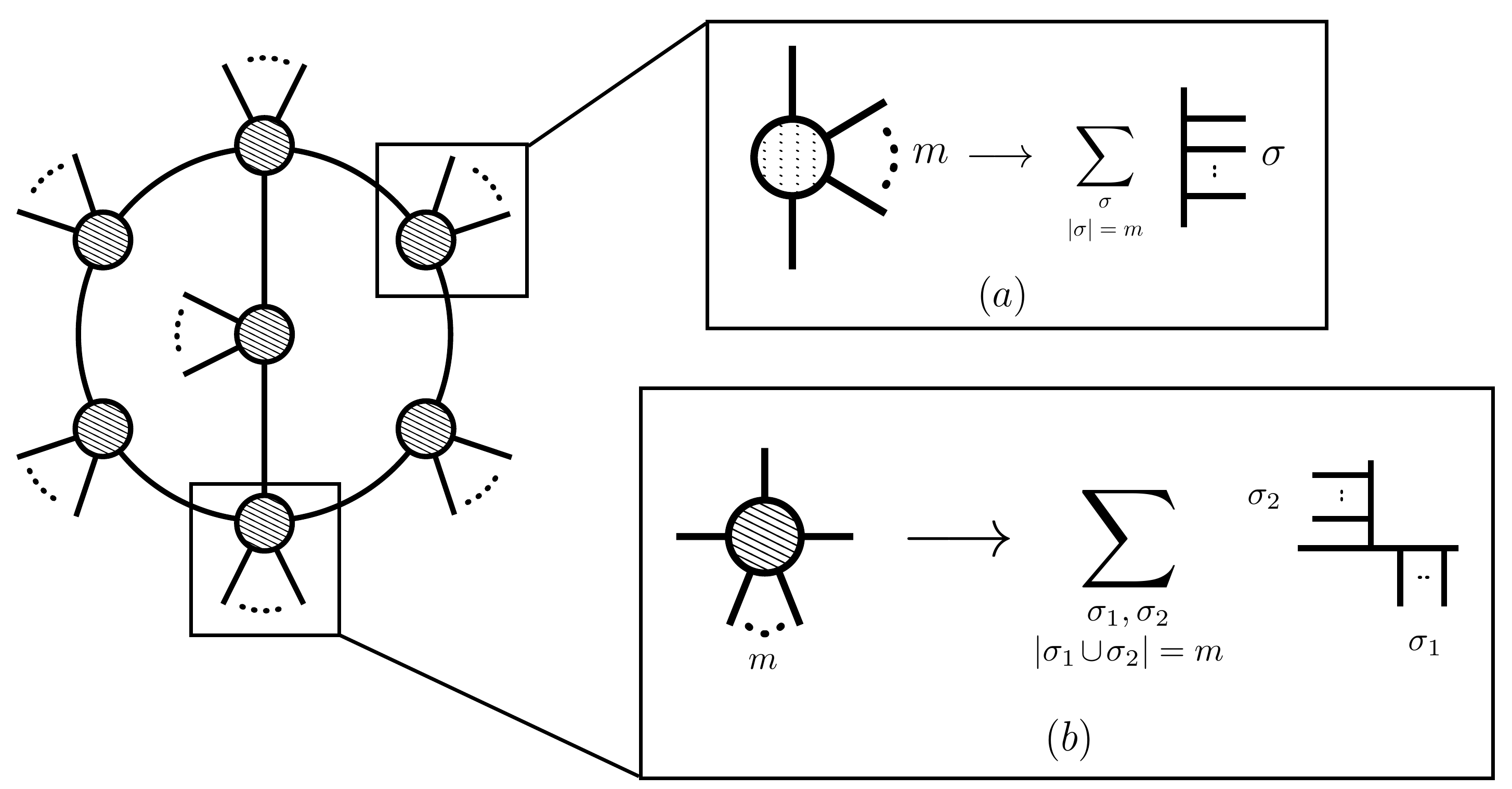}
  \caption{\small Inserting the DDM tree basis into the colour dressed cuts of a two-loop amplitude. The upper insert~$(a)$ shows the simple case of two loop
  propagators, while the lower insert~$(b)$ shows the case of three loop propagators. The sums run over the permutations of the external
  legs in the tree-level amplitude.}
  \label{fig:coltopo1}
\end{figure}

\begin{figure}[t]
  \centering
  \includegraphics[width=0.95\textwidth]{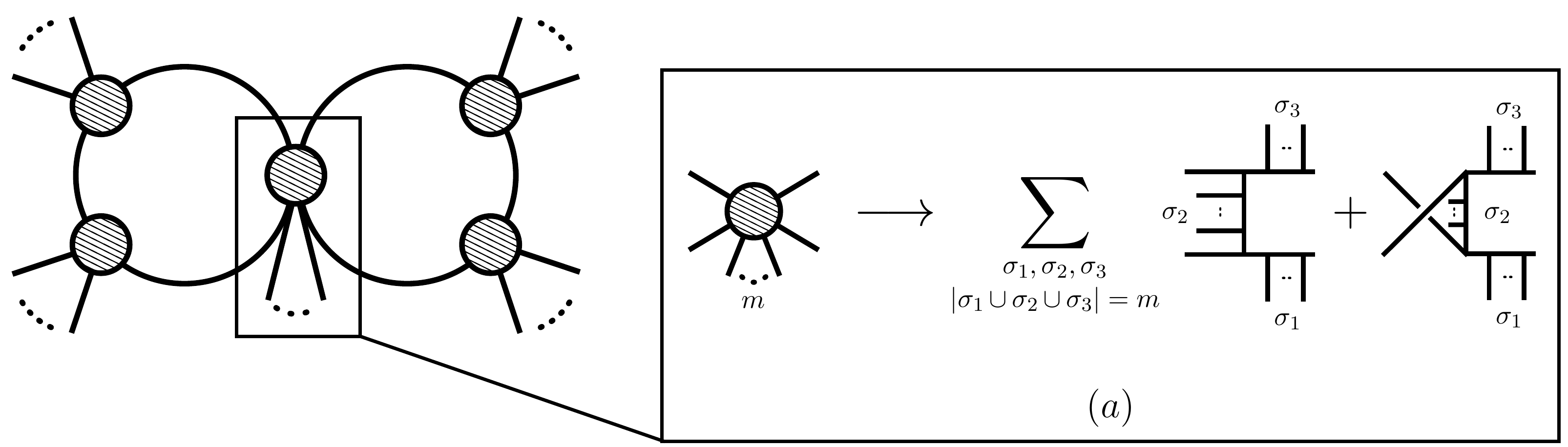}
  \caption{\small Inserting the DDM tree basis into a colour dressed cut of a butterfly topology at two-loops. There are four loop propagators
  in this case, and the insert~$(a)$ shows the result of inserting the DDM tree decomposition fixing the two right legs. The sums run over the permutations of the external legs in the tree-level amplitude.}
  \label{fig:coltopo2}
\end{figure}

\begin{itemize}
\item Two propagators

In this case, it is natural to ``stretch'' using the two propagators as the
special legs. Thus we build the colour structure by pasting a DDM
multi-peripheral colour structure between the two propagators. We must sum over
every ordering of the external legs. Pictorially, the operation is show in
the upper insert $(a)$ of Figure~\ref{fig:coltopo1}.

\item Three propagators

In the case of three propagators, we select two out of the three propagators to
be the special lines in the DDM presentation. Notice that this choice hides
some of the full symmetry of the diagram. In constructing the DDM tree, the
propagators we have selected must be at the end of the multi-peripheral
structure; we must sum over the positions of the other legs. The result is a
sum of diagrams, as shown in the lower insert $(b)$ of Figure~\ref{fig:coltopo1}.

\item Four propagators

We again choose two propagators to ``stretch'' the cut amplitude into a DDM
tree. At two loops, we only encounter this case in the butterfly topologies.
We choose upper and lower propagators on the right side of the diagram as special; by symmetry, the result is the same as if we chose upper and lower propagators on the left of the diagram.
The insert of Figure~\ref{fig:coltopo2} sketches out the procedure.

\end{itemize}

In this way, we build a set of colour structures. The kinematic structure
associated with each colour structure is easily understood. Each time we insert
a particular DDM colour trace, we also pick up a factor of the associated
colour-ordered tree amplitude. Thus, the orientation of the legs in the kinematic
diagram, associated to an irreducible numerator, is the same as in the colour
structure; of course, the ``stretching'' procedure does not produce new
propagators in the irreducible numerator.

One advantage of using the DDM presentation of the amplitude at tree level is
that the KK relations are automatically satisfied. Our procedure recycles this
property to loop level: we automatically generate a set of colour diagrams
that is KK-independent. Along the way, we generate ordered diagrams for the
kinematics. The same procedure works at $L$ loops;
the amplitude is expressed as
\begin{align}
\cA^{(L)}_n=
i^{L-1}g^{n+2L-2}\!\!\!\!\!
\sum_{\substack{\text{KK-independent}\\\text{1PI graphs }T}}
\int\left(\prod_{j=1}^{L}\frac{\d^D \ell_j}{(2\pi)^D}\right)\frac{1}{S_T}
\frac{{c}_T\,\Delta_T(p_i,\ell_i)}{\prod_{\alpha\in T}\cQ_{\alpha}(p_i,\ell_i)} ,
\label{Areduced}
\end{align}
where $S_T$ are the symmetry factors of the graphs and the $\Delta_T$ are irreducible numerators for appropriate colour factors $c_T$ generated through our algorithm.

\subsection{Review: four-point, two-loop amplitude}

To illustrate the above points we now briefly consider the four-gluon amplitude.  Our discussion follows closely that given by Ochirov and Page~\cite{Ochirov:2016ewn}.  Based on Bern, Dixon and Kosower's (BDK's) presentation~(\ref{eq:4ptamplitude}) we should only expect three nonzero numerators: the double box~(\ref{eq:331bern}), the nonplanar double box~(\ref{eq:delta332np}) and the double triangle~(\ref{eq:330bern}).  The full-colour amplitude is
\begin{align}\label{eq:4ptcolourdressedamp}
\cA^{(2)}_{++++}\!=\!
ig^6\cT\sum_{\sigma\in S_4}\!I^D\!\Bigg[\frac{1}{4}\tilde{\Delta}\bigg(\usegraph{9}{331}\bigg)\!\!+\!\frac{1}{4}\tilde{\Delta}\bigg(\!\usegraph{9}{322}\bigg)\!\!+\!\frac{1}{8}\tilde{\Delta}\bigg(\usegraph{9}{330}\bigg)\Bigg],
\end{align}
where we have introduced appropriate symmetry factors and extracted the permutation-invariant factor $\cT$~(\ref{eq:cT}).

We seek a colour decomposition for each diagram.  As the first two diagrams have only cubic vertices their colour structures follow trivially by dressing with structure constants.  To find the colour factor of the double triangle we consider its corresponding colour-dressed cut:
\begin{align}
&{\cal C}\text{ut}\bigg(\usegraph{10}{330}\bigg)=
c\bigg(\usegraph{10}{331i}\bigg)\text{Cut}\bigg(\usegraph{9}{330}\bigg)+
c\bigg(\usegraph{9}{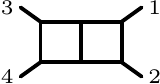}\bigg)\text{Cut}\bigg(\usegraph{9}{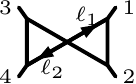}\bigg),
\end{align}
where, as shown in Figure~\ref{fig:coltopo2}, we have expanded the four-point tree in the DDM decomposition and the legs on the right-hand side have been fixed.  This suggests that we should decompose the colour-dressed double-triangle numerator as
\begin{align}
&\tilde{\Delta}\bigg(\usegraph{10}{330}\bigg)=
c\bigg(\usegraph{10}{331i}\bigg)\Delta\bigg(\usegraph{9}{330}\bigg)+
c\bigg(\usegraph{9}{331i1243}\bigg)\Delta\bigg(\usegraph{9}{330o1243}\bigg).
\end{align}
When inserted into the amplitude~(\ref{eq:4ptcolourdressedamp}) the sum on permutations $S_4$ symmetrises over these two terms; the double-triangle numerator is left with the colour factor of a double box and a symmetry factor of $1/4$.  The amplitude becomes
\begin{align}
&\mathcal{A}^{(2)}_{++++}
=ig^6\mathcal{T}\sum_{\sigma \in S_4}
I^D\Bigg[\frac{1}{4}c\bigg(\usegraph{9}{331i}\bigg)
\Bigg(\Delta\bigg(\usegraph{9}{331}\bigg)
+\Delta\bigg(\usegraph{9}{330}\bigg)\Bigg)\nnl
&\qquad\qquad\qquad\qquad\qquad\qquad\qquad\qquad\!\!
+\frac{1}{4}c\bigg(\!\usegraph{9}{322i}\bigg)
\Delta\bigg(\!\usegraph{9}{322}\bigg)\Bigg],
\end{align}
which is BDK's presentation~(\ref{eq:4ptamplitude}).

\subsection{Five-point, two-loop amplitude}
\label{sec:5point2loop}

Now we describe the colour structure of the previously-unknown five-point, two-loop amplitude.  We concentrate on the diagrams that do not vanish in the all-plus case according to Badger, Frellesvig and Zhang's (BFZ's) planar numerators~(\ref{eq:2L5gbfz}) and our calculations in \sec{sec:kinematic}.  An expression for the generic five-point two-loop amplitude's colour structure may be found in ref.~\cite{Ochirov:2016ewn}; it is given by a straightforward extension of the present discussion.

Let us first write the amplitude and then explain its content in more detail.  The complete formula for the amplitude is:
\begin{align}
&\cA^{(2)}_{+++++}\nn\\
&\,\,\,=ig^7\sum_{\sigma\in S_5}I^D\Bigg[\,
c\bigg(\usegraph{9}{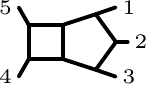}\!\!\!\:\bigg) 
\Bigg(\frac{1}{2}\Delta\bigg(\usegraph{9}{431i}\!\!\!\:\bigg)
\!+\Delta\bigg(\usegraph{9}{331M1i}\bigg)
\!+\frac{1}{2} \Delta\bigg(\usegraph{13}{3315Li}\bigg)\nn\\
&\qquad\qquad+\frac{1}{2}\Delta\bigg(\usegraph{13}{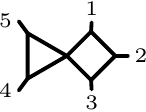}\!\!\!\:\bigg)
\!+\Delta\bigg(\usegraph{10}{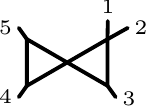}\bigg)
\!+\frac{1}{2} \Delta\bigg(\usegraph{13}{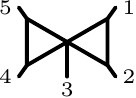}\bigg) 
\!\Bigg) \!\!\!\!\!\!\!\!\!\! \nn \\
&\qquad+c\bigg(\usegraph{9}{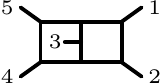}\bigg)\Bigg(
\frac{1}{4}\Delta\bigg(\usegraph{9}{332i}\bigg)
\!+\frac{1}{2} \Delta\bigg(\usegraph{9}{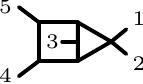}\!\bigg)
\!+\frac{1}{2} \Delta\bigg(\usegraph{9}{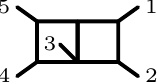}\bigg)\nn\\
&\qquad\qquad-\Delta\bigg(\!\!\!\;\usegraph{16}{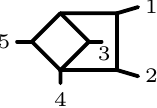}\bigg)
\!+\frac{1}{4} \Delta\bigg(\usegraph{9}{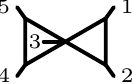}\bigg)\!\Bigg)\nn\\
&\qquad+c\bigg(\!\!\!\:\usegraph{9}{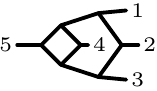}\!\!\!\:\bigg)\Bigg(
\frac{1}{4}\Delta\bigg(\!\!\!\:\usegraph{9}{422i}\!\!\!\:\bigg)
\!+\frac{1}{2}\Delta\bigg(\!\!\!\:\usegraph{8.8}{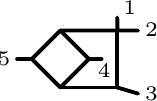}\bigg)\!\Bigg)\!
+\dots \,\Bigg] ,
\label{A5point2loop}
\end{align}
where our integral convention for $I^D$ was given in eq.(\ref{IntMeasure}).  The explicit symmetry factors compensate for the overcounts introduced by the overall sum over permutations of external legs.  For convenience, we recapitulate all these $\Delta_i$ in \tab{tab:AllGraphs}, where for each irreducible numerator $\Delta_i$ we also give its colour factor and the set of its non-equivalent permutations.

\begin{table*}
\centering
\begin{tabular}{| l | c | c | l|}
\hline
& Graphs & Colour factors & Permutation sum \\
\hline
$ (a) \bigg. $ &
\includegraphics[scale=1.0,trim=0 5 0 -5]{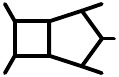} &
\includegraphics[scale=1.0,trim=0 5 0 -5]{graphs/431c} &
$S_5/$Vertical flip \\
\hline
$ (b) \bigg. $ &
\includegraphics[scale=1.0,trim=0 5 0 -5]{graphs/332c} &
\includegraphics[scale=1.0,trim=0 5 0 -5]{graphs/332c} &
$S_5/$Vertical \& horizontal flip \\
\hline
$ (c) \bigg. $ &
\includegraphics[scale=1.0,trim=0 5 0 -5]{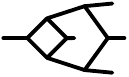} &
\includegraphics[scale=1.0,trim=0 5 0 -5]{graphs/422c} &
$S_5/$Vertical \& diamond flip  \\
\hline
$ (d) \bigg. $ &
\includegraphics[scale=1.0,trim=0 5 0 -5]{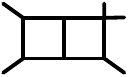} &
\includegraphics[scale=1.0,trim=0 5 0 -5]{graphs/431c} &
$S_5$ \\
\hline
$ (e) \bigg. $ &
\includegraphics[scale=1.0,trim=0 5 0 -5]{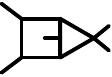} &
\includegraphics[scale=1.0,trim=0 5 0 -5]{graphs/332c} &
$S_5/$Vertical flip \\
\hline
$ (f) \bigg. $ &
\includegraphics[scale=1.0,trim=0 5 0 -5]{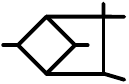} &
\includegraphics[scale=1.0,trim=0 5 0 -5]{graphs/422c} &
$S_5/$Diamond flip \\
\hline
$ (g) \bigg. $ &
\includegraphics[scale=1.0,trim=0 5 0 -5]{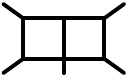} &
\includegraphics[scale=1.0,trim=0 5 0 -5]{graphs/431c} &
$S_5/$Horizontal flip \\
\hline
$ (h) \bigg. $ &
\includegraphics[scale=1.0,trim=0 5 0 -5]{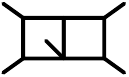} &
\includegraphics[scale=1.0,trim=0 5 0 -5]{graphs/332c} &
$S_5/$Horizontal flip \\
\hline
$ (i) \bigg. $ &
\includegraphics[scale=1.0,trim=0 5 0 -5]{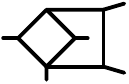} &
\!\!\!\!\!\!$~^-$\includegraphics[scale=1.0,trim=0 5 0 -5]{graphs/332c} &
$S_5$ \\
\hline
$ (j) \bigg. $ &
\includegraphics[scale=1.0,trim=0 5 0 -5]{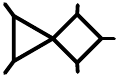} &
\includegraphics[scale=1.0,trim=0 5 0 -5]{graphs/431c} &
$S_5/$Vertical flip \\
\hline
$ (k) \bigg. $ &
\includegraphics[scale=1.0,trim=0 5 0 -5]{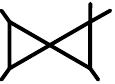} &
\includegraphics[scale=1.0,trim=0 5 0 -5]{graphs/431c} &
$S_5$ \\
\hline
$ (l) \bigg. $ &
\includegraphics[scale=1.0,trim=0 5 0 -5]{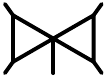} &
\includegraphics[scale=1.0,trim=0 5 0 -5]{graphs/431c} &
$S_5/$Horizontal flip \\
\hline
$ (m) \bigg. $ &
\includegraphics[scale=1.0,trim=0 5 0 -5]{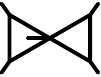} &
\includegraphics[scale=1.0,trim=0 5 0 -5]{graphs/332c} &
$S_5/$Horizontal \& vertical flip \\
\hline
\end{tabular}
\caption{\small The irreducible numerators that are nonzero
         for the all-plus five-point two-loop amplitude,
         along with their colour factors and reduced permutation sums.}
\label{tab:AllGraphs}
\end{table*}

The first three graphs (a)-(c) in \tab{tab:AllGraphs} are the master diagrams corresponding to the maximal cuts.  They are purely trivalent, thus their colour factors are unambiguously defined by their proper graphs.

The next three graphs (d)-(f) have four-point vertices with two external and two internal edges.  The two external legs automatically enter in the permutation sum with two possible orderings, hence multi-peripheral subgraphs are naturally obtained by fixing the internal lines, as in the insert $(a)$ of Figure~\ref{fig:coltopo1}.  ``Stretching'' the four-point vertex by these lines gives a master graph for each colour factor.

The following two diagrams (g) and (h) share the same graph structure, up to the ordering of the four-point vertex.  The apparent asymmetry introduced by our selecting these two diagrams,  and omitting the graph with the external leg in the right loop, is an artefact of our colour decomposition.  One could make other choices; the KK relations satisfied by the trees and their symmetries ensure that any choice is valid.

To expand the four-point vertex in graph (i) we fixed the internal lines of the ``diamond'' subdiagram, hence its colour factor is $c(\includegraphics[scale=0.5,trim=0 5 0 -5]{graphs/332c})$, but with a minus sign due to one flipped vertex.  The other permutation of the four-point vertex corresponds to the same topology and is present in the overall permutation sum with the right permuted colour diagram.

The colour factor of the planar graph (j) follows in a straightforward manner from our algorithm (see Figure~\ref{fig:coltopo2}), yielding $c(\includegraphics[scale=0.5,trim=0 5 0 -5]{graphs/431c})$ as its colour factor.  Its descendant (k) is more interesting, since it is the only graph in the all-plus case with two four-point vertices.  They can be treated independently by linearity of colour decomposition.

The last two graphs, (l) and (m), share a five-point vertex. To explain their colour factors, let us consider the corresponding colour-dressed cut:
\begin{align}
&{\cal C}\text{ut}\bigg(\usegraph{13}{3305Li}\bigg)=\label{Cut3305L1}\\
&\qquad c\bigg(\usegraph{9}{431i}\!\!\!\:\bigg)
\text{Cut}\bigg(\usegraph{13}{3305Li}\bigg)
+ c\bigg(\usegraph{9}{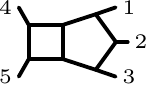}\!\!\!\:\bigg)
\text{Cut}\bigg(\usegraph{13}{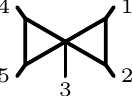}\bigg) \nn \\
&\qquad+ c\bigg(\usegraph{9}{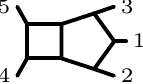}\!\!\!\:\bigg)
      \text{Cut}\bigg(\usegraph{9}{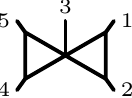}\bigg)
    + c\bigg(\usegraph{9}{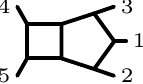}\!\!\!\:\bigg)
      \text{Cut}\bigg(\usegraph{9}{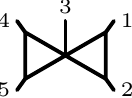}\bigg) \nn \\
&\qquad+ c\bigg(\usegraph{9}{332i}\bigg)
      \text{Cut}\bigg(\usegraph{9}{3305L2i}\bigg)
    + c\bigg(\usegraph{9}{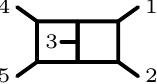}\bigg)
      \text{Cut}\bigg(\usegraph{9}{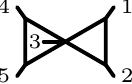}\bigg) .\nn
\end{align} 
We obtain the multi-peripheral decomposition of the five-point vertex
by fixing the two right-hand loop edges
and permuting the other three edges. 
The graphs in the second line can be vertically flipped
to put leg $3$ downstairs to match the presentation in \tab{tab:AllGraphs}.
Obviously, an equivalent decomposition could be achieved
by fixing the loop edges on the left, which would change
the orientation of leg $3$ in the superficially nonplanar graphs.
The $S_5$-summation in \eqn{A5point2loop}
effectively symmetrises the colour structure
over the two choices of multi-peripheral decompositions.

In the present work we can avoid lower levels of the graph hierarchy
thanks to the simplicity of the fully symmetric helicity configuration,
but it already incorporates all the key elements
of the general colour structure~\cite{Ochirov:2016ewn}.

\section{Kinematic structure}
\label{sec:kinematic}

With our colour decomposition in hand, we turn our attention to the kinematic structure of the amplitude.  We need to compute an irreducible numerator associated to each diagram in \eqn{A5point2loop}.  For the planar numerators, BFZ's original expressions~(\ref{eq:2L5gbfz}) and the local integrands~(\ref{eq:2L5g}) work equally well.  Our task is now to determine the nonplanar numerators. Of course, these numerators can be computed directly from their cuts. However, as we will see, it is easy to determine the complete set of nonplanar irreducible numerators for this amplitude from its planar numerators and the knowledge of tree-level amplitude relations.

\subsection{Nonplanar from planar}\label{sec:npfromp}

The nonplanar numerator $\Delta(\includegraphics[scale=0.5,trim=0 5 0 -5]{graphs/332c})$ can, of course, be obtained directly from its cut using the procedure outlined in chapter~\ref{ch:localintegrands}.  However, we can avoid calculating  this nonplanar cut explicitly by relating it to a planar cut. We do so in two steps: first, we coalesce two (ordered) three-point trees into a limit of an ordered four-point tree; then we use the BCJ relations~\cite{Bern:2008qj} satisfied by the ordered four-point tree to reorder the legs until the complete diagram becomes planar.  The BCJ tree identities will be explored further in chapter~\ref{ch:allplusbcj}.

In more detail, we use the following well-known relation, which is satisfied by on-shell amplitudes in the cuts:
\begin{align}
   A^{(0)}(1,2,-(1\!+\!2)) \, A^{(0)}(1\!+\!2,3,4) \,
  & = \, \big\{ s_{13}\,A^{(0)}(1,3,2,4) \big\} \big|_{s_{12}=0} , \nonumber \\
      \includegraphics[scale=1.0,trim=0 20 0 0]{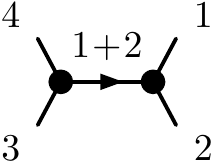}
  & = \, \left. \!\!\!\: \left\{ s_{13}
      \includegraphics[scale=1.0,trim=0 20 0 0]{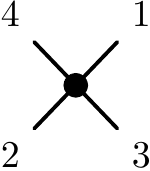}
      \right\} \right|_{s_{12} = 0}.
\label{BCFW4limit}
\end{align}
Since this identity is of central importance for us, we present a short proof.
A four-point tree amplitude can be constructed from two three-point amplitudes
using the BCFW recursion relation~\cite{Britto:2004ap,Britto:2005fq}:
\begin{align}\label{eq:BCFW4}
   A^{(0)}(1,2,3,4) = \frac{1}{s_{12}} \hat{A}^{(0)}(1,2,-(1\!+\!2))
                                 \hat{A}^{(0)}(1\!+\!2,3,4) ,
\end{align}
where hat signs on the right-hand side indicate that the amplitudes
are evaluated on complex kinematics for some BCFW shift of external legs.
The exact complex value of the shifted internal momentum $(\widehat{1\!+\!2})$
is defined by the on-shell condition
\begin{align}
   \hat{s}_{12} = s_{12} + \alpha z = 0 ,
\end{align}
where the precise expression for $\alpha$ depends on the particular BCFW shift.
The key point is that $\hat{s}_{12}$ is a linear function of $z$,
with the property that in the limit $s_{12} \rightarrow 0$,
$z \rightarrow 0$. In this limit eq.~(\ref{eq:BCFW4}) becomes
\begin{align}\label{BCFW4real}
   \big\{ s_{12} A^{(0)}(1,2,3,4) \big\} \Big|_{s_{12}=0}
      = A^{(0)}(1,2,-(1\!+\!2)) A^{(0)}(1\!+\!2,3,4) .
\end{align}
Notice that the left-hand side contains a nonzero contribution due to
the pole in $s_{12}$.
Now we can remove the factor of $s_{12}$
on the left-hand side of \eqn{BCFW4real}
by using the BCJ amplitude relation~\cite{Bern:2008qj},
\begin{align}\label{BCJ4su}
   s_{12} A^{(0)}(1,2,3,4) = s_{13} A^{(0)}(1,3,2,4) .
\end{align}
This proves the identity~\eqref{BCFW4limit}.

We proceed by applying our identity~\eqref{BCFW4limit} to tree amplitudes inside the nonplanar cut, rearranging the diagram until it becomes planar.  It is simplest to begin with maximal diagrams, and then to continue to topologies with fewer propagators.  We will work through the calculation of $\Delta(\includegraphics[scale=0.5,trim=0 5 0 -5]{graphs/332c})$ as an example; we computed all non-planar irreducible numerators using the same technique.

The calculation starts at the level of the cuts:
\begin{align}\label{cut332NPdiagram}
\!\!\!\!\!\!\!
&\text{Cut}\bigg(\usegraph{8}{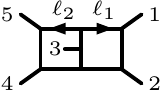}\bigg)
=\includegraphics[scale=0.7,trim=0 27 0 0]{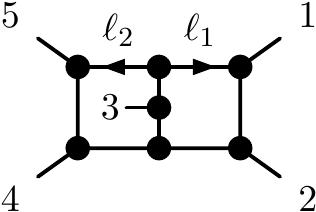}
=(\ell_1+p_{45})^2 ~\left.
\includegraphics[scale=0.7,trim=0 27 0 0]{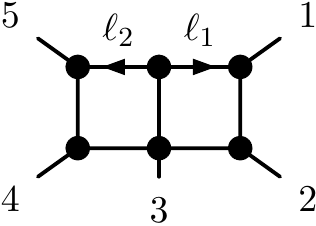}
~~\right|_{(\ell_1+\ell_2+p_3)^2=0}\nn\\
&\qquad=\left.(\ell_1+p_{45})^2\,\text{Cut}\bigg(\usegraph{13}{3315L}\bigg)
\right|_{(\ell_1+\ell_2+p_3)^2=0} ,
\!\!\!\!\!\!\!\!\!\!\!\!\!\!\!\!\!
\end{align}
where we understand that all internal helicities are implicitly summed over while all exposed propagators are cut.  These cuts are decomposed into irreducible numerators as
\begin{subequations}\label{eq:delta332NPBCJ}
\begin{align}
&\text{Cut}\bigg(\usegraph{8}{332}\bigg)
=\Delta\bigg(\usegraph{8}{332}\bigg),
\label{cut332NP}\\
&\text{Cut}\bigg(\usegraph{13}{3315L}\bigg)\label{eq:cuteqn3315L}\\
&\qquad=\Delta\bigg(\usegraph{13}{3315L}\bigg)
+\frac{1}{(\ell_1+p_{45})^2}\Delta\bigg(\usegraph{9}{431}\bigg)
+\frac{1}{(\ell_2+p_{12})^2}\Delta\bigg(\!\usegraph{9}{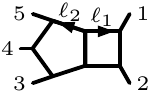}\bigg).\nn
\end{align}
\end{subequations}
Using the fact that $(\ell_1+p_{45})^2=-(\ell_2+p_{12})^2$ on this cut, we see that
\begin{align}\label{eq:delta332NPos}
\Delta\bigg(\usegraph{8}{332}\bigg)
=(\ell_1+p_{45})^2\Delta\bigg(\usegraph{13}{3315L}\bigg)
+\Delta\bigg(\usegraph{9}{431}\bigg)
-\Delta\bigg(\!\usegraph{9}{431o34512}\bigg) &.
\end{align}
A similar calculation for the other nonplanar master leads to
\begin{align}\label{eq:delta422NPos}
\Delta\bigg(\!\usegraph{10}{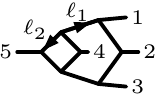}\bigg)=\Delta\bigg(\usegraph{9}{431}\bigg).
\end{align}
So far the obtained nonplanar numerators are valid only on their cuts, so it is equally valid to write them in terms of BFZ's irreducible numerators~(\ref{eq:2L5gbfz}) or the local integrands~(\ref{eq:2L5g}).

\subsection{Off-shell symmetries}

We would now like to extend the expressions off-shell.  As in chapter~\ref{ch:localintegrands}, we express the numerators in terms of a given set of ISPs and then define off-shell numerators unambiguously through these ISPs.  The value of a given numerator depends on the choice of ISP basis off-shell (in contrast to the situation on shell, of course).  In this way, we determine a valid set of non-planar irreducible numerators.  Notice that the ISP monomial choices made in the planar sector, such as the higher powers of $\mu_{ij}$ preferred over high powers of $\ell_i\cdot p_j$, will then be easily translated to the non-planar numerators.

When choosing ISPs, we might consider applying the guidelines given in section~\ref{sec:unitarity}: focus on the infrared structure and elimination of spurious singularities at the integrand level.  For now, however, we will satisfy ourselves with a more basic approach; the open question of whether nonplanar local integrands can be found will be discussed at the end of this thesis.  Our focus is on maintaining the symmetries of the graphs in the off-shell continuation, which we achieve by choosing ISPs on a graph-by-graph basis. We engineer the ISP basis such that the loop momentum-dependence of each irreducible numerator is captured by a set of ISPs, which map into one another under the graph symmetries without using any cut conditions.

\begin{table*}[t]
\centering
\begin{tabular}{| l | l | l |} 
\hline
Graphs & ISPs & RSPs \\
\hline
& $\ell_1\cdot(p_5-p_4)$, & $\ell_1^2$, $(\ell_1-p_1)^2$, $(\ell_1-p_{12})^2$, \\
\includegraphics[scale=1.0,trim=0 5 0 10]{graphs/332c} &
$\ell_2\cdot(p_1-p_2)$, & $\ell_2^2$, $(\ell_2-p_5)^2$, $(\ell_2-p_{45})^2$, \\
& $(\ell_1-\ell_2)\cdot p_3$ & $(\ell_1+\ell_2+p_3)^2$, $(\ell_1+\ell_2)^2$ \\
\hline
& $(\ell_1+2\ell_2)\cdot(p_1-p_3)$, & $\ell_1^2$, $(\ell_1-p_1)^2$, $(\ell_1-p_{12})^2$, $(\ell_1-p_{123})^2$, \\
\includegraphics[scale=1.0,trim=0 5 0 10]{graphs/422c} &
$(\ell_1+2\ell_2)\cdot p_2$, & $\ell_2^2$, $(\ell_2-p_5)^2$,\\
& $\ell_1\cdot(p_5-p_4)$ & $(\ell_1+\ell_2+p_4)^2$, $(\ell_1+\ell_2)^2$ \\
\hline
\end{tabular}
\caption{\small The choices of ISPs and RSPs for the two nonplanar masters, 
where the RSPs are chosen as the propagators of the respective graphs.
Additionally, the higher-dimensional ISPs $\mu_{ij}$
are shared by all topologies.}
\label{tab:mastersps}
\end{table*}

The symmetries of the maximal nonplanar graphs, for example, are
\begin{subequations}\label{eq:mastersymmetries}
\begin{align}
\Delta\bigg(\usegraph{8}{332}\bigg)
&=-\Delta\bigg(\usegraph{10}{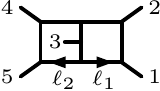}\bigg)
=\Delta\bigg(\usegraph{8}{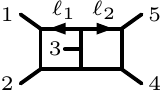}\bigg),
\label{eq:delta332NPsymmetry} \\
\Delta\bigg(\!\usegraph{10}{422}\bigg)
&=-\Delta\bigg(\!\usegraph{12}{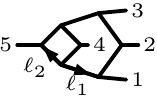}\bigg)=
\Delta\bigg(\!\usegraph{10}{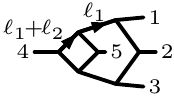}\bigg).
\label{eq:delta422NPsymmetry}
\end{align}
\end{subequations}
These symmetries motivate our choices of ISPs, given in Table \ref{tab:mastersps}.  For instance, the second symmetry in eq.~(\ref{eq:delta332NPsymmetry}) leads to a map of ISPs
\begin{align}
  \ell_1\cdot(p_5-p_4) &\leftrightarrow \ell_2\cdot(p_1-p_2) ,\nonumber\\
  (\ell_1-\ell_2)\cdot p_3&\leftrightarrow -(\ell_1-\ell_2)\cdot p_3.
\label{eq:332NPispsym}
\end{align}
After we express loop-momentum dependence in \eqns{eq:delta332NPos}{eq:delta422NPos} in terms of the ISPs of \Tab{tab:mastersps}, using the fact that the RSPs (cut propagators) are zero on shell, we are left with appropriate off-shell irreducible numerators.  These are listed in the next section.  Note that $\Delta(\includegraphics[scale=0.5,trim=0 5 0 -5]{graphs/422c})$, given in eq.~(\ref{eq:422}), is not the same as the pentabox irreducible numerator (either of them) despite the on-shell equation (\ref{eq:delta422NPos}): different ISPs are chosen to make different off-shell symmetries manifest.

We obtained irreducible numerators for lower-level non-planar diagrams in the same way, using the BCJ relations on cuts and extending the results off shell.  For the all-plus amplitude at hand we find that many lower-level irreducible numerators vanish.  In other words, the higher-level numerators capture the unitarity cut structure of the full amplitude, which is given below.

\subsection{Nonplanar irreducible numerators}\label{sec:npnumerators}

Here we present a complete summary of all kinematic numerators contributing to the colour decomposition~(\ref{A5point2loop}).  As the relevant planar numerators have already been provided, here we list only the nonplanars.  They may then be assembled into the full amplitude~(\ref{A5point2loop}).

The two maximal nonplanar graphs are
\begin{subequations}\label{Level0}
\begin{align}
\Delta\bigg(\usegraph{9}{332}\bigg)
&=\frac{s_{12}s_{45}F_1}{4\SpDenom5\trfive}\label{eq:332}\\
&\qquad\times\Big(s_{23}\trp(1345)\big(2s_{12}-4\ell_1\cdot(p_5-p_4)
+2(\ell_1-\ell_2)\cdot p_3\big)\nnl
&\qquad\qquad-s_{34}\trp(1235)\big(2s_{45}-4\ell_2\cdot(p_1-p_2)
-2(\ell_1-\ell_2)\cdot p_3\big)\nnl
&\qquad\qquad-4s_{23}s_{34}s_{15}(\ell_1-\ell_2)\cdot p_3\Big),\nnl
\Delta\bigg(\!\!\!\:\usegraph{9}{422}\!\!\!\:\bigg)
&=-\frac{s_{12}s_{23}s_{45}F_1}{\SpDenom5\trfive}\label{eq:422}\\
&\qquad\times\Big(\trp(1345)\Big(\ell_1\cdot(p_5-p_4)-\frac{s_{45}}{2}\Big)
+s_{15}s_{34}s_{45}\Big),\nn
\end{align}
\end{subequations}
where $\trfive=\trfive(1234)$.  The nonplanar graphs at level 1 are
\begin{subequations}\label{Level1}
\begin{align}
\Delta\bigg(\usegraph{9}{3315L2i}\bigg)
&=-\frac{s_{12}s_{45}F_1}{4\SpDenom5\trfive}\\
&\qquad\times\big(2s_{23}s_{34}s_{15}-s_{23}\trp(1345)+s_{34}\trp(1235 \big),\nnl
\Delta\bigg(\!\!\!\;\usegraph{16}{3225Li}\bigg)
&=-\frac{s_{12}F_1}{2 \SpDenom5\trfive}\\
&\qquad\times\big(s_{23}s_{45}\trp(1435)-s_{15}s_{34}\trp(2453)\big),\nnl
\Delta\bigg(\!\!\!\:\usegraph{8.8}{322M1i}\bigg)
&=\Delta\bigg(\usegraph{9}{232i}\!\bigg)
=-\frac{s_{34}s_{45}^2\trp(1235)F_1}{\SpDenom5\trfive},
\end{align}
\end{subequations}
and there is only one nonplanar graph at level 2:
\begin{align}
&\Delta\bigg(\usegraph{9}{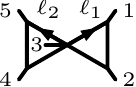}\bigg)=
\frac{F_3}{2\SpDenom5 s_{12}}\\
&\qquad\times
\Bigg((s_{45}-s_{12})\trp(1245)
-\left(\trp(1245)-\frac{\trp(1345)\trp(1235)}{s_{13}s_{35}}\right)
2\ell_1\cdot p_3\nn\\&
             \qquad \qquad \qquad \qquad \qquad
             -\frac{s_{45} \trp(1235)}{s_{35}}
             \left(2\ell_2\!\cdot\!\omega_{543}
                   +\frac{s_{12}-s_{45}}{s_{45}}(\ell_2-p_5)^2
             \right) \nn \\ &
             \qquad \qquad \qquad \qquad \qquad
             +\frac{s_{12} \trp(1345)}{s_{13}}
             \left(2\ell_1\!\cdot\!\omega_{123}
                   +\frac{s_{45}-s_{12}}{s_{12}}(\ell_1-p_1)^2
             \right)
      \Bigg). \nn
\end{align}

We have found a representation of the full amplitude with no topologies with fewer than six propagators.
We note that there are nonzero cuts at the integrand level,
but the resulting integrals are scaleless and hence zero in dimensional regularisation.
We have checked additional cuts at levels 2 and 3 to ensure that
no nonzero topologies remain.
To find an integrand with this property, the ISPs $\ell_1\!\cdot\!\omega_{123}$ and $\ell_2\!\cdot\!\omega_{543}$ in the level 2 numerator were upgraded to include terms proportional to $(\ell_1-p_1)^2$ and $(\ell_2-p_5)^2$.

\section{Checking the soft divergences}\label{sec:colourdressedsoft}

Now that we have the complete five-point amplitude we would like to check its infrared (IR) decomposition.  We return to the full colour-dressed decomposition given in section~\ref{sec:irreview}: in this case we require that
\begin{align}\label{eq:IRfull}
\mathcal{A}^{(2)}_{+++++}=
g^2\sum_{i\neq j}\frac{c_\Gamma}{\eps^2}\left(\frac{1}{-s_{ij}}\right)^\eps\!
\mathbf{T}_i\cdot\mathbf{T}_j \circ \mathcal{A}^{(1)}_{+++++}
+\mathcal{O}(\eps^{0}).
\end{align}
Just like in the last chapter, reproducing this behaviour requires us to find the IR divergences up to $\cO(\epsilon^{-1})$ of all our two-loop integrals.  As we already know the planar ones, our task is therefore to compute the nonplanar divergences.  We follow the same approach developed in section~\ref{sec:softlimits}: factorise them into products of one-loop integrals.  Unfortunately, as was mentioned previously, the procedure only correctly predicts subleading $\eps^{-1}$ poles when local integrands are used as numerators.  As we do not have local-integrand based nonplanar numerators we will have to be satisfied with the $\eps^{-2}$ terms.

As before, all of the poles of our amplitude are contained in the topologies whose numerators are proportional to $F_1$.  For instance, the necessary soft integrals for the pentabox topology are
\begin{subequations}
\begin{align}
I^D\bigg(\usegraph{9}{431i}\!\!\!\:\bigg)[F_1]&=
\frac{1}{(4\pi)^4} \frac{D_s-2}{3 s_{12}s_{23}\eps^2}+\mathcal{O}(\eps^{-1}),\\
I^D\bigg(\usegraph{9}{431}\!\!\!\:\bigg)[F_1\,(\ell_1\!\cdot\!p_5)]&=
\frac{1}{(4\pi)^4}\frac{(D_s-2)(2 s_{15}+s_{25})}{12 s_{12}s_{23}\eps^2}
+\mathcal{O}(\eps^{-1}),
\end{align}\label{I431}
\end{subequations}
where both integrals have three soft regions.  The remaining integrals are listed in appendix~\ref{app:2lsoftintegrals}.

We can rewrite eq.~(\ref{eq:IRfull}) in the leading soft limit as
\begin{equation}
  \mathcal{A}^{(2)}_{+++++} =
  -\frac{5N_c c_\Gamma}{\eps^2}g^2\mathcal{A}^{(1)}_{+++++}
  + \mathcal{O}(\eps^{-1}),
  \label{eq:IRsoft}
\end{equation}
where we have used
\begin{align}
\sum_{i\neq j}\mathbf{T}_i\cdot\mathbf{T}_j=\bigg(\sum_i\mathbf{T}_i\bigg)^2-\sum_i\mathbf{T}_i^2
\end{align}
together with $\sum_i\mathbf{T}_i=0$ (colour conservation) and $\mathbf{T}_i^2=N_c$.  As we only need the one-loop amplitude to leading order in $\eps$ we can use its integrated expression~(\ref{eq:oneloopintegrated}) together with the one-loop DDM colour decomposition~(\ref{eq:1lDDM}).  We find that this decomposition is indeed satisfied: clearly this is a weaker check than the full IR poles, but it does require non-trivial properties of the nonplanar sector.

We could also consider checking the subleading $\eps^{-1}$ poles in \eqn{eq:IRfull} numerically, as was done in the original planar calculation \cite{Badger:2013gxa}.  But this is computationally prohibitive for two reasons. Firstly, the full colour expansion contains a large number of dimension-shifted integrals ($\sim\mathcal{O}(1000)$) -- an order of magnitude more than the leading colour terms. Secondly, there is no Euclidean region for the complete amplitude, and so contour deformation must be performed for many of these integrals, making them more complicated than the planar cases. This task is probably achievable using \textsc{FIESTA}~\cite{Smirnov:2013eza} or newer versions of \textsc{SecDec}~\cite{Borowka:2017idc}.

\section{Discussion}

In this chapter we have explored the impact of tree-level amplitude relations
in multi-loop integrand computations. There were two major aspects to our work. Firstly, we exploited the Kleiss-Kuijf relations
to find a compact colour decomposition for the five-point two-loop all-plus amplitude in terms of multi-peripheral colour factors in an analogous way
to the tree-level and one-loop decompositions of Del Duca, Dixon and Maltoni~\cite{DelDuca:1999ha,DelDuca:1999rs}.

Secondly, we applied the BCJ amplitude relations~\cite{Bern:2008qj}
to relate all non-planar generalised unitarity cuts
to the previously computed planar ones. This allowed us to easily generate
a compact representation of the full-colour two-loop, five-gluon, all-plus integrand building on previous planar work~\cite{Badger:2013gxa}.
The soft infrared poles of the full amplitude were checked
against the well-known universal pole structure.

As with the local-integrand technology, we hope that the computational methods developed here will be of good use in the
necessary extension to more general helicity configurations and other $2\to3$
scattering processes at two loops. They highlight some advantages of relating
two-loop integrands to tree-level amplitudes via generalized unitarity cuts.
As well as avoiding the large intermediate steps that make Feynman diagram computations
at this loop order and multiplicity extremely computationally intensive, we are able to
build known on-shell symmetries and relations into the amplitude by construction.

So far in this chapter we have not discussed the continuing connection between all-plus Yang Mills and $\cN=4$ SYM.  Using the five-point two-loop supersymmetric numerators given by Carrasco and Johansson in ref.~\cite{Carrasco:2011mn} we can easily confirm that the genuine two-loop nonplanar numerators given in section \ref{sec:npnumerators} are given by their supersymmetric counterparts under the replacement $\delta^8(Q)\to F_1$.  It is simply necessary to reduce the supersymmetric numerators onto the same basis of ISPs as we used for the all-plus.  As for the nonplanar butterfly graphs, there is still no obvious connection to $\cN=4$.

In the next chapter this connection will become important as we study colour-dual representations.  The additional numerator structure is enough to make off-shell BCJ symmetries nontrivial to satisfy, even though the $\cN=4$ integrands have already been cast in such a form.

\chapter{Colour-dual representations}\label{ch:allplusbcj}

\section{Introduction}

We now turn to another recent insight in (supersymmetric) Yang-Mills theory: colour-kinematics duality and the associated double copy relation between gauge theory and gravity, discovered by Bern, Carrasco and Johansson (BCJ)~\cite{Bern:2008qj, Bern:2010ue, Bern:2010yg}.  The duality offers our third and final way of presenting of two-loop all-plus integrands.

Colour-kinematics duality is the statement that numerators of trivalent Feynman-like diagrams may be chosen such that they satisfy the same algebraic relations as their respective colour factors --- we will review the duality in the next section.  At tree level, this fact leads to the BCJ identities~\cite{Bern:2008qj} that we used in the last chapter; these have been proven using a variety of techniques~\cite{BjerrumBohr:2009rd,Stieberger:2009hq,Feng:2010my,BjerrumBohr:2012mg,Cachazo:2012uq}. The existence of the numerators themselves has also been proven~\cite{Mafra:2011kj,BjerrumBohr:2012mg}. The double copy relation states that gravitational scattering amplitudes can be deduced from gauge scattering amplitudes, expressed in a form where colour-kinematics duality holds, by simply replacing the colour factors with a second set of gauge theoretic kinematic numerators. At tree level, this is equivalent to the celebrated KLT relations~\cite{Kawai:1985xq, Bern:2010yg}. 

At loop level, however, the situation is less clear.  The existence of numerators which satisfy the requirements of colour-kinematics duality (for brevity, we will call these colour-dual numerators in what follows) remains a conjecture in general.  Several infinite families of numerators exist at one loop; for example, in the case of pure Yang-Mills scattering amplitudes with all helicities equal, or only one different helicity~\cite{Boels:2013bi} and, more recently, for MHV scattering amplitudes at one loop in $\cN=4$ SYM~\cite{He:2015wgf}.

A major reason for our interest in colour-dual numerators is that they provide a promising route towards understanding the ultraviolet structure of supergravity.  This has motivated calculations of colour-dual numerators at four points in $\mathcal{N}=4$ SYM, which are now available at up to four loops~\cite{Bern:2012uf}. There is some flexibility in the structure of the double copy---although the double copy requires two sets of gauge theory numerators, they may be from different gauge theories and only one set of numerators needs to be colour-dual. Therefore the availability of a selection of colour-dual numerators in gauge theory has also allowed rapid progress in our understanding of non-maximally supersymmetric gravity~\cite{Bern:2011rj, BoucherVeronneau:2011qv, Bern:2012cd, Bern:2012gh, Carrasco:2012ca, Bern:2013yya, Nohle:2013bfa, Bern:2013qca, Bern:2013uka, Carrasco:2013ypa, Bern:2014lha, Bern:2014sna, Bern:2015xsa}. The flexibility of the double-copy allows the construction of a range of interesting different theories of gravity; understanding the structure of this set of gravity theories has developed into a vigorous area of research~\cite{Broedel:2012rc, Chiodaroli:2013upa, Chiodaroli:2014xia, Johansson:2014zca, Johansson:2015oia, Chiodaroli:2015rdg}.

There is intense interest in colour-kinematics duality at more than four loops~\cite{Bern:2011qn}.  However, notwithstanding recent developments involving contact terms~\cite{Bern:2017yxu}, construction of a set of numerators for the five-loop, four-point, $\mathcal{N} = 4$ SYM amplitude has proven to be difficult. An expression for the integrand of the amplitude is known~\cite{Bern:2012uc}, but finding an equivalent set of colour-dual numerators has been problematic. Given the large scale of the problem, it has also been difficult to locate the precise nature of the obstruction. This has motivated interest in finding scattering amplitudes which are simple enough to understand, but complicated enough that the colour-dual numerators are elusive. The idea is to understand the nature of obstructions to the existence of colour-dual numerators, with a view to identifying methods for overcoming these obstructions. For example, one recent suggestion is that the requirements of colour-kinematics duality can be relaxed so that they only hold on unitarity cuts~\cite{Bern:2015ooa}.

In this chapter we compute colour-dual numerators for the same five-gluon, two-loop amplitude that we have studied throughout this thesis.  Our starting point is, once again, the corresponding $\cN=4$ supersymmetric amplitude.  However, unlike in chapter~\ref{ch:localintegrands} where we used the local-integrand presentation of the supersymmetric amplitude~\cite{ArkaniHamed:2010kv,ArkaniHamed:2010gh}, in this chapter we use Carrasco and Johansson's colour-dual presentation~\cite{Carrasco:2011mn}.  As we will see, the problem of computing colour-dual numerators for the all-plus amplitude is surprisingly complicated. We find an obstruction to the existence of a set of colour-dual numerators containing at most 7 powers of loop momenta (as one would expect from power counting the Feynman rules.) This obstruction can be described as a tension between the value of one cut, and the symmetry properties of one of our graphs.

We resolve this tension by introducing extra powers of loop momentum into our numerators, obtaining in the end a set of numerators with 12 powers of loop momentum. It is typically a dangerous idea to consider such high powers of loop momenta, for the practical reason that a general ansatz with such high power counting will contain many terms. We circumvent this problem by identifying a desirable symmetry property of our BCJ master numerator. This symmetry is highly constraining, which made it quite feasible for us to increase the amount of loop momenta in our numerators.

\section{Review: colour-kinematics duality}

Here we review BCJ's colour-kinematics duality for tree- and loop-level amplitudes.  The duality is applicable to both supersymmetric and non-supersymmetric amplitudes in arbitrary external helicity configurations.  We also discuss a simple one-loop all-plus example.

\subsection{Tree level}

We begin by decomposing tree-level gauge-theory amplitudes into sums of diagrams containing only cubic vertices:
\begin{align}\label{eq:bcjtreedecomp}
\mathcal{A}_n^{(0)}(a_j,p_j)=
g^{n-2}\sum_{\text{cubic graphs }\Gamma_i}
\frac{c_i(a_j)n_i(p_j)}{D_i(p_j)}.
\end{align}
As we are dealing with pure-adjoint particle content the colour factors $c_i$ are given by contractions of structure constants $\tf^{abc}$ with external colour indices $a_i$ (an extension of colour-kinematics duality to include fundamental matter has been discussed in refs.~\cite{Johansson:2014zca,Johansson:2017bfl}).  The Feynman propagators are $D_i=\prod_{\alpha\in\Gamma_i}\cQ_\alpha(p_i)$ and $n_i$ are kinematic numerators containing the remaining kinematic information.

As a conventional Feynman diagram expansion includes four-point vertices, it does not generally conform to the above structure.  Four-point vertices can be removed by introducing additional propagators and adding compensating factors to the numerators.  For example,
\begin{align}
\includegraphics[scale=0.3,trim=0 60 0 0]{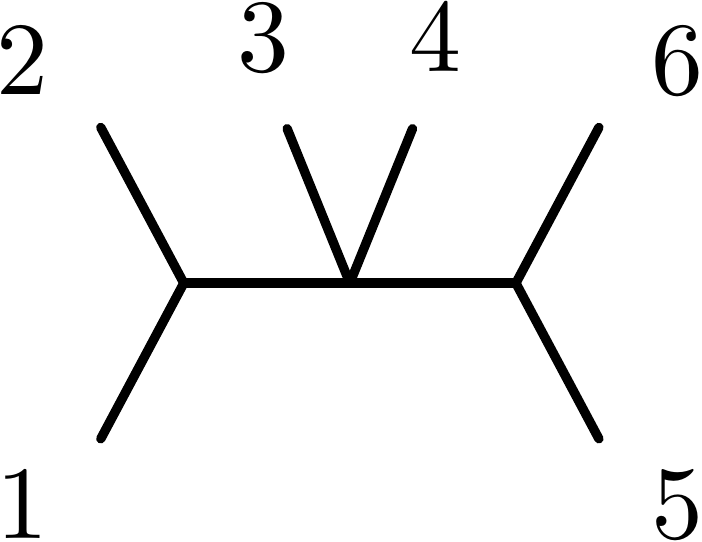}
=s_{34}\includegraphics[scale=0.3,trim=0 60 0 0]{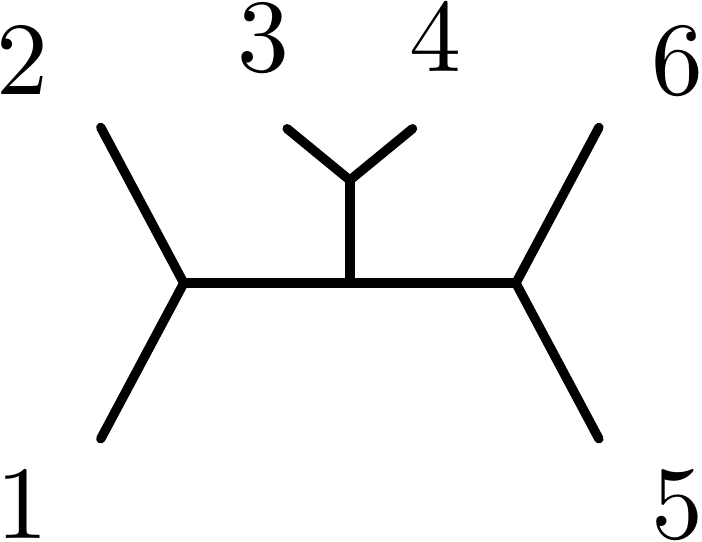}
=s_{123}\includegraphics[scale=0.3,trim=0 60 0 0]{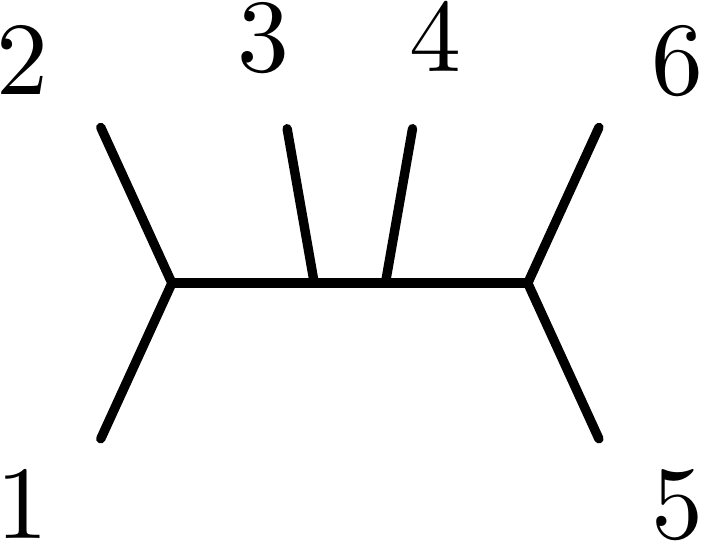},
\end{align}
which illustrates that this procedure is not unique.  As colour factors are built from structure constants $\tf^{abc}$, they are cubic in nature.  So we look at the relevant colour structure and choose additional propagators which ensure that the propagator structure matches the colour structure.

Structure constants satisfy the Jacobi identity $\tf^{a_1a_2b}\tf^{ba_3a_4}=\tf^{a_4a_1b}\tf^{ba_2a_3}-\tf^{a_4a_2b}\tf^{ba_1a_3}$.  Diagrammatically, this corresponds to a relation between the colour factors of the s-, t- and u-channels:
\begin{align}
c\bigg(\usegraph{8}{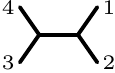}\bigg)
=c\bigg(\usegraph{13}{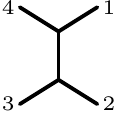}\bigg)-c\bigg(\usegraph{13}{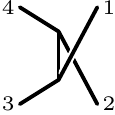}\bigg).
\end{align}
Associated with each propagator in a cubic diagram there is therefore a Jacobi relation of the form $c_i=c_j-c_k$, where $i$, $j$ and $k$ are identical diagrams except for the given propagator.  The kinematic numerators $n_i$, $n_j$ and $n_k$ are not uniquely determined: letting $n_i\to n_i+D_i\chi$, $n_j\to n_j-D_j\chi$ and $n_k\to n_k+D_k\chi$, for some arbitrary (and possibly non-local) function of external kinematics $\chi$, leaves $\mathcal{A}^{(0)}_n$ unchanged.  The freedom to do this on all internal propagators for all diagrams is known as generalised gauge invariance~\cite{Bern:2008qj}.

The statement of colour-kinematics duality when all particle content is pure-adjoint is that it is always possible to make a choice of kinematic numerators such that
\begin{align}\label{eq:jacobitree}
c_i=c_j-c_k \iff n_i=n_j-n_k
\end{align}
for all possible Jacobi triples $\{i,j,k\}$.  It is also a requirement that any diagrammatic symmetries of the colour factors are manifested in the kinematic numerators:
\begin{align}
c_i\to\pm c_i \iff n_i\to\pm n_i.
\end{align}
The kinematic structure then mirrors the algebraic structure of the colour factors.  The existence of such colour-dual numerators has been proven at tree level~\cite{Mafra:2011kj,BjerrumBohr:2012mg}.  The representations are not, in general, unique.

The existence of tree-level colour-dual representations has implications for colour-ordered amplitudes.  They must satisfy BCJ tree identities~\cite{Bern:2008qj}:
\begin{align}
\sum_{i=3}^n\left(\sum_{j=3}^is_{2j}\right)A^{(0)}\left(1,3,...,i,2,i+1,...,n\right)=0.
\end{align}
This formula has been proven using a variety of different techniques, including monodromy relations from string theory~\cite{BjerrumBohr:2009rd,Stieberger:2009hq}.  When $n=4$ it takes an especially simple form:
\begin{align}
s_{12}A^{(0)}(1,2,3,4)=s_{13}A^{(0)}(1,3,2,4),
\end{align} 
which we used in chapter~\ref{ch:allplusfull}~(\ref{BCJ4su}).

Furthermore, once a colour-dual representation has been found, replacing the colour factors $c_i$ in eq.~(\ref{eq:bcjtreedecomp}) with a second set of kinematic numerators $\tilde{n}_i$ leads to a wide variety of gravity amplitudes~\cite{Bern:2010ue,Bern:2010yg}
\begin{align}\label{eq:doublecopytree}
\mathcal{M}^{(0)}_n(p_j)=i\left(\frac{\kappa}{2}\right)^{n-2}\sum_{\text{cubic graphs }\Gamma_i}\frac{n_i(p_j)\tilde{n}_i(p_j)}{D_i},
\end{align}
where $\kappa$ is the gravitational coupling.  The second set of numerators $\tilde{n}_i$ need not be colour-dual.  They also need not correspond to the same theory; double copying different gauge theories enables the calculation of a wide variety of (supersymmetric) gravity amplitudes.

\subsection{Loop level}

The loop-level analogue of~(\ref{eq:bcjtreedecomp}) is~\cite{Bern:2010ue}
\begin{align}\label{eq_colourdecomploop}
\mathcal{A}_n^{(L)}(a_j,p_j)=i^{L-1}g^{n-2+2L}\sum_{\text{cubic graphs }\Gamma_i}
\int\left(\prod_{k=1}^L\frac{\d^d\ell_k}{(2\pi)^d}\right)\frac{1}{S_i}\frac{c_i(a_j)n_i(p_j,\ell_j)}
{D_i(p_j,\ell_j)},
\end{align}
where $S_i$ are symmetry factors and the dimensionality $d$ is undetermined (we may choose to use dimensional regularisation).  The kinematic Jacobi identities (\ref{eq:jacobitree}) generalise naturally, but with the extra complication that the numerators may now be functions of loop momenta,
\begin{align}\label{eq:jacobiloop}
c_i=c_j-c_k \iff n_i(\ell)=n_j(\ell)-n_k(\ell).
\end{align}
One must choose the loop momenta of individual Feynman diagrams so that the momenta between their shared propagators match.

The kinematic Jacobi identities can be used to define all graph numerators in terms of a small set of numerators known as masters.  The set of masters is not always unique, but given a set of masters the numerators of all other graphs, known as descendants, are uniquely determined via Jacobi identities.  For example, the triangle graph can be evaluated as a difference of boxes:
\begin{align}\label{eq:bcjtriangle}
n\bigg(\usegraph{10}{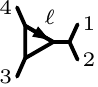}\bigg)
=n\bigg(\usegraph{10}{box}\bigg)-n\bigg(\usegraph{10}{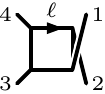}\bigg).
\end{align}
At one loop $n$ points one can always use the $n$-gon diagram as a unique master.  The problem of finding a colour-dual representation for a given amplitude therefore amounts to finding colour-dual master numerators.  The ability to do this in general remains a conjecture.

Once suitable numerators have been found one can write down corresponding gravity amplitudes.  The loop-level analogue of eq.~(\ref{eq:doublecopytree}) is
\begin{align}\label{eq:doublecopyloop}
\mathcal{M}_n^{(L)}(p_j)=i^{L-1}\left(\frac{\kappa}{2}\right)^{n-2+2L}\!\!\!\!\!\!\sum_{\text{cubic graphs }\Gamma_i}\int\left(\prod_{k=1}^L\frac{\d^d\ell_k}{(2\pi)^d}\right)\frac{1}{S_i}\frac{n_i(p_j,\ell_j)\tilde{n}_i(p_j,\ell_j)}{D_i(p_j,\ell_j)},
\end{align}
where we only require that one of the sets of numerators $n_i$ or $\tilde{n}_i$ be colour dual.  It is generally much easier to calculate loop-level gravity amplitudes using this formula than using traditional Feynman diagrams --- this is one of the main motivations for finding colour-dual numerators.

\subsection{One-loop all-plus example}\label{sec:oneloopbcj}

To illustrate these basic concepts of colour-kinematics duality we briefly revisit the four-gluon one-loop all-plus amplitude introduced in section~\ref{sec:1lreview}.  There we showed that the simple replacement $\delta^8(Q)\to(D_s-2)\mu^4$ can be used to obtain all-plus integrands from their supersymmetric counterparts.  The extra-dimensional factor $\mu^4$ and the supersymmetric delta function $\delta^8(Q)$ are both invariant under all shifts of external and loop momenta.  So this one-loop replacement rule works equally well for colour-dual numerators.  Given a one-loop colour-dual representation in $\cN=4$ we can trivially obtain one for the all-plus.

With four external legs the only nonzero numerator is the box:
\begin{align}
n\bigg(\usegraph{10}{box}\bigg)=(D_s-2)\mu^4,
\end{align}
which trivially satisfies all its symmetries.  As with the two-loop irreducible numerators in section~\ref{sec:2lreview} we have extracted a factor of the permutation-invariant factor $\cT$, previously defined as
\begin{align}
\cT=\frac{[12][34]}{\braket{12}\braket{34}}.
\end{align}
The triangles vanish by the Jacobi identity (\ref{eq:bcjtriangle}), as do the bubbles and tadpoles.  So we can write down the complete amplitude as
\begin{align}
\cA_{++++}^{(1)}
=g^4\cT(D_s-2)\sum_{\sigma\in S_4}\frac{1}{8}c\bigg(\usegraph{10}{boxi}\bigg)I^D\bigg(\usegraph{10}{box}\bigg)[\mu^4],
\end{align}
which is the presentation that we gave in chapter~\ref{ch:review}, eq.~(\ref{eq:1l4gddm}).

\section{Four points, two loops}\label{sec:4points}

We will now discuss the calculation \cite{Bern:2013yya} of a set of colour-dual numerators for the four-point, two-loop, all-plus case.  This will be an important warm up to the five-point, two-loop calculation.  As we will see, some aspects of the five-point system are closely analogous to the four-point case.

\subsection{Master numerators}

This BCJ system requires two masters; appropriate expressions were found to be~\cite{Bern:2013yya}
\begin{subequations}\label{eq:4ptmastersold}
\begin{align}
\label{eq:n331old}
n\bigg(\usegraph{9}{331}\bigg)
&=s\,F_1+\frac{1}{2}(\ell_1+\ell_2)^2F_3+(\ell_1+\ell_2)^2F_2,\\
\label{eq:n322old}
n\bigg(\!\usegraph{9}{322}\bigg)
&=s\,F_1,
\end{align}
\end{subequations}
where $F_1$, $F_2$ and $F_3$ are the functions of $\mu_1$ and $\mu_2$ introduced in chapter~\ref{ch:review}:
\begin{subequations}
\begin{align}
&F_1(\mu_1,\mu_2)\\
&\qquad
=(D_s-2)(\mu_{11}\mu_{22}+(\mu_{11}+\mu_{22})^2
+2\mu_{12}(\mu_{11}+\mu_{22}))+16(\mu_{12}^2-\mu_{11}\mu_{22}),\nn\\
&F_2(\mu_1,\mu_2)
=4(D_s-2)\mu_{12}(\mu_{11}+\mu_{22}),\\
&F_3(\mu_1,\mu_2)
=(D_s-2)^2\mu_{11}\mu_{22}.
\end{align}
\end{subequations}
We observe that the function $F_2$ can also be defined in terms of $F_1$ as
\begin{align}
F_2(\mu_1,\mu_2)&=
F_1(\mu_1,\mu_2)-F_1(\mu_1,-\mu_2).
\end{align}
As in previous chapters we have extracted a factor of $\cT$ which carries the little-group weight of the amplitude.

All other numerators in this system are descendents, so can be computed from these two masters using appropriate Jacobi identities (a complete list was provided in ref.~\cite{Bern:2013yya}).  For example, at four points the double-triangle graph is a difference of double-box graphs:
\begin{equation}\label{eq:n330example}
n\bigg(\usegraph{9}{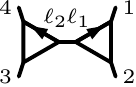}\bigg)
=n\bigg(\usegraph{9}{331}\bigg)-n\bigg(\usegraph{9}{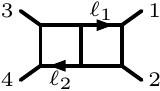}\bigg).
\end{equation}
Since the Jacobi relations are all simple linear combinations of graphs in various different orderings (perhaps with shifted loop momenta), the descendent numerators are all local functions of external and loop momenta since the master numerators have this property.

We will determine a set of colour-dual master numerators for the five-point, two-loop, all-plus Yang-Mills amplitude below. In structure, our numerators will be very similar to the numerators given in eqs.~\eqref{eq:4ptmastersold}. The similarity can be made even closer by observing that the final term in the double box, eq. (\ref{eq:n331old}), namely $(\ell_1+\ell_2)^2F_2$, is not necessary.
To see this, notice that the set of Jacobi equations is a set of linear equations. Consider setting the double box to this term alone and setting the nonplanar master to zero --- that is, take
\begin{align}
\label{eq:fakemasters}
n\bigg(\usegraph{9}{331}\bigg)
=(\ell_1+\ell_2)^2F_2(\mu_{1},\mu_{2}),\qquad
n\bigg(\!\usegraph{9}{322}\bigg)
=0.
\end{align}
We will now show that the resulting amplitude contribution vanishes
and that all descendent symmetries and automorphisms are satisfied. 
Linearity of the system then allows us to conclude that we may
omit the term $(\ell_1+\ell_2)^2F_2$ from eq. \eqref{eq:n331old}.

Starting from the masters in eq. \eqref{eq:fakemasters}, the only diagram that gives a nonzero contribution
upon integration besides the double box is the double triangle,
determined by the Jacobi identity in eq. (\ref{eq:n330example}) to be
\begin{align}\label{eq:n330}
n\bigg(\usegraph{9}{n330}\bigg)
&=n\bigg(\usegraph{9}{331}\bigg)-n\bigg(\usegraph{9}{331o1243}\bigg) \\
&= (\ell_1^2 + \ell_2^2 + (\ell_1 - p_{12})^2 + (\ell_2+p_{12})^2 -s)F_2.
\end{align}
Thus, the complete contribution to the full colour-dressed amplitude is
\begin{align}\label{eq:4ptnullcontributio}
\mathcal{A}&=ig^6\mathcal{T}\sum_{\sigma\in S_4}
\bigg\{\frac{1}{4}c\bigg(\usegraph{9}{331i}\bigg)I\bigg(\usegraph{9}{331}\bigg)\left[(\ell_1+\ell_2)^2F_2\right]\nonumber\\
&+\frac{1}{8}c\bigg(\usegraph{9}{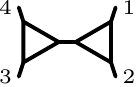}\bigg)I\bigg(\usegraph{9}{330}\bigg)\left[s^{-1}(\ell_1^2+\ell_2^2+(\ell_1-p_{12})^2+(\ell_2+p_{12})^2-s)F_2\right]\bigg\}\nonumber \\
&=ig^6\mathcal{T}\sum_{\sigma\in S_4}
\bigg\{\frac{1}{4}c\bigg(\usegraph{9}{331i}\bigg)I\bigg(\usegraph{9}{330}\bigg)\left[F_2\right]
\nonumber\\
&\qquad
-\frac{1}{8}c\bigg(\usegraph{9}{n330i}\bigg)I\bigg(\usegraph{9}{330}\bigg)\left[F_2\right]\bigg\},
\end{align}
where we cancelled propagators in numerators and denominators and disposed of scaleless integrals.  Since the integrands in the two terms are the same, we may combine them and use the Jacobi identity (\ref{eq:n330}) on the colour factors to find
\begin{align}
\mathcal{A}&=
ig^6\mathcal{T}\sum_{\sigma\in S_4}\left\{\frac{1}{8}\left(c\bigg(\usegraph{9}{331i}\bigg)+c\bigg(\usegraph{9}{331i1243}\bigg)\right)I\bigg(\usegraph{9}{330}\bigg)\left[F_2\right]\right\}\nonumber\\
&=ig^6\mathcal{T}\sum_{\sigma\in S_4}\left\{\frac{1}{8}c\bigg(\usegraph{9}{331i}\bigg)I\bigg(\usegraph{9}{330}\bigg)\left[F_2(\mu_1,\mu_2)+F_2(\mu_1,-\mu_2)\right]\right\} \nonumber\\
&=0,
\end{align}
where we exploited the sum on $S_4$ permutations to relabel $p_3\leftrightarrow p_4$
while also shifting $\ell_2\to-\ell_2-p_{12}$ in the second integral.
Finally, we have used the fact that $F_2(\mu_1,\mu_2)=-F_2(\mu_1,-\mu_2)$.

It remains to check that symmetries and automorphisms of all descendent graphs are satisfied: we have done this exhaustively for all graphs; while this is not difficult since many are vanishing, we also found private code written by Tristan Dennen to be very helpful~\cite{tristanCode}.  Consequently, an equivalent set of master numerators is
\begin{subequations}\label{eq:4ptmasters}
\begin{align}
\label{eq:n331}
n\bigg(\usegraph{9}{331}\bigg)
&=s\,F_1+\frac{1}{2}(\ell_1+\ell_2)^2F_3,\\
\label{eq:n322}
n\bigg(\!\usegraph{9}{322}\bigg)
&=s\,F_1.
\end{align}
\end{subequations}
We will find a closely related set of five-point two-loop masters below.

\subsection{Bubbles and tadpoles}\label{sec:bubblestadpoles}

\begin{figure}
\centering
\begin{subfigure}{0.25\textwidth}
  \centering
  \includegraphics{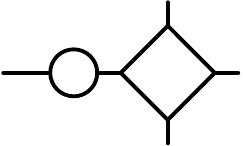}
  \caption{Bubble}
\end{subfigure}
\begin{subfigure}{0.25\textwidth}
  \centering
  \includegraphics{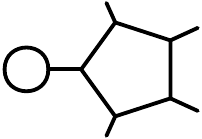}
  \caption{Tadpole}
\end{subfigure}
\caption{\small Example bubble and tadpole diagrams.  In each case there is an ill-defined intermediate propagator between the loops - for the bubble this is due to the on-shell condition $p^2=0$ for external legs.}
\label{fig:bubbletadpole}
\end{figure}

This BCJ system presents us with a new problem: the inclusion of nonzero numerators whose graphs contain bubbles on external legs or tadpoles, such as those displayed in Figure~\ref{fig:bubbletadpole}.  The issue was discussed in ref.~\cite{Nohle:2013bfa}: such diagrams are a problem for any BCJ system at the integrand level due to singularities coming from ill-defined propagators.  In previous approaches, such as the one-loop example discussed in ref.~\cite{Bern:2013yya}, the solution has been to demand vanishing of these numerators; however, in the four-point example currently under consideration we cannot find a set of colour-dual numerators that satisfy this property.

Bubble-on-external-leg and tadpole integrals vanish in dimensional regulation for massless external states as they lack a mass scale.  This zero has the potential to ``cancel'' singularities coming from the ill-defined propagators, leaving a finite result after integration.  To check this we perform a simple bubble integral with numerator $\mu^{2r}$ in $D=4-2\eps$ dimensions, taking the external leg $p$ off shell:
\begin{align}\label{eq:bubbleintegral}
I^D\bigg(\usegraph{6}{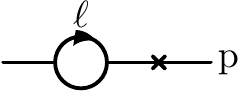}\bigg)[\mu^{2r}]&=
\frac{1}{p^2}\int\!\!
\frac{\d^D\ell}{(2\pi)^{D}}\frac{\mu^{2r}}{\ell^2(\ell-p)^2}\nnl
&=-\eps(1-\eps)\cdots(r-1-\eps)\frac{(4\pi)^r}{p^2}\int\!\!
\frac{\d^{D+2r}\ell}{(2\pi)^{D+2r}}\frac{1}{\ell^2(\ell-p)^2}\nnl
&\sim(p^2)^{-\eps+r-1}.
\end{align}
In the first line we used the one-loop dimension shift (\ref{eq:1ldimshift}) and in the second a simple Feynman parametrisation.  We see how the result scales with $p^2$: as $\eps<0$ and we take $r\geq1$ this always vanishes in the on-shell limit $p^2\to0$.

So bubble-on-external-leg integrals are safe after integration, provided their numerators contain an overall factor of $\mu^2$.  This is always true at four points and we shall see that it is also true at five points.  As for the tadpoles, one can perform a similar analysis by giving the internal propagator a small mass - one arrives at the same conclusion.  Therefore we will ignore these graphs, safe in the knowledge that they vanish after integration.

\subsection{The maximally-supersymmetric subsector}

A set of BCJ master numerators is also available~\cite{Bern:2008qj} for the two-loop, $\mathcal{N} =4$ four-point amplitude. They are simply given by
\begin{subequations}\label{eq:4ptN4numerators}
\begin{align}
n^{[\mathcal{N}=4]}\bigg(\usegraph{9}{331i}\bigg) &= s\,\delta^8(Q), \\
n^{[\mathcal{N}=4]}\bigg(\!\usegraph{9}{322i}\bigg) &=s\,\delta^8(Q).
\end{align}
\end{subequations}
Comparing these master numerators to the simplified all-plus master numerators given in eqs. (\ref{eq:4ptmasters}) reveals an $\mathcal{N}=4$ SYM subsector of the all-plus amplitude, generated by terms containing up to one power of $(D_s-2)$.  As discussed in section~\ref{sec:2lreview}, the function $F_1$ of extra-dimensional components plays the role of the supersymmetric delta function $\delta^8(Q)$.  However, unlike the factor of $(D_s-2)\mu^4$ used at one-loop order, $F_1$ is not invariant under all shifts in loop momentum.  So this subsector is not closed under Jacobi relations.

In any given Jacobi relation, all propagators but one are the same among the three diagrams. We can therefore associate each relation with one propagator.  For our purposes, it us useful to divide the set of Jacobi relations into two categories: those that preserve the $(\ell_1+\ell_2)^2$ propagator in (\ref{eq:n331}), and those that act on it.  It is easy to see that Jacobi moves of the first kind leave $F_1$ unaffected, so descendent diagrams which can be formed using only this category of Jacobi relations belong to the $\mathcal{N}=4$ subsector.  An example of the second kind of move was given in eq. (\ref{eq:n330example}), where if we evaluate the numerator we find
\begin{align}\label{eq:n3302}
n\bigg(\usegraph{9}{n330}\bigg)
=s\,F_2+\frac{1}{2}\left((\ell_1+\ell_2)^2-(\ell_1-\ell_2-p_{12})^2\right)F_3,
\end{align}
using $F_2(\mu_1, \mu_2) = F_1(\mu_1, \mu_2) - F_1(\mu_1, -\mu_2)$.  This numerator vanishes in the supersymmetric case as it contains triangles; the fact that $F_1$ does not have the full symmetries of $\delta^8(Q)$ now prevents this from happening.  These butterfly diagrams are generated by BCJ moves from the masters that take them outside the $\mathcal{N}=4$ subsector.

\subsection{Spanning cuts}

\begin{figure}[t]
\centering
\begin{subfigure}{0.25\textwidth}
  \centering
  \includegraphics[width=0.8\textwidth]{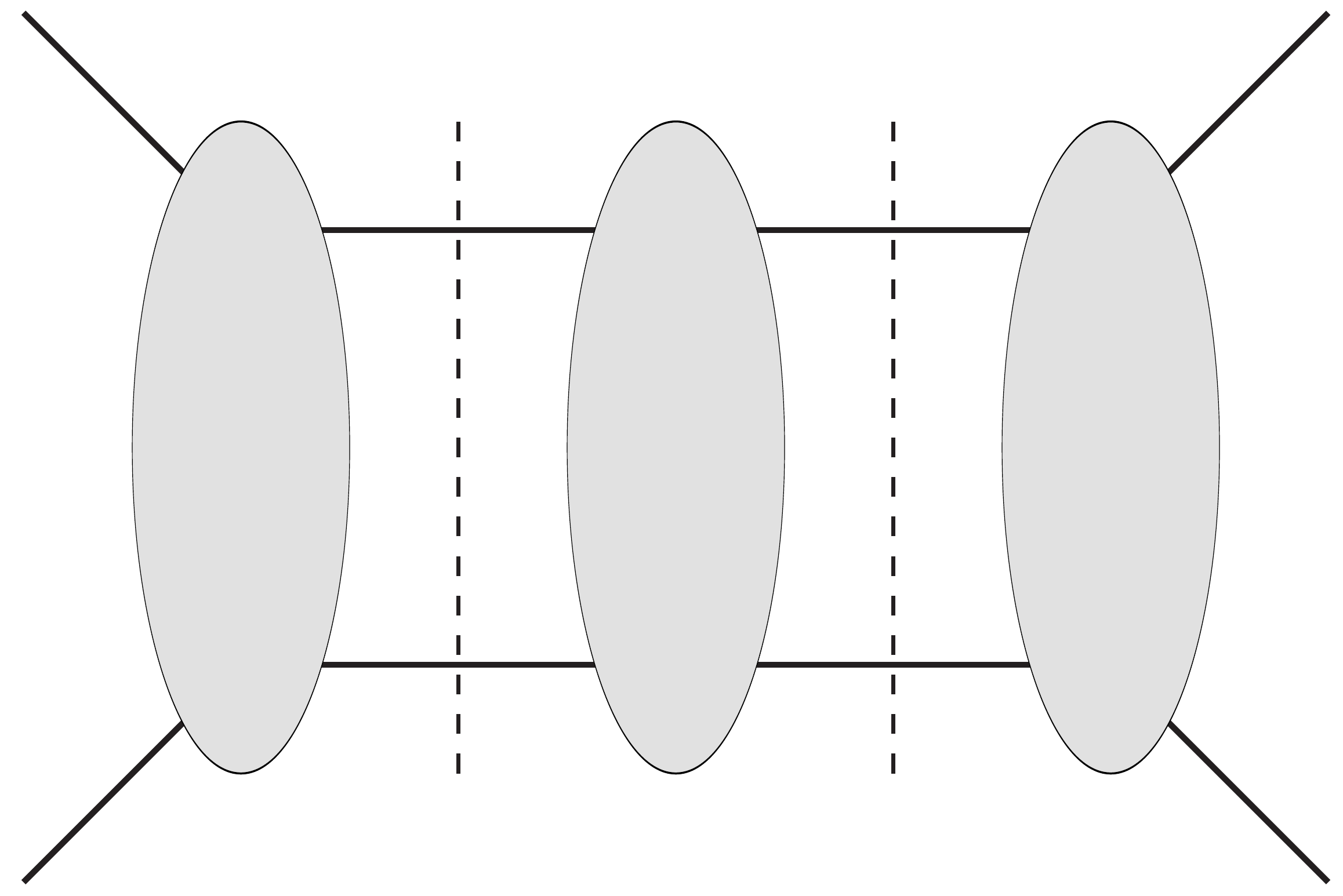}
  \caption[a]{}
\end{subfigure}
\begin{subfigure}{0.25\textwidth}
  \centering
  \includegraphics[width=0.8\textwidth]{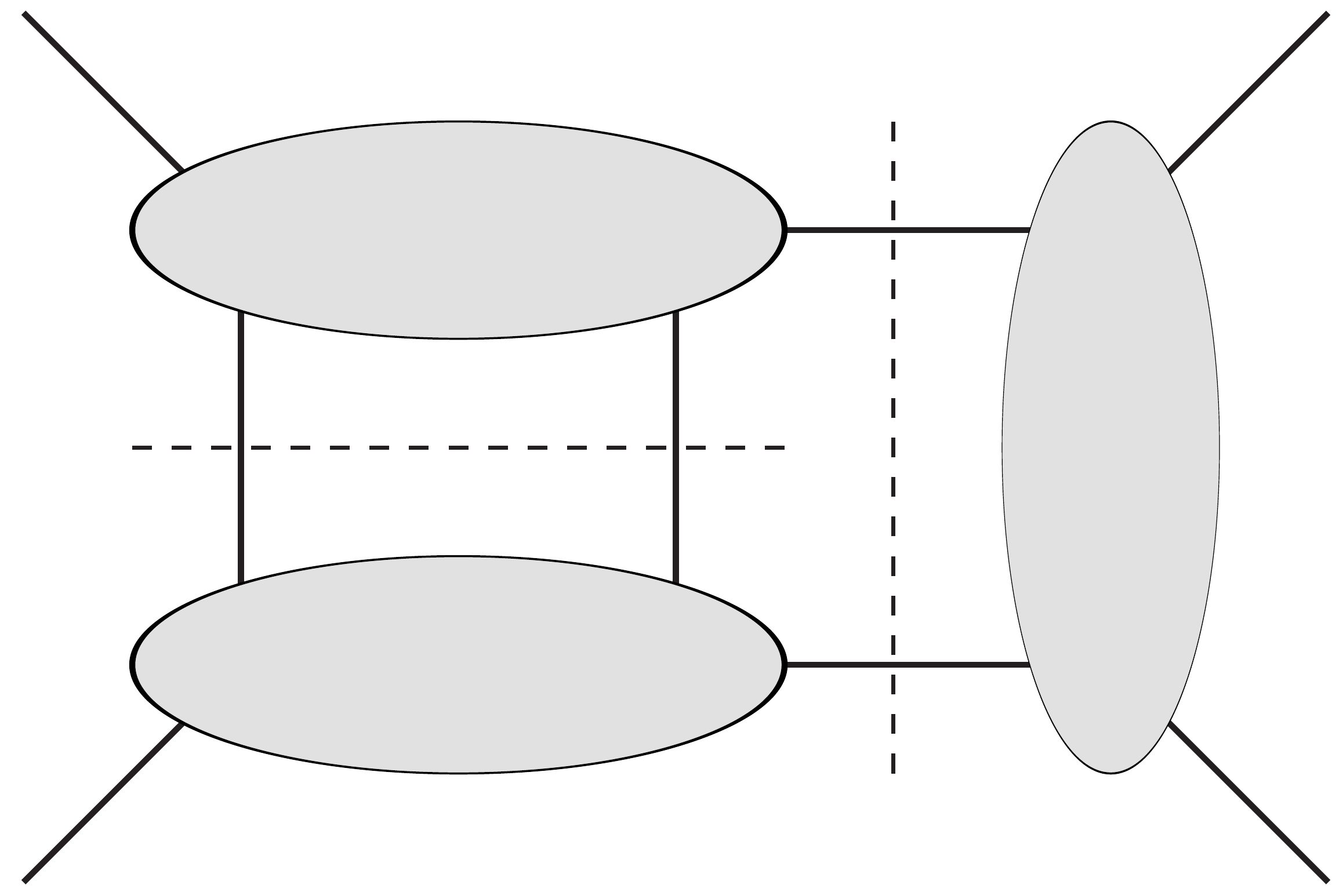}
  \caption[b]{}
\end{subfigure}
\begin{subfigure}{0.25\textwidth}
  \centering
  \includegraphics[width=0.8\textwidth]{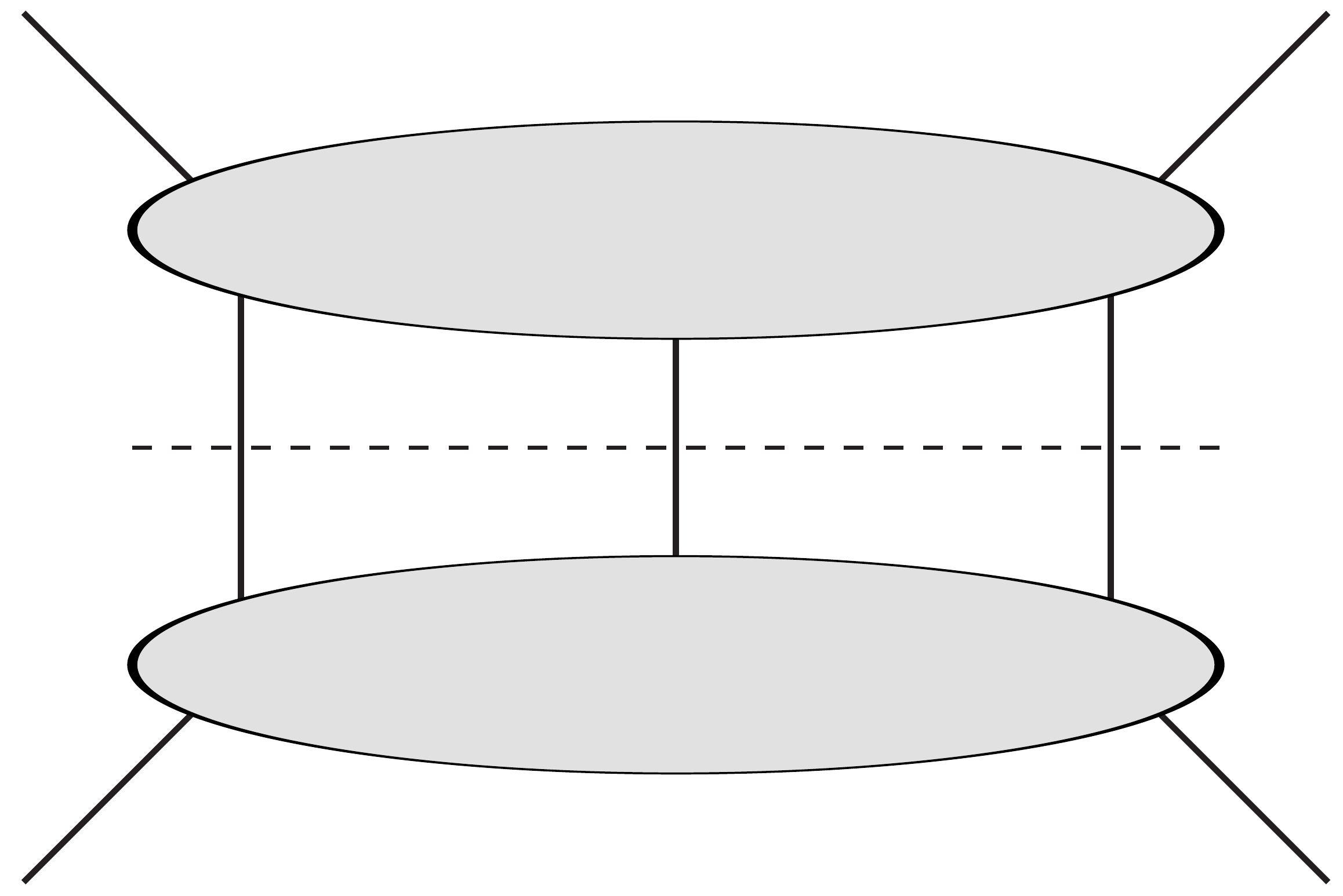}
  \caption[c]{}
\end{subfigure}
\caption{\small The three two-loop spanning cuts required at four points. The first is a butterfly cut; the 				other two are satisfied automatically by the $\mathcal{N}=4$ subsector.
         \label{fig:4ptspanningcuts}}
\end{figure}

In order to show that the BCJ presentation of the amplitude, generated by the two masters in eq. (\ref{eq:4ptmasters}), reproduces the full amplitude, it is sufficient to show that it reproduces the three spanning cuts displayed in Figure~\ref{fig:4ptspanningcuts}.  While one could calculate these using tree-level amplitudes, we find it simpler to start from the full presentation of the amplitude given in section~\ref{sec:4g2lreview}.  As we also remarked in this section, a subset of the cuts of $\mathcal{N}=4$ and all-plus two-loop amplitudes are related by replacing the supersymmetric delta function $\delta^8(Q)$ with the function $F_1$.  Cuts involving butterfly topologies have a different structure.

With this in mind, we consider cuts (b) and (c).  The only diagrams contributing to these (or any) cuts are those that include all of the cut propagators.  For cuts (b) and (c), cutting the central $(\ell_1+\ell_2)^2$ propagator ensures that all constituent diagrams from the BCJ presentation of the amplitude belong to the $\mathcal{N}=4$ subsector.  In other words, the only contributing diagrams are also present in the five-point, two-loop $\cN=4$ SYM supersymmetric amplitude.  As this amplitude does not contain any butterflies, terms proportional to $(D_s-2)^2$ vanish.  Therefore, on these cuts the all-plus colour-dual numerators are precisely equal to their $\mathcal{N}=4$ counterparts, with the replacement $\delta^8(Q)\to F_1$.  This is consistent with the BCJ presentation of the amplitude.

Cut (a) contains butterflies, and therefore requires a little more work.
We need only consider the planar colour-stripped form:
it is known that nonplanar information is encoded in the planar cut,
as discussed by BCJ in ref.~\cite{Bern:2008qj}.\footnote{Indeed, this point was essential for finding the nonplanar numerators in chapter~\ref{ch:allplusfull}.}  A simple check confirms that this cut is satisfied for terms up to a single power of $(D_s-2)$,
despite the new factors of $F_2$.
For terms proportional to $(D_s-2)^2$,
the full expression for the planar colour-stripped form of cut (a),
given in terms of irreducible numerators, is
\begin{align}\label{eq:cut220}
\left.(\ell_1-p_1)^2(\ell_2-p_4)^2\text{Cut}\bigg(\usegraph{9}{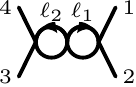}\bigg)\right|_{(D_s-2)^2}
&=\left.\Delta\bigg(\usegraph{9}{330}\bigg)\right|_{(D_s-2)^2}\nonumber\\
&=\mu_{11}\mu_{22}\left(\frac{s+(\ell_1+\ell_2)^2}{s}\right),
\end{align}
where the relevant cut conditions are $\ell_1^2=\ell_2^2=(\ell_1-p_{12})^2=(\ell_2+p_{12})^2=0$.
In terms of BCJ numerators,
the only nonzero numerators that contribute are the double box, eq. (\ref{eq:n331}),
and the double triangle, eq. (\ref{eq:n3302}).
The cut in this presentation becomes
\begin{align}\label{eq:cut220bcj}
&\left.(\ell_1-p_1)^2(\ell_2-p_5)^2\text{Cut}\bigg(\usegraph{9}{220}\bigg)\right|_{(D_s-2)^2}\nonumber\\
&\qquad=\left.\left(\frac{1}{(\ell_1+\ell_2)^2}n\bigg(\usegraph{9}{331}\bigg)
+\frac{1}{s}n\bigg(\usegraph{9}{n330}\bigg)\right)\right|_{(D_s-2)^2}\nn\\
&\qquad=\mu_{11}\mu_{22}\left(\frac{s+(\ell_1+\ell_2)^2}{s}\right)
\end{align}
as required.
At five points,
cut equations similar to this one will be important for us.

\section{Five points, two loops}\label{sec:5points}

At five points and two loops,
the BCJ presentation of the $\mathcal{N}=4$  amplitude,
previously computed by Carrasco \& Johansson (CJ) in ref.~\cite{Carrasco:2011mn},
forms our starting point for the all-plus calculation.  CJ found it useful to introduce a set of prefactors $\gamma_{ij}$ that generalise the four-point prefactor $\mathcal{T}$~(\ref{eq:cT}) to five points.  These prefactors encode all external state dependence on helicity: we find them to be equally applicable to the all-plus calculation as to the supersymmetric one.\footnote{These objects also prove useful for the five-point tree amplitude~\cite{Broedel:2011pd}.}

The kinematic prefactors $\gamma_{ij}$ are defined in terms of the objects $\beta_{ijklm}$ as
\begin{align}\label{eq:cjgammas}
\beta_{12345}=i\frac{[12][23][34][45][51]}{\trfive(1234)},\qquad
\gamma_{12}=\beta_{12345}-\beta_{21345}=i\frac{[12]^2[34][45][35]}{\trfive(1234)}
\end{align}
in the standard spinor-helicity formalism.\footnote{CJ also include the supersymmetric delta function $\delta^8(Q)$ in their definition of $\beta_{12345}$; for notational convenience we include this delta function elsewhere.
}  As $\gamma_{12}$ is totally symmetric on external legs $p_3$, $p_4$ and $p_5$,
we drop these last three subscripts.
The prefactors satisfy linear relations
\begin{align}
\sum_{i=1}^5\gamma_{ij}=0,\qquad
\gamma_{ij}=-\gamma_{ji},
\end{align}
so only six $\gamma_{ij}$ are linearly independent,
though they satisfy more complex relationships when kinematic factors $s_{ij}$ are involved.
We are also able to write
\begin{align}\label{eq:betagamma}
\beta_{12345}=\frac{1}{2}(\gamma_{12}+\gamma_{23}+\gamma_{13}+\gamma_{45}).
\end{align}

The all-plus pentabox numerator by itself is a master for the five-point, two-loop all-plus amplitude,
in the same way that the $\mathcal{N}=4$ pentabox is for the corresponding supersymmetric amplitude.
By analogy to the double-box numerator given in eq. (\ref{eq:n331}),
we propose that
\begin{align}\label{eq:n431}
n\bigg(\usegraph{9}{431}\!\bigg)=
F_1\,\tilde {n}^{[\mathcal{N}=4]}\bigg(\usegraph{9}{431}\!\bigg)
+(\ell_1+\ell_2)^2F_3X(12345;\ell_1,\ell_2),
\end{align}
where $X$ is an unknown function of external and loop momenta  which we must determine,
while $\tilde {n}^{[\mathcal{N}=4]}$ is the coefficient of the supersymmetric delta function in CJ's $\mathcal{N}=4$ supersymmetric pentabox numerator:
\begin{align}\label{eq:n431N4}
&n^{[\mathcal{N}=4]}\bigg(\usegraph{9}{431}\!\bigg) = \delta^{(8)}(Q) \tilde {n}^{[\mathcal{N}=4]}\bigg(\usegraph{9}{431}\!\bigg) \nonumber\\
&\qquad=\frac{1}{4}\delta^8(Q)\big(\gamma_{12}(2s_{45}-s_{12}+2\ell_1\cdot(p_2-p_1))
+\gamma_{23}(s_{45}+2s_{12}+2\ell_1\cdot(p_3-p_2))\nonumber\\
&\qquad\qquad+4\gamma_{45}\,\ell_1\cdot(p_5-p_4)
+\gamma_{13}(s_{12}+s_{45}+2\ell_1\cdot(p_3-p_1))\big).
\end{align}
For notational convenience below, we will drop the tilde on $\tilde {n}^{[\mathcal{N}=4]}$.

The all-plus pentabox numerator shares many of the properties of the all-plus double box.
There is a non-closed $\mathcal{N}=4$ subsector generated by terms containing up to one power of $(D_s-2)$,
where once again the extra-dimensional function $F_1$, given in eq. (\ref{eq:F1}),
plays the role of the supersymmetric delta function $\delta^8(Q)$.
Jacobi relations are again divided into two categories:
those that preserve both $F_1$ and $(\ell_1+\ell_2)^2$,
and those that act upon them.
The unknown function $X$ gives us information about butterfly topologies.
However, unlike the situation at four points,
we will show that $X$ is necessarily nonlocal in external kinematics.

\subsection{Symmetries and automorphisms}
\label{sec:5ptsymmetry}

We choose to recycle another desirable property of the four-point solution:
the absence of terms proportional to $(D_s-2)^2$ in all nonplanar numerators.
Although there is no five-point equivalent of the four-point nonplanar master, eq. (\ref{eq:n322}),
we can still impose this condition using the following nonplanar numerator:
\begin{align}\label{eq:n332}
n\bigg(&\usegraph{9}{332}\bigg)
=n\bigg(\usegraph{9}{431}\!\bigg)-n\bigg(\!\usegraph{9}{431o34512}\bigg)
=F_1(\mu_1,\mu_2)n^{[\mathcal{N}=4]}\bigg(\usegraph{9}{332}\bigg) \nonumber \\
&+(\ell_1+\ell_2)^2F_3(\mu_1,\mu_2)
(X(12345;\ell_1,\ell_2)-X(34512;-p_{12}-\ell_2,p_{12}-\ell_1)).
\end{align}

All nonplanar numerators at five points can be generated from this graph alone using Jacobi identities.
In this sense, it is a master of nonplanar graphs.
Therefore, vanishing of terms proportional to $(D_s-2)^2$ in this numerator is necessary and sufficient
to guarantee the same property for all nonplanar graphs.
Imposing this requirement, we learn that
$X(12345;\ell_1,\ell_2)=X(34512;-p_{12}-\ell_2,p_{12}-\ell_1)$.

Another property of $X$ follows from the overall flip symmetry of the pentabox through a horizontal axis,
namely $X(12345;\ell_1,\ell_2)=-X(32154;-p_{45}-\ell_1,p_{45}-\ell_2)$.
Both of these properties are functional identities,
holding for any permutation of external legs and any shift in the loop momenta.
By applying the two conditions to each other repeatedly and performing relabellings,
we can refine them into three simple properties of $X$:
\begin{align}\label{eq:Xsymmetries}
X(12345;\ell_1,\ell_2)=\left\{ 
  \begin{array}{l}
    X(23451;\ell_1-p_1,\ell_2+p_1), \\
    -X(54321;\ell_2,\ell_1), \\ 
    X(12345;-\ell_2,-\ell_1).
  \end{array} \right. 
\end{align}
These identities will be important below.\footnote{At four points, the double box satisfies an analogous set of symmetries, with the constant $1/2$ playing the role of $X$. As the number of vertices (six) is even, the sign on the second equation is positive.}

We have exhaustively checked that these three conditions are sufficient to guarantee
all symmetries and automorphisms for all descendent BCJ numerators in the entire system.
For instance, the nonplanar numerator in eq. (\ref{eq:n332}) now equals its $\mathcal{N}=4$ term alone,
so it satisfies its symmetries by virtue of the $\mathcal{N}=4$ numerator having exactly the same properties.

From the planar sector,
a more nontrivial example of these symmetries in action comes from the ``hexatriangle'' diagram:
\begin{align}\label{eq:n521}
n\bigg(\!\!\usegraph{13}{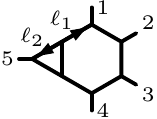}\bigg)
&=n\bigg(\usegraph{9}{431}\!\bigg)-n\bigg(\!\usegraph{9}{422}\!\bigg)\nonumber\\
&=(\ell_1+\ell_2)^2X(12345;\ell_1,\ell_2)F_3.
\end{align}
Here we have used the fact that
\begin{align}\label{eq:N4planarnonplanar}
n^{[\mathcal{N}=4]}\bigg(\usegraph{9}{431}\!\bigg)
=n^{[\mathcal{N}=4]}\bigg(\!\usegraph{9}{422}\!\bigg),
\end{align}
and that the nonplanar numerator in eq. (\ref{eq:n521}) has no terms carrying powers of $(D_s-2)^2$.
As expected, the hexatriangle has no $\mathcal{N}=4$ component (it contains an internal triangle).
Its symmetry through a horizontal axis implies
$X(12345;\ell_1,\ell_2)=-X(43215;-p_5-\ell_1,p_5-\ell_2)$,
which follows from the three conditions in eq. (\ref{eq:Xsymmetries}).
The Jacobi identity in eq. (\ref{eq:n521}) partitions the pentabox into its $\mathcal{N}=4$
and pure YM components.

Returning to the issue of bubble diagrams discussed in section~\ref{sec:bubblestadpoles}, we can calculate the ``master'' of bubbles as
\begin{align}
n\bigg(\!\usegraph{16}{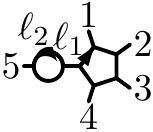}\bigg)
&=n\bigg(\!\!\usegraph{13}{521}\bigg)-n\bigg(\!\!\usegraph{13}{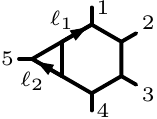}\bigg)\\
&=F_3\big((\ell_1+\ell_2)^2X(12345;\ell_1,\ell_2)\nn\\
&\qquad-(\ell_1+p_5-\ell_2)^2X(12345;\ell_1,p_5-\ell_2)\big).\nn
\end{align}
All other bubble numerators can be generated using Jacobi identities from this numerator, and we see that they will always carry a factor of $\mu_{22}$ on the bubble loop.  So they are safe under integration; again, a similar logic can be applied to the tadpoles.

\subsection{Spanning cuts}
\label{sec:5ptcuts}

\begin{figure}[t]
\centering
\begin{subfigure}{0.25\textwidth}
  \centering
  \includegraphics[width=0.8\textwidth]{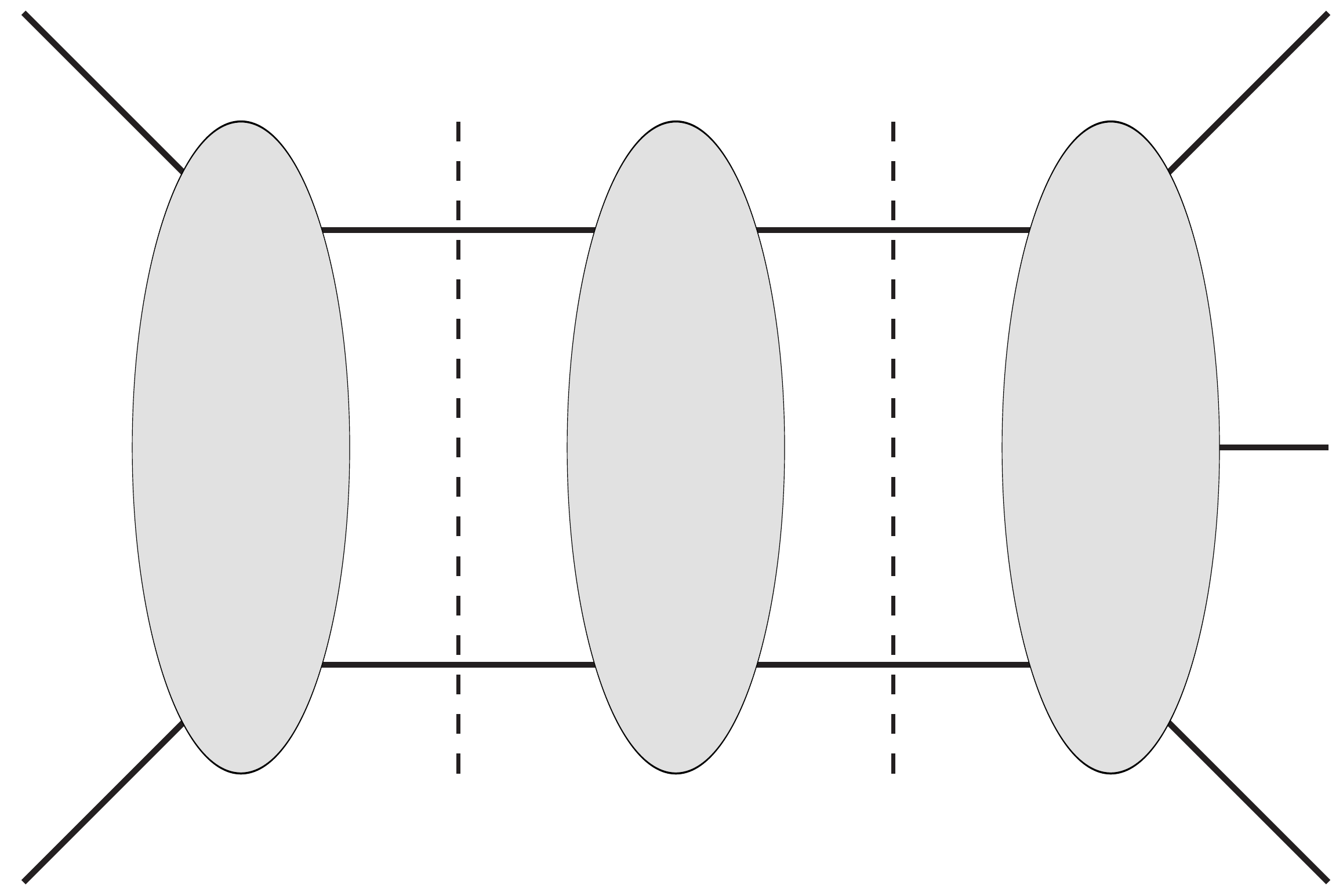}
  \caption[a]{}
\end{subfigure}
\begin{subfigure}{0.25\textwidth}
  \centering
  \includegraphics[width=0.8\textwidth]{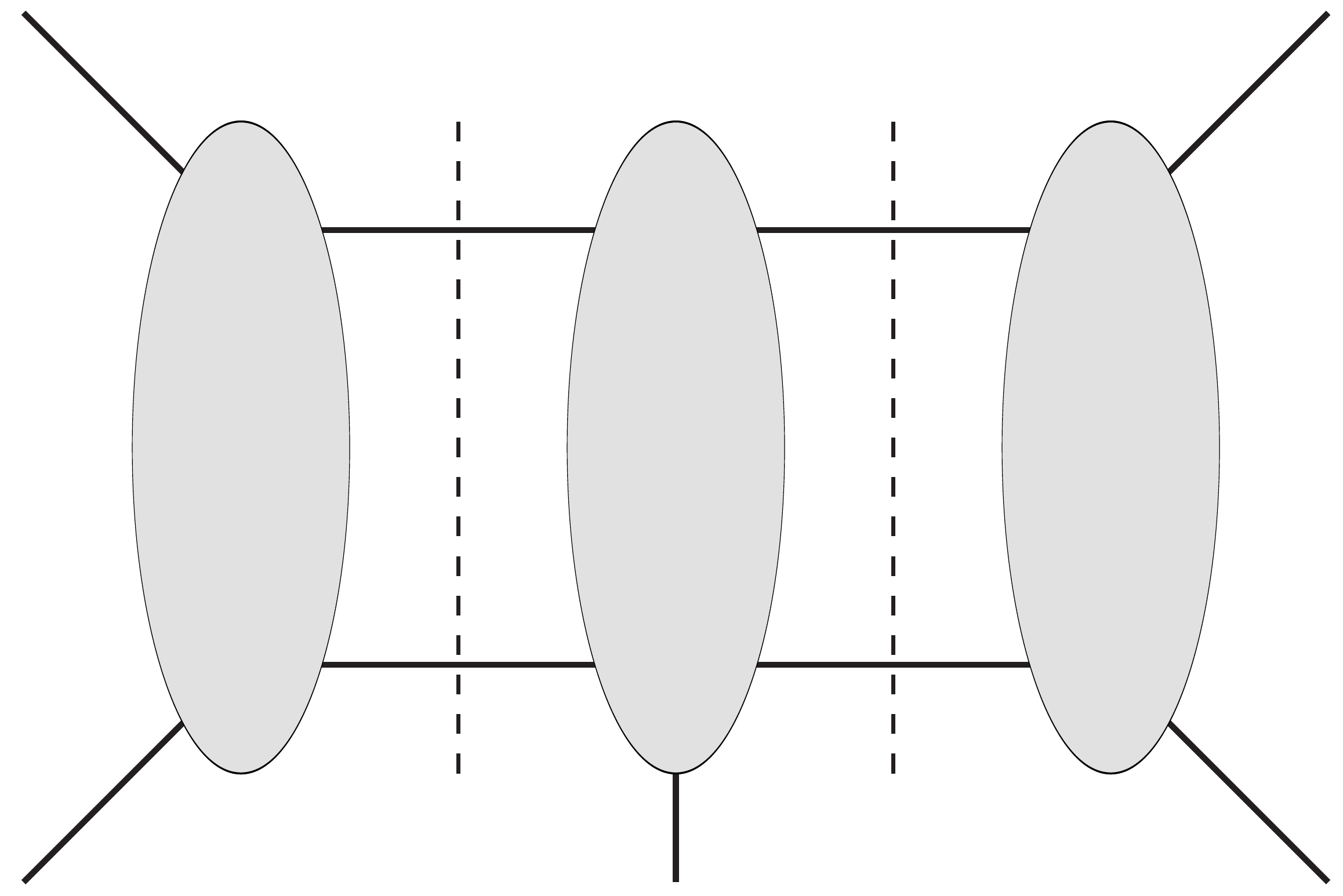}
  \caption[b]{}
\end{subfigure}
\begin{subfigure}{0.25\textwidth}
  \centering
  \includegraphics[width=0.8\textwidth]{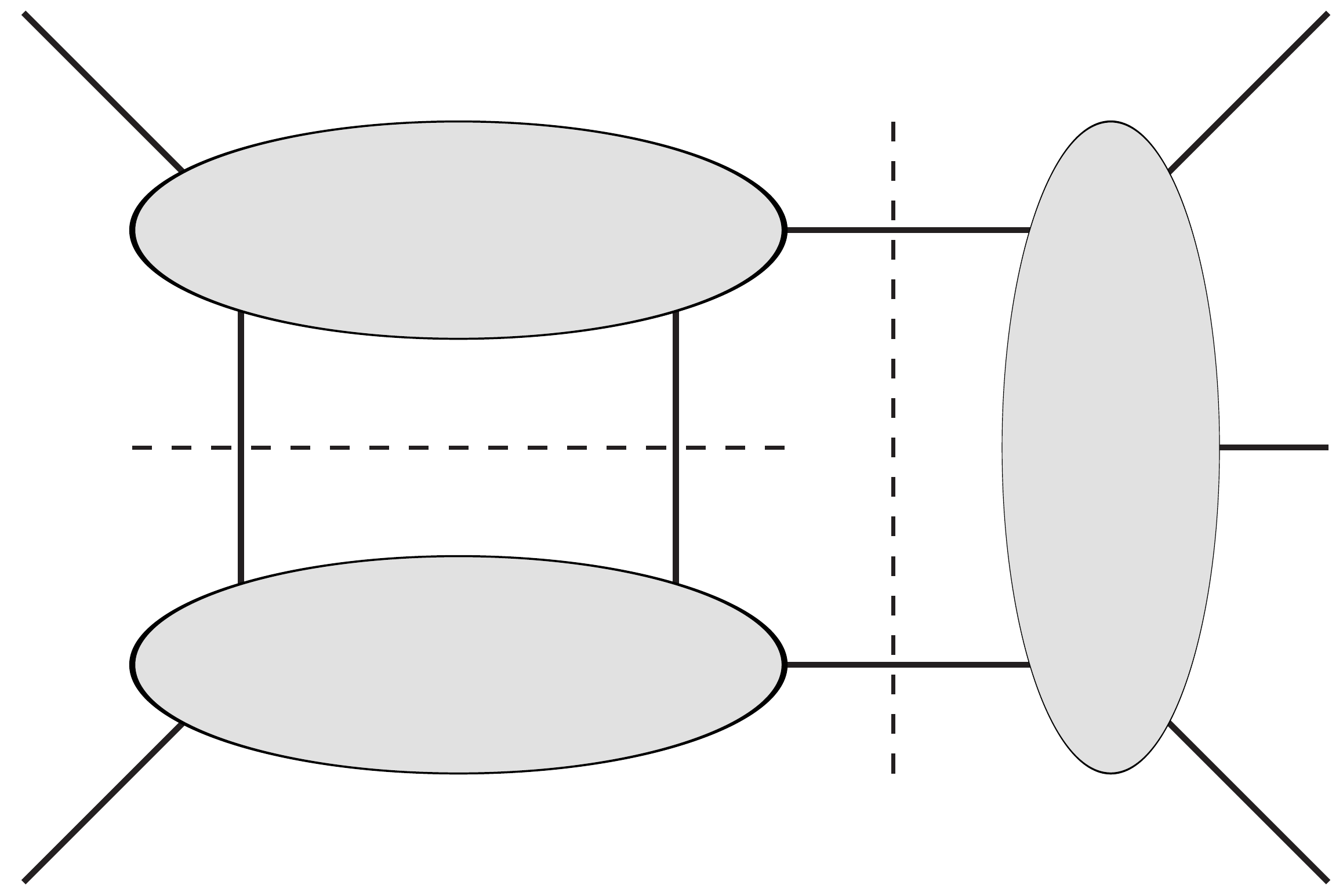}
  \caption[c]{} 
\end{subfigure}
\begin{subfigure}{0.25\textwidth}
  \centering
  \includegraphics[width=0.8\textwidth]{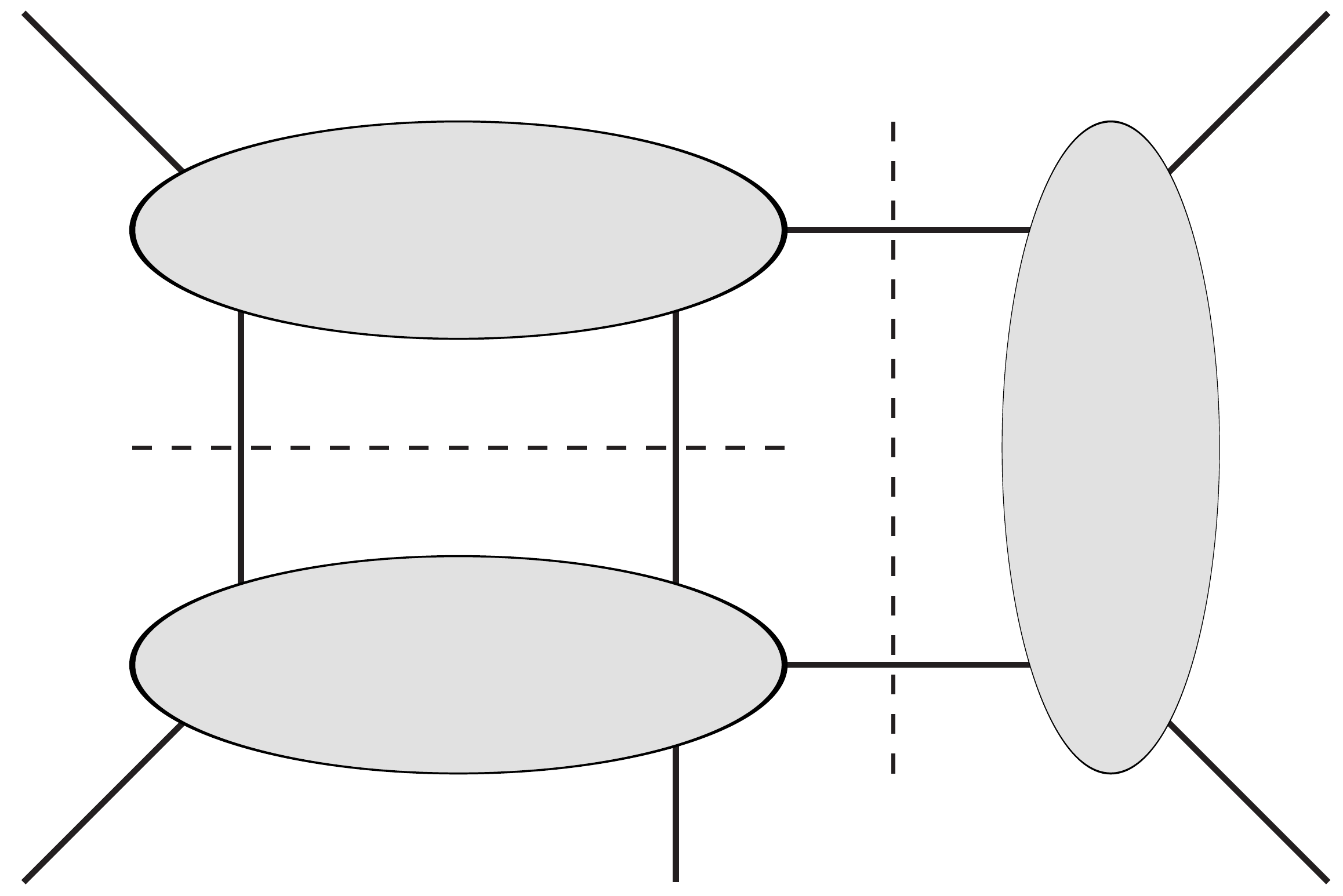}
  \caption[d]{}
\end{subfigure}
\begin{subfigure}{0.25\textwidth}
  \centering
  \includegraphics[width=0.8\textwidth]{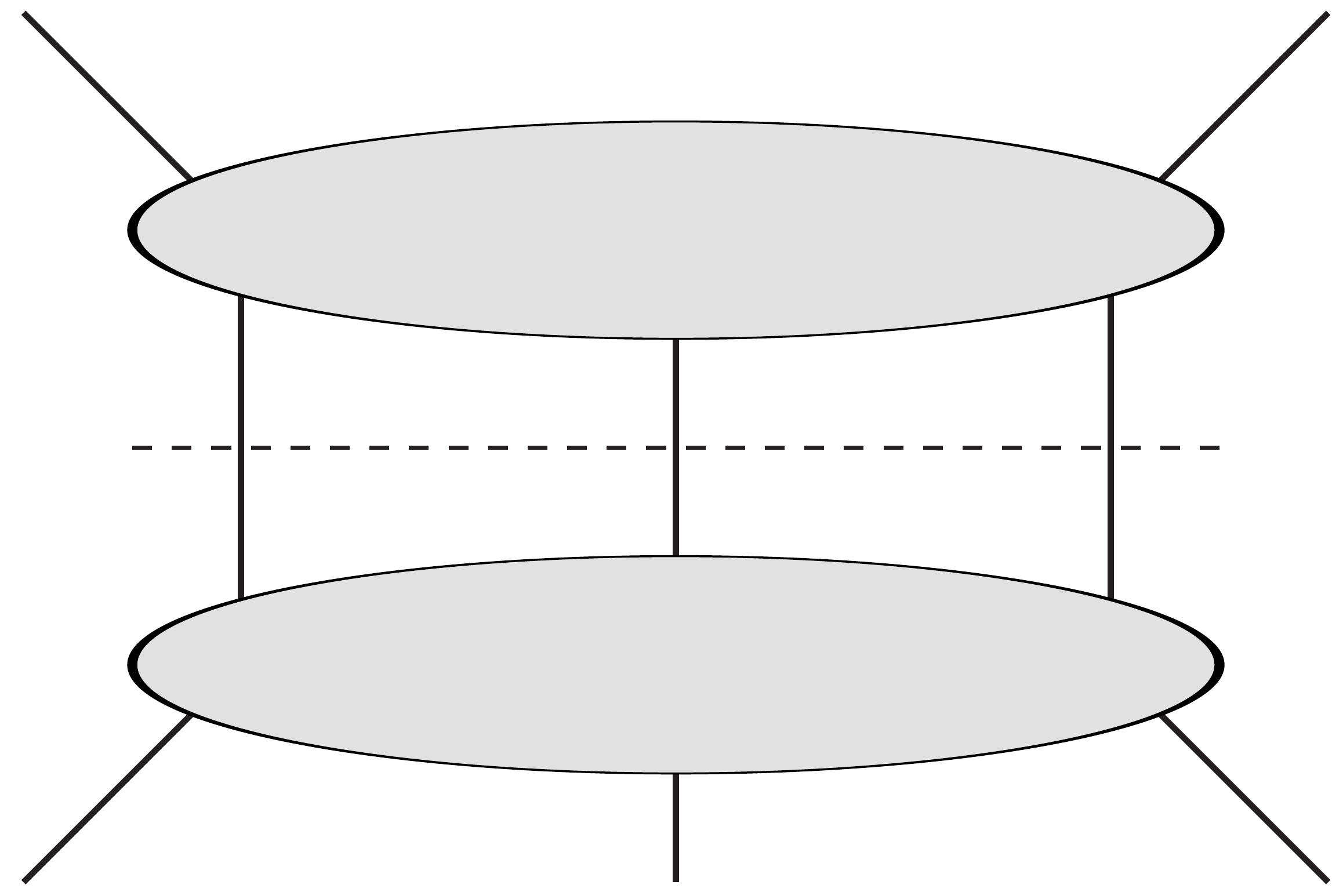}
  \caption[e]{}
\end{subfigure} 
\begin{subfigure}{0.25\textwidth}
  \centering
  \includegraphics[width=0.8\textwidth]{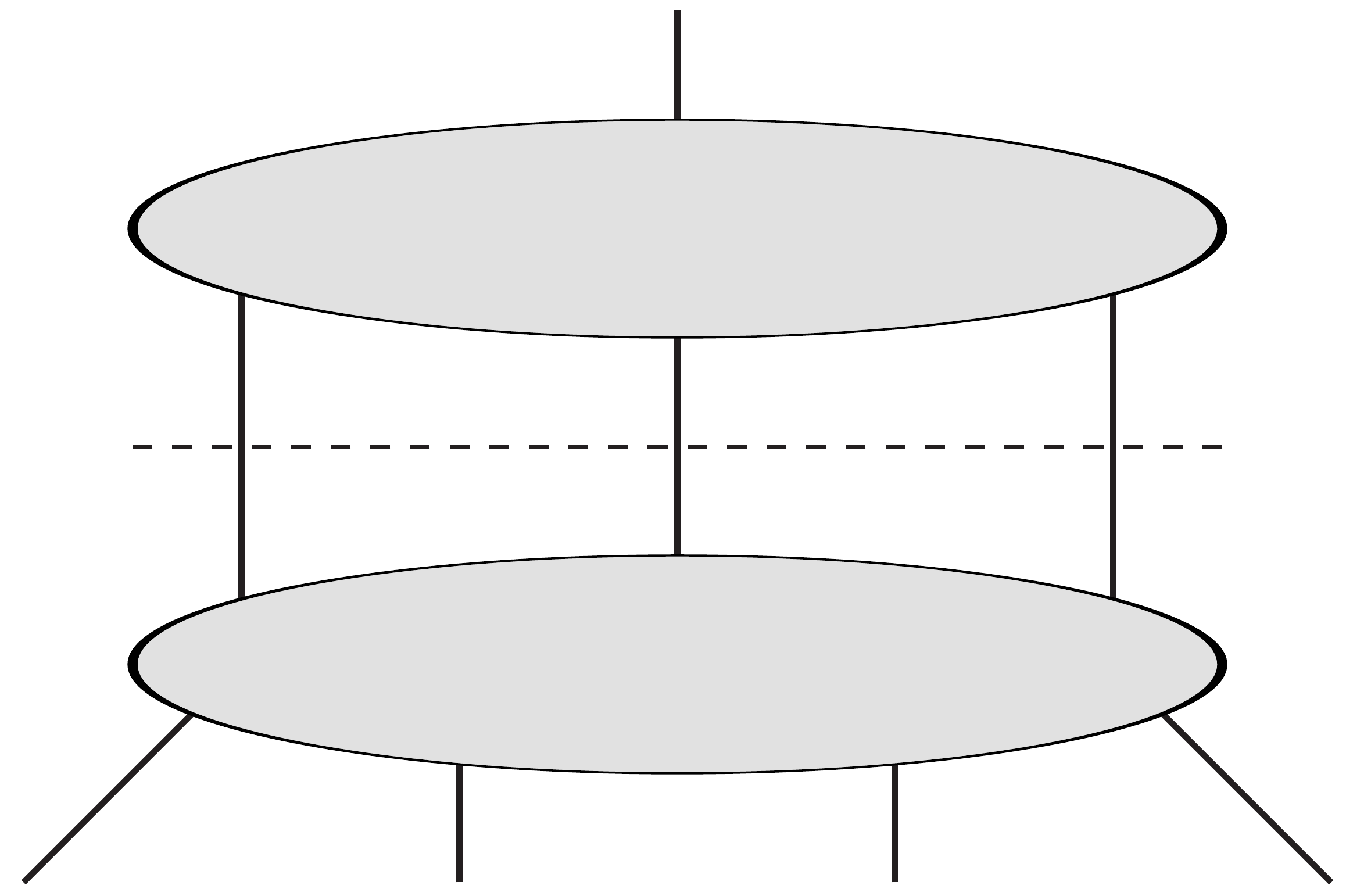}
  \caption[f]{}
\end{subfigure}
\caption{\small A spanning set of cuts at  two loops and five points. The first two are butterfly cuts; the rest are satisfied automatically by the $\mathcal{N}=4$ subsector.
         \label{fig:5ptspanningcuts}}
\end{figure}

To determine the unknown function $X$, 
we now match to physical information using the relevant set of spanning cuts,
displayed in Figure~\ref{fig:5ptspanningcuts} (these can be obtained from the $n$-point spanning cuts given in Figure~\ref{fig:nptspanningcuts}).
We compare to the irreducible presentation of the amplitude given in eqs.~(\ref{eq:2L5gbfz}) (or equivalently eqs.~(\ref{eq:2L5g})).
As in the four points example, 
spanning cuts that involve cutting the central $(\ell_1+\ell_2)^2$ propagator
do not provide any information on $X$:
in this case, the first two cuts in Figure~\ref{fig:5ptspanningcuts} contain the relevant new physics.
Again, it suffices to reproduce them in the planar sector as the nonplanar sector follows through
tree-level relations on colour-ordered amplitudes.

The first of these two cuts, using the BFZ amplitude, is
\begin{align}\label{eq:cut2205BFZ}
&(\ell_1-p_1)^2(\ell_1-p_{12})^2(\ell_2-p_5)^2
\left.\text{Cut}\bigg(\usegraph{9}{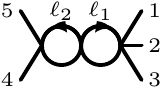}\bigg)\right|_{F_3}\\
&=\frac{s_{45}+(\ell_1+\ell_2)^2}{s_{45}}\left(\beta_{12345}
+\frac{\gamma_{45}}{s_{45}}(\ell_1+p_5)^2
+\frac{\gamma_{12}}{s_{12}}(\ell_1-p_1)^2
+\frac{\gamma_{23}}{s_{23}}(\ell_1-p_{12})^2\right).\nn
\end{align}
We may also evaluate this cut using the BCJ master numerator, eq. \eqref{eq:n431}. For brevity,
we define $X'(12345;\ell_1,\ell_2)\equiv X(12345;\ell_1,\ell_2)+X(12354;\ell_1,p_{45}-\ell_2)$
and we also introduce antisymmetric brackets,
e.g. $X'([12]345;\ell_1,\ell_2)=X'(12345;\ell_1,\ell_2)-X'(21345;\ell_1,\ell_2)$. Using this
notation, and the on-shell conditions $\ell_1^2=(\ell_1+p_{45})^2=\ell_2^2=(\ell_2-p_{45})^2=0$, we find
\begin{align}\label{eq:cut2205BCJ}
&(\ell_1-p_1)^2(\ell_1-p_{12})^2(\ell_2-p_5)^2
\left.\text{Cut}\bigg(\usegraph{9}{220M}\bigg)\right|_{F_3}\\
&\qquad=\frac{s_{45}+(\ell_1+\ell_2)^2}{s_{45}}\bigg(X'(12345;\ell_1,\ell_2)
+\frac{X'([12]345;\ell_1,\ell_2)}{s_{12}}(\ell_1-p_1)^2\nn\\
&\qquad\qquad+\frac{X'(1[23]45;\ell_1,\ell_2)}{s_{23}}(\ell_1-p_{12})^2
+\frac{X'(123[45];\ell_1,\ell_2)}{s_{45}}(\ell_2-p_5)^2\nn\\
&\qquad\qquad\qquad+\frac{X'([[12]3]45;\ell_1,\ell_2)}{s_{12}s_{45}}(\ell_1-p_1)^2(\ell_1-p_{12})^2\nn\\
&\qquad\qquad\qquad+\frac{X'([1[23]]45;\ell_1,\ell_2)}{s_{23}s_{45}}(\ell_1-p_1)^2(\ell_1-p_{12})^2\nn\\
&\qquad\qquad\qquad+\frac{X'([12]3[45];\ell_1,\ell_2)}{s_{12}s_{45}}(\ell_1-p_1)^2(\ell_2-p_5)^2\nn\\
&\qquad\qquad\qquad+\frac{X'(1[23][45];\ell_1,\ell_2)}{s_{23}s_{45}}(\ell_1-p_{12})^2(\ell_2-p_5)^2\nn\\
&\qquad\qquad\qquad+\frac{X'([[12]3][45];\ell_1,\ell_2)}{s_{12}s_{45}^2}(\ell_1-p_1)^2(\ell_1-p_{12})^2(\ell_2-p_5)^2\nn\\
&\qquad\qquad\qquad+\frac{X'([1[23]][45];\ell_1,\ell_2)}{s_{23}s_{45}^2}(\ell_1-p_1)^2(\ell_1-p_{12})^2(\ell_2-p_5)^2\bigg).\nn
\end{align}
This cut bears a strong similarity to its four-point counterpart,
given in eqs. (\ref{eq:cut220}) and (\ref{eq:cut220bcj}),
preserving an overall flip symmetry through the horizontal axis.
The prefactor $\mathcal{T}$ has now been generalised to the five-point kinematic prefactors $\gamma_{ij}$.

The other planar cut has a somewhat more complicated structure. Using the irreducible presentation of the amplitude,
we find
\begin{align}\label{eq:cut2205LBFZ}
&(\ell_1-p_1)^2(\ell_2-p_5)^2(\ell_1+p_{45})^2(\ell_2+p_{12})^2
\left.\text{Cut}\bigg(\usegraph{9}{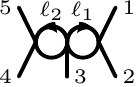}\bigg)\right|_{F_3}
\nonumber\\
&=(\ell_2+p_{12})^2\left(\frac{s_{45}+(\ell_1+\ell_2)^2}{s_{45}}\right)
\left(\beta_{12345}+\frac{\gamma_{45}}{s_{45}}(\ell_1+p_5)^2+\frac{\gamma_{12}}{s_{12}}(\ell_1-p_1)^2\right)
\nonumber\\
&\;\;\;\;\;\;
+\frac{(\ell_1+p_{45})^2(\ell_2+p_{12})^2}{s_{12}s_{45}}\bigg[
\frac{\beta_{12345}}{2}(\ell_1+\ell_2)^2\nonumber\\
&\;\;\;\;\;\;\;\;\;\;\;\;
+\left(s_{45}-s_{12}+(\ell_2+p_{12})^2\right)
\left(\frac{\gamma_{45}}{s_{45}}(\ell_1+p_5)^2+\frac{\gamma_{12}}{s_{12}}(\ell_1-p_1)^2\right)
\nonumber\\
&\;\;\;\;\;\;\;\;\;\;\;\;
-\left(s_{12}(\gamma_{12}+\gamma_{45}-\beta_{12345})+s_{13}\gamma_{12}\right)\times
\nonumber\\
&\;\;\;\;\;\;\;\;\;\;\;\;\;\;\;\;\;\;
\left(\frac{(\ell_1+\ell_2)^2}{s_{12}}
+\frac{1}{2}\left(\frac{s_{45}-(\ell_1+p_{45})^2}{s_{45}}\right)
\left(\frac{s_{12}-(\ell_2+p_{12})^2}{s_{12}}\right)\right)\bigg]
\nonumber\\
&\;\;\;\;\;\;
-(p_1\leftrightarrow p_5,p_2\leftrightarrow p_4,\ell_1\leftrightarrow\ell_2).
\end{align}
Meanwhile, using the BCJ presentation of the amplitude, we find
\begin{align}\label{eq:cut2205LBCJ}
&(\ell_1-p_1)^2(\ell_2-p_5)^2(\ell_1+p_{45})^2(\ell_2+p_{12})^2
\left.\text{Cut}\bigg(\usegraph{9}{2205L}\bigg)\right|_{F_3}
\\
&=\bigg\{\!(\ell_2+p_{12})^2\!\left(\frac{s_{45}+(\ell_1+\ell_2)^2}{s_{45}}\right)\!
\bigg(\!X'(12345;\ell_1,\ell_2)\!\!+\!\frac{X'([12]345;\ell_1,\ell_2)}{s_{12}}(\ell_1-p_1)^2
\nonumber\\
&\;\;\;\;\;\;
+\frac{X'(123[45];\ell_1,\ell_2)}{s_{45}}(\ell_2-p_5)^2
+\frac{X'([12]3[45];\ell_1,\ell_2)}{s_{12}s_{45}}(\ell_1-p_1)^2(\ell_2-p_5)^2\bigg)
\nonumber\\
&\;\;\;\;\;\;\;\;\;\;\;\;
-(p_1\leftrightarrow p_5,p_2\leftrightarrow p_4,\ell_1\leftrightarrow\ell_2)\bigg\}\nonumber\\
&\;\;\;\;\;\;
+\frac{(\ell_1+p_{45})^2(\ell_2+p_{12})^2}{s_{12}s_{45}}\bigg(
X''(12345;\ell_1,\ell_2)
+\frac{X''([12]345;\ell_1,\ell_2)}{s_{12}}(\ell_1-p_1)^2
\nonumber\\
&\;\;\;\;\;\;\;
+\frac{X''(123[45];\ell_1,\ell_2)}{s_{45}}(\ell_2-p_5)^2
+\frac{X''([12]3[45];\ell_1,\ell_2)}{s_{12}s_{45}}(\ell_1-p_1)^2(\ell_2-p_5)^2\bigg),\nn
\end{align}
where this time we have used the on-shell conditions
$\ell_1^2=(\ell_1-p_{12})^2=\ell_2^2=(\ell_2-p_{45})^2=0$,
and also the three symmetries on $X$ introduced in eq. (\ref{eq:Xsymmetries}).
We have also introduced
\begin{align}\label{eq:xdoubleprimed}
&X''(12345;\ell_1,\ell_2)\equiv(\ell_1+\ell_2)^2X(12345;\ell_1,\ell_2)\\
&\qquad
+(s_{45}-s_{12}+(\ell_1+\ell_2)^2-(\ell_1+p_{45})^2)X(12354;\ell_1,p_{45}-\ell_2)\nonumber\\
&\qquad
+(s_{12}-s_{45}+(\ell_1+\ell_2)^2-(\ell_2+p_{12})^2)X(21345;p_{12}-\ell_1,\ell_2)\nonumber\\
&\qquad
+((\ell_1+\ell_2)^2-(\ell_1+p_{45})^2-(\ell_2+p_{12})^2)X(21354;p_{12}-\ell_1,p_{45}-\ell_2).\nn
\end{align}
This cut preserves a symmetry through the vertical axis.

Our task is now to find a solution for $X$ that satisfies these two cut equations,
while at the same time enjoying the off-shell symmetries in eq. (\ref{eq:Xsymmetries}).

\subsection{Nonlocality properties}
\label{sec:nonlocal}

Before we describe our strategy for finding $X$, let us pause to comment on what kind
of structure we may expect. We claim that $X$ must be 
nonlocal in the kinematic factors $s_{ij}$. 
The basis of our claim is simple.
We input physical information into our calculation using the irreducible set of numerators given in eqs.~(\ref{eq:2L5gbfz}), and
we further insist on writing our results in terms of the prefactors $\gamma_{ij}$.
Consider making a local ansatz for the full pentabox numerator
in terms of the six linearly-independent kinematic prefactors $\gamma_{ij}$:
\begin{align}\label{eq:localansatz}
n\bigg(\usegraph{9}{431}\!\bigg)
=\gamma_{12}m_1+\gamma_{13}m_2+\gamma_{14}m_3+\gamma_{23}m_4+\gamma_{24}m_5+\gamma_{34}m_6,
\end{align}
where dimensional analysis tells us that the local objects $m_i$ may carry up to six powers of loop momentum,
including factors of extra-dimensional components $\mu_{ij}$.
As the pentabox is the master for the five-point, two-loop system,
linearity of the Jacobi identities implies that \emph{all} numerators
are limited to six powers of loop momentum.

BFZ's box-triangle irreducible numerator was given in eq.~(\ref{eq:430bfz}) as
\begin{align}\label{eq:delta430}
\Delta\bigg(\usegraph{13}{430}\!\!\!\:\bigg)
&=-\frac{i\,s_{12}\text{tr}_+(1345)}{2 \SpDenom5 s_{13}}(2\ell_1\!\cdot\!\omega_{123}+s_{23})\\
& \qquad \qquad \,\, \times\left(F_2+F_3\,
\frac{s_{45}+(\ell_1+\ell_2)^2}{s_{45}} \right);\nn
\end{align}
it has a maximum of seven powers of loop momentum.
This continues to be true even if we move terms to other irreducible graphs with
fewer propagators using integrand reduction.
It is also still true if we choose to rewrite the regulator components $\mu_{ij}$ in terms of
conventional scalar products $\ell_i\cdot p_j$ and $\ell_i\cdot\ell_j$.
We conclude that $X$ must be a nonlocal object. Notice that our argument applies to master
numerators with a more general structure than we assumed in eq.~\eqref{eq:n431}; we merely used the
BFZ irreducible numerators in combination with an ansatz which is linear in $\gamma_{ij}$.

\subsection{Attempting a minimal solution}

Our first attempt at a solution for $X$ is motivated by power counting from the Feynman rules.
As all diagrams have seven vertices,
one would expect up to seven powers of loop momentum in a solution for the pentabox numerator:
the factor of $\mu_{11}\mu_{22}(\ell_1+\ell_2)^2$ in eq. (\ref{eq:n431}) already accounts for six,
so we anticipate $X$ having a single power of loop momentum.
Although this proposal will not prove successful,
the resulting calculation conveniently demonstrates the obstacle to colour-kinematics duality.
We shall also use it to introduce concepts and notation that will help us find a solution for $X$
in the next section.

Dependence on the two $D$-dimensional loop momenta $\ell_i$ accounts for a total of eleven degrees of freedom:
eight coming from the components of $\bar{\ell}_1$ and $\bar{\ell}_2$,
plus an extra three from $\mu_{11}$, $\mu_{22}$ and $\mu_{12}$.
This freedom can be parametrised by eleven independent scalar products $z_i$:
\begin{align}\label{eq:z}
z_1&=\ell_1^2,  & z_2&=(\ell_1-p_1)^2,  & z_3&=(\ell_1-p_{12})^2, & z_4&=(\ell_1+p_{45})^2,\nonumber\\
z_5&=\ell_2^2,  & z_6&=(\ell_2-p_5)^2,  & z_7&=(\ell_2-p_{45})^2, & z_8&=(\ell_1+\ell_2)^2,\nonumber\\
z_9&=(\ell_1+p_5)^2,  & z_{10}&=(\ell_2+p_1)^2, & z_{11}&=(\ell_2+p_{12})^2,
\end{align}
of which the first eight are the propagators of the pentabox.
Any polynomial expression depending on scalar products of the form $\ell_i\cdot\ell_j$,
$\ell_i\cdot p_j$ or $\mu_{ij}$ can be uniquely expressed in terms of these objects,
e.g. $\ell_1\cdot\ell_2=(z_8-z_1-z_5)/2$.
This particular choice is convenient because the $z_i$ always transform directly
into each other under the symmetries in eq. (\ref{eq:Xsymmetries}).
For instance,
under the third of these symmetries which acts as $\ell_1\leftrightarrow-\ell_2$,
\begin{align}\label{eq:zmap}
z_1\leftrightarrow z_5, \qquad z_2\leftrightarrow z_{10}, \qquad z_3\leftrightarrow z_{11},
\qquad z_4\leftrightarrow z_7, \qquad z_6\leftrightarrow z_9,
\end{align}
and $z_8$ remains unchanged.

To find a solution for $X$ with minimal dependence on loop momentum,
we make an ansatz of the form
\begin{align}\label{eq:minansatz}
X(12345;\ell_1,\ell_2)=A(12345)+\sum_{i=1}^{11}A_i(12345)z_i,
\end{align}
where the unknown new functions $A$ and $A_i$ are functions of the external momenta $p_i$.
They are ``relabelling friendly'',
in the sense that knowledge of a function acting on a given ordering of its arguments
suffices to reproduce any other ordering through relabelling of the external momenta $p_i$.
However, for the reasons outlined in section \ref{sec:nonlocal},
we do not expect these functions to be local functions of external momenta.

With this setup, solving the three symmetries in eq. (\ref{eq:Xsymmetries}) is straightforward.
If one substitutes the ansatz presented in eq. (\ref{eq:minansatz}),
matching individual powers of $z_i$ gives a set of simple functional identities,
e.g. $A_1(23451)=A_2(12345)=A_{10}(12345)$, etc.
The relabelling property ensures the validity of these identities for any ordering of the external momenta,
so such identities can be used to eliminate unnecessary functions $A_i$ entirely.
Once all such identities are solved,
\begin{align}\label{eq:minsoln}
&X(12345;\ell_1,\ell_2)=A(12345)+A_1(12345)(\ell_1^2+\ell_2^2)\nonumber\\
&\qquad
+A_1(23451)((\ell_1-p_1)^2+(\ell_2+p_1)^2)+A_1(34512)((\ell_1-p_{12})^2+(\ell_2+p_{12})^2)\nonumber\\
&\qquad
+A_1(45123)((\ell_1+p_{45})^2+(\ell_2-p_{45})^2)+A_1(51234)((\ell_1+p_5)^2+(\ell_2-p_5)^2)\nonumber\\
&\qquad
+A_8(12345)(\ell_1+\ell_2)^2,
\end{align}
along with the additional requirements that
\begin{subequations}\label{eq:minsymmetries}
\begin{align}
A(12345)&=-A(54321)=A(23451),\label{eq:Asymmetry}\\
A_1(12345)&=-A_1(54321),\label{eq:A1symmetry}\\
A_8(12345)&=-A_8(54321)=A_8(23451).\label{eq:A8symmetry}
\end{align}
\end{subequations}

It now suffices to consider a box-triangle cut,
an expression for which may be deduced from either of the
two double-bubble cuts in section \ref{sec:5ptcuts}.
The cut conditions are
$\ell_1^2=(\ell_1-p_1)^2=(\ell_1-p_{12})^2=(\ell_1+p_{45})^2=\ell_2^2=(\ell_2-p_5)^2=(\ell_2-p_{45})^2=0$,
yielding a condition on $X$:
\begin{align}\label{eq:cut430}
\left.\text{Cut}\bigg(\usegraph{13}{430}\!\!\!\:\bigg)\right|_{F_3}
&=\frac{s_{45}+(\ell_1+\ell_2)^2}{s_{45}}\left(\beta_{12345}
+\frac{\gamma_{45}}{s_{45}}(\ell_1+p_5)^2\right)\nonumber\\
&=\frac{s_{45}+(\ell_1+\ell_2)^2}{s_{45}}X'(12345;\ell_1,\ell_2),
\end{align}
where $X'(12345;\ell_1,\ell_2)\equiv X(12345;\ell_1,\ell_2)+X(12354;\ell_1,p_{45}-\ell_2)$ as before.
Rewriting this in terms of the new variables $z_i$,
on which the cut conditions are $z_i=0$ for $i\leqslant7$,
\begin{align}
\alpha_{12345}+\frac{\gamma_{45}}{s_{45}}z_9
&=A(123\{45\})+s_{12}A_1(35412)+s_{23}A_1(23541)\nonumber\\
&\qquad-s_{45}(A_1(35412)+A_1(41235)+A_1(23541)+A_8(12354))\nonumber\\
&\qquad+A_8(123[45])z_8+(A_1(51234)-A_1(41235))z_9\nonumber\\
&\qquad+A_1(23[45]1)z_{10}+A_1(3[45]12)z_{11}.
\end{align}
Here we have chosen to explore a more general equation
with the unknown object $\alpha_{12345}$ playing the role of $\beta_{12345}$.
Symmetry of the box-triangle cut implies that $\alpha_{12345}=-\alpha_{32154}$.
We have also introduced an anticommutator, $A(123\{45\})\equiv A(12345)+A(12354)$.

We proceed by matching powers of the four nonzero scalar products $z_i$.
The coefficient of $z_8$ reveals that $A_8(123[45])=0$;
together with the requirement from eq. (\ref{eq:A8symmetry}) that $A_8(12345)=A_8(23451)$,
we see that $A_8$ is totally permutation symmetric on its five arguments.
The antisymmetry property in eq. (\ref{eq:A8symmetry}) therefore requires $A_8$ to vanish.\footnote{This
	is to be expected: were $A_8$ nonzero,	it would necessitate more than one power of loop momentum in the final solution for $X$ as $\ell_1\cdot\ell_2$ cannot be cancelled elsewhere.}
The coefficients of $z_{10}$ and $z_{11}$ tell us that $A_1(1[23]45)=A_1(12[34]5)=0$.
The coefficient of $z_9$ then provides a unique solution:
\begin{align}\label{eq:A1}
A_1(12345)=\frac{\gamma_{51}}{2s_{15}}, \qquad A_8(12345)=0.
\end{align}

The condition one would use to solve for $A$, with zero powers of $z_i$, presents an obstacle.
Substituting the results for $A_1$ and $A_8$ found above,
\begin{align}\label{eq:Acond}
A(123\{45\})
=\alpha_{12345}
-\frac{s_{23}}{2s_{12}}\gamma_{12}-\frac{s_{12}}{2s_{23}}\gamma_{23}
+\frac{s_{45}}{2}\left(\frac{\gamma_{12}}{s_{12}}+\frac{\gamma_{23}}{s_{23}}-\frac{\gamma_{45}}{s_{45}}\right),
\end{align}
which can be used to generate a consistency condition on $\alpha_{12345}$.
One cycles
\begin{align}\label{eq:leg5cycle}
A(123\{45\})&=A(12345)+A(12354),\nonumber\\
A(412\{35\})&=A(12354)+A(12534),\nonumber\\
A(341\{25\})&=A(12534)+A(15234),\nonumber\\
A(234\{15\})&=A(15234)+A(12345),
\end{align}
where we have used the cyclic condition from eq. (\ref{eq:Asymmetry})
to pull the $p_5$ leg around $A$.\footnote{The
	same technique was used in \cite{Bjerrum-Bohr:2013iza} to explore loop-momentum dependence of colour-dual $n$-gons in $\mathcal{N}=4$ SYM.  The cyclic nature of (\ref{eq:Xsymmetries}) suggests a similar $n$-gon structure in $X$.}
As a result,
\begin{align}
0&=A(123\{45\})-A(412\{35\})+A(341\{25\})-A(234\{15\})\nonumber\\
&=\alpha_{12345}-\alpha_{41235}+\alpha_{34125}-\alpha_{23415}\nonumber\\
&\qquad-\gamma_{25}-\gamma_{45}
-\frac{1}{2}\left(s_{15}+s_{35}\right)
\left(\frac{\gamma_{12}}{s_{12}}+\frac{\gamma_{23}}{s_{23}}+\frac{\gamma_{34}}{s_{34}}+\frac{\gamma_{41}}{s_{14}}\right).
\end{align}
Because $\beta_{12345}$ is not a valid solution for $\alpha_{12345}$, this is inconsistent.
A solution with minimal power counting in loop momentum is therefore not possible.

One might question whether a solution with minimal power counting is possible
if the three symmetry conditions on $X$ in eq. (\ref{eq:Xsymmetries}) are relaxed.
In this case, an intriguing solution to the box-triangle cut is
\begin{align}\label{eq:oldsoln}
X(12345;\ell_1,\ell_2)=\frac{1}{2s_{45}}n^{[\mathcal{N}=4]}\bigg(\usegraph{9}{431}\!\bigg),
\end{align}
where we have used the fact that
\begin{align}
n^{[\mathcal{N}=4]}\bigg(\usegraph{9}{431}\!\bigg)
=s_{45}\,\beta_{12345}+\gamma_{45}(\ell_1+p_5)^2
\end{align}
when $\ell_1^2=(\ell_1-p_1)^2=(\ell_1-p_{12})^2=(\ell_1+p_{45})^2=0$.\footnote{A
	similar structure appears in the double-box numerator (\ref{eq:n331}): the $\mathcal{N}=4$ numerator cancels away the nonlocal factor $s^{-1}$.}
However, this solution, or any other adhering to the form (\ref{eq:minansatz}),
fails to reproduce an off-shell symmetry of the nonplanar graph introduced in eq. (\ref{eq:n332}):
\begin{align}
n\bigg(\usegraph{9}{332}\bigg)=-n\bigg(\usegraph{9}{332o21354}\bigg).
\end{align}
As explained in section \ref{sec:5ptsymmetry},
the symmetries (\ref{eq:Xsymmetries}) automatically imply this condition.

\subsection{Solving the ansatz}

Given the failure of a minimal solution,
we now explore a more general class of solutions depending on higher powers of loop momenta.
Still requiring that $X$ should be local in loop momenta,
we generalise eq. (\ref{eq:minansatz}) to a degree $\text{deg}(X)$ polynomial of the scalar products $z_i$:
\begin{align}\label{eq:Xansatz}
X(12345;\ell_1,\ell_2)
=\sum_{k=0}^{\text{deg}(X)}\sum_{1\leqslant i_1\leqslant\ldots\leqslant i_k\leqslant11}
A_{i_1\ldots i_k}(12345)z_{i_1}\ldots z_{i_k},
\end{align}
where the loop-momentum-independent functions $A_{i_1\cdots i_k}$
are the coefficients of the linearly independent monomials $z_{i_1}\ldots z_{i_k}$,
carrying total degree $k$.
For a given value of $k$,
we require a total of $11(11+1)\ldots(11+k)/k!$ functions $A_{i_1\ldots i_k}$ to account for all
possible monomials.
Loop-momentum dependence in our solution is now explicit:
it remains to identify dependence on the external momenta
which lives inside the functions $A_{i_1\ldots i_k}$.

Following the procedure used in the previous section,
we make considerable progress by solving two-term functional identities for
$A_{i_1\ldots i_k}$ that come from the symmetries in eq. (\ref{eq:Xsymmetries}).
Again, the relabelling property ensures the validity of these for any ordering of external momenta.
It is also possible to solve some identities coming from the
two cuts eqs. (\ref{eq:cut2205BFZ}) and (\ref{eq:cut2205LBFZ}).
For the former, the cut conditions are $z_1=z_4=z_5=z_7=0$;
for the latter, $z_1=z_3=z_5=z_7=0$.
However, these techniques are not sufficient to provide a unique solution for $X$,
so we make an ansatz for the remaining independent functions.

In view of our discussion in section~\ref{sec:nonlocal},
we know that the functions $A_{i_1\cdots i_k}$ may be nonlocal in the kinematic invariants.
We therefore define bases of linearly independent nonlocal monomials,
with degree of nonlocality $n$ and dimension $1+2m-2n$, as
\begin{align}\label{eq:nlbasis}
\Gamma^{[m,n]}=\left\{\left.\gamma_{ij}\frac{\prod_{k=1}^{m}s_k}{\prod_{l=1}^ns_l}\right|s_k\in\left\{s_{ij}\right\}\right\},
\end{align}
carrying linear dependence on $\gamma_{ij}$.
The lengths of these bases for various values of $m$ and $n$ are summarised in table \ref{tab:nlbases}.
Linear relations between such objects are generated by the identity
\begin{align}\label{eq:gammaidentity}
0=(\gamma_{12}+\gamma_{13})(s_{23}-s_{45})+\gamma_{23}(s_{12}-s_{13})+\gamma_{45}(s_{14}-s_{15}),
\end{align}
as well as the ability to cancel $s_{ij}$ terms between numerators and denominators.

\begin{table*}
\centering
\begin{tabular}{| c | c c c c |}
\hline
& $m=0$ & $m=1$ & $m=2$ & $m=3$ \\
\hline
$n=0$ & 6 	& 25 	& 66 	& 140 \\
$n=1$ & 60  & 196 	& 435 	& 806 \\
$n=2$ & 300 & 790 	& 1536 	& 2585 \\
$n=3$ & 966	& 2185	& 3885	& 6131 \\
\hline
\end{tabular}
\caption{\small
The number of linearly independent basis elements of the nonlocal $\gamma$-bases $\Gamma^{[m,n]}$.
When $n=0$, there are $\frac{1}{12}(1+m)(2+m)(4+m)(9+m)$ elements (this is always integer-valued for integer $m$).}
\label{tab:nlbases}
\end{table*}

These bases of nonlocal monomials allow us to make nonlocal ans\"{a}tze,
with degree of nonlocality $n$, for $A_{i_1\ldots i_k}$:
\begin{align}\label{eq:gammaansatz}
A_{i_1\ldots i_k}(12345)&=\sum_ja_{i_1\ldots i_k;j}\,\Gamma_j^{[n-k,n]},
\end{align}
where $a_{i_1\ldots i_k;j}$ are rational numbers and $j$ spans the length of the nonlocal basis.
Of course,
$A_{i_1\ldots i_k}$ applied to other orderings of the external legs
generates nonlocal monomials outside of the bases prescribed above.
It is necessary to rewrite these monomials in terms of the linearly independent basis elements
using identities such as the one in eq. (\ref{eq:gammaidentity}).

The procedure for solving for $X$ is now clear.
Having eliminated as many functions $A_{i_1\ldots i_k}$ as possible using functional identities,
one directly applies the ansatz decompositions given in eq. (\ref{eq:gammaansatz}) to the remaining functions.
Identifying coefficients of these bases leaves a large system of linear equations
for the rational numbers $a_{i_1\ldots i_k;j}$.
We find that a valid solution requires $\text{deg}(X)\geqslant3$:
this implies up to 6 powers of loop momentum in $X$,
so a total of 12 powers in the final expression for the pentabox numerator.
An initial ansatz with 10,004 parameters left 1260 undetermined once all conditions were accounted for.

\begin{table*}
\centering
\begin{tabular}{| c | c | c |}
\hline
Function    & Symmetries                                      & Freedom \\
\hline
$A$         & $A(12345)=A(23451)=-A(54321)$                         & 139 \\
$A_1$       & $A_1(12345)=-A_1(54321)$                              & 139 \\
$A_{1,2}$   & $A_{1,2}(12345)=-A_{1,2}(15432)$                      & 139 \\
$A_{1,3}$   & $A_{1,3}(12345)=-A_{1,3}(21543)$                      & 139 \\
$A_{1,5}$   & $A_{1,5}(12345)=-A_{1,5}(54321)$                      & 215 \\
$A_{1,6}$   & $A_{1,6}(12345)=-A_{1,6}(43215)$                      & 139 \\
$A_{1,7}$   & $A_{1,7}(12345)=-A_{1,7}(32154)$                      & 139 \\
$A_{1,1,6}$ & $A_{1,1,6}(12345)=A_{1,1,6}(52341)$                   & 139 \\
$A_{1,2,5}$ &                                                       & 214 \\
$A_{1,2,6}$ &                                                       & 139 \\
$A_{1,2,7}$ & $A_{1,2,7}(12345)=-A_{1,2,7}(15432)=A_{1,2,7}(12435)$ & 139 \\
$A_{1,3,5}$ &                                                       & 139 \\
$A_{1,3,6}$ &                                                       & 215 \\
\hline
\end{tabular}
\caption{\small
  Remaining degrees of freedom and known symmetries in the 13 independent functions specifying our solution for $X$.  Complete expressions with the extra degrees of freedom set to zero may be found in the ancillary file attached to the \texttt{arXiv} submission of ref~\cite{Mogull:2015adi}.}
\label{tab:Adof}
\end{table*}

We found it possible to further constrain the system by setting $A_{1,3,3}=A_{1,3,10}=A_{1,7,7}=0$,
as well as extracting an overall factor of $s_{12}^{-1}$ in $A_{1,3}(12345)$:
this left 215 undetermined parameters.
The final solution is completely specified by 13 nonzero functions $A_{i_1\ldots i_k}$,
given with their known symmetry properties and remaining degrees of freedom in table \ref{tab:Adof}.
The full solution for $X$ may be found in the ancillary file attached to the \texttt{arXiv} submission of ref~\cite{Mogull:2015adi}; in this file the remaining degrees of freedom $a_{i_1\ldots i_k;j}$ have been set to zero.

\section{Discussion}\label{sec:bcjdiscussion}

In contrast to previous chapters where we managed to cast the two-loop five-point amplitude in a compact form, in this chapter we have shown that the amplitude is sufficiently complex that the problem of constructing a set of BCJ numerators is non-trivial. The numerators we found contain more powers of loop momentum than one would expect on the basis of the Feynman rules. They also contain spurious singularities in kinematic invariants.

Of course, it is always possible to add more loop-momentum dependence and additional spurious singularities to numerators of diagrams. The physical requirement is that any cut must take on its physical value. There is a large space of numerators which satisfy this condition. We exploited the freedom of adding more terms in order to build a numerator which satisfies the requirements of colour-kinematics duality. Given the generous freedom available, it is tempting to speculate that one can always find colour-dual numerators this way.

In this chapter, we chose to deal with the obstructions we encountered to the existence of colour-dual numerators with more traditional power counting by adding higher powers of loop momenta. Introducing large amounts of loop momenta in numerators at first seems like a bad idea, because the size of numerator ans\"atze grow quickly as more loop momentum dependence is allowed. We maintained control of our ansatz by imposing a powerful symmetry on the unknown function $X$. This symmetry allowed us to consider numerators with much more loop momentum than is typically possible. Finding a deeper understanding of the importance of this symmetry may help to find similar symmetry requirements in other cases.

It is interesting that the two-loop connection between all-plus amplitudes and MHV amplitudes in $\mathcal N = 4$ SYM survives even in the context of colour-kinematics duality.  Whereas previously we thought the butterfly topologies had no connection with $\cN=4$, now we see that terms proportional to $F_2$ can be extracted from the Jacobi identities.  It is only the $F_3$ terms, carrying $(D_s-2)^2$, that are genuinely new contributions.

Finally, a family of colour-dual MHV numerators in the maximally supersymmetric theory at one loop was recently discovered~\cite{He:2015wgf}, built on our understanding of colour-kinematics duality in the self-dual theory~\cite{Monteiro:2011pc, Boels:2013bi}. 
Even though two-loop all-plus amplitudes are not self dual, it may be that insight into the nature of the residual link at two loops will allow for progress on one side to be recycled into the other.

\chapter{Conclusions}\label{ch:conclusions}

This thesis began with a simple question: what is the best way of writing two-loop scattering amplitudes before integration?  Indeed, what does it mean for a presentation to be the ``best''?  The answer, of course, is that it depends on which physical properties of the amplitude one is most interested in.  We focused on the all-plus sector of pure Yang-Mills theory, giving three different ways of expressing the integrands for various numbers of scattered gluons.  Each was designed to emphasise different physical properties of the amplitudes without the need for explicit integration.

The five-gluon integrand in particular was presented differently in each of the three main chapters.  While BFZ's original presentation~\cite{Badger:2013gxa} was designed with ease of integration in mind, the local-integrand presentation of chapter~\ref{ch:localintegrands}, benefitting from being the most compact, emphasised the amplitude's physical poles and IR structure.  The multi-peripheral colour decomposition of chapter~\ref{ch:allplusfull} gave us subleading colour structures, emphasising the connection to generalised unitarity cuts through Kleiss-Kuijf relations and off-shell symmetries.  Finally, the colour-dual presentation of chapter~\ref{ch:allplusbcj}, beyond its applications in future work on the double copy, emphasised the all-plus sector's continuing connection with $\cN=4$ SYM at two loops.

If we could combine elements of the three presentations, might we see more of these physical properties at play together?  The two most obvious candidates --- which we have already hinted could be combined --- are the local-integrand and multi-peripheral presentations.  For each of these we checked Catani's IR decomposition (\ref{eq:catani}), but with limitations.  The local-integrand presentation is planar, so our check was limited to the large-$N_c$ limit.  In the multi-peripheral presentation we lacked analytic expressions for the integrals' subleading-$\eps$ poles, so we could only check the leading $\eps^{-2}$ terms.  A combined presentation would have neither of these limitations: we would be able to see, like we did at four points in section~\ref{sec:4ptirexample}, full five- (or higher-) point IR structure analytically.

The challenge here would be to find nonplanar analogues of the local-integrand-based planar numerators given in section~\ref{sec:twoloops}.  Although the original $\cN=4$ supersymmetric all-loop local integrand~\cite{ArkaniHamed:2010gh,ArkaniHamed:2010kv} was limited to planar amplitudes, recent nonplanar work by Bern, Herrmann, Litsey, Stankowicz and Trnka~\cite{Bern:2015ple} may provide a good starting point.  The main requirement on the new nonplanar numerators would be that, as with the planars, their subleading-IR behaviour should be predictable without the need for explicit integration.  Checking this is difficult: as we explained in section~\ref{sec:colourdressedsoft}, nonplanar integrals are more computationally expensive to evaluate numerically as Euclidean phase-space points cannot be chosen.

A less obvious pair of candidates are the local-integrand and colour-dual presentations.  Our ability to eliminate spurious singularities from the planar two-loop five-gluon integrand in chapter~\ref{ch:localintegrands} begs the question whether a colour-dual representation free of spurious singularities also exists.  While we ruled out this possibility in section~\ref{sec:nonlocal}, the argument presented there assumed our use of Carrasco and Johansson's $\gamma_{ij}$ monomials~(\ref{eq:cjgammas}).  If we used a different monomial basis with, say, explicit Parke-Taylor functions multiplying the numerators, it might be possible to reproduce the desired loop-momentum dependence without the need for unphysical $s_{ij}$ or $\trfive$ poles.  Local integrands could be an interesting starting point.

It is also worth mentioning that we have we have not exhausted the possible ways of writing all-plus integrands.  For instance, the Q-cut representation of loop-level integrands~\cite{Baadsgaard:2015twa} shares elements of both traditional unitarity cuts and Britto, Cachazo, Feng and Witten's (BCFW's) recursive procedure~\cite{Britto:2005fq}.  The representation, which has already been applied to a variety of one-loop helicity amplitudes~\cite{Huang:2015cwh}, is related to Cachazo, He and Yuan's (CHY's) arbitrary-dimensional tree-level gluon and graviton scattering amplitudes~\cite{Cachazo:2013hca}.  This connection has been made evident by loop-level extensions of CHY's formulae found using ambitwistor string theories \cite{Geyer:2015jch,Geyer:2016wjx}; the resulting expressions match those generated using Q-cuts at one-loop order.  It could be interesting to reconsider the amplitudes discussed in this thesis using Q-cuts.

Our hope in doing this work was that the all-plus sector could be used as a bridge, helping to connect new developments in supersymmetric Yang-Mills theory with ongoing phenomenological studies of pure Yang-Mills amplitudes.  While the multi-peripheral colour decomposition is already applicable to more general cases, the local-integrand picture needs further development if it is to be used as a general framework.  The colour-dual picture, also, needs further simplification.  Nevertheless, we hope that this thesis will provide valuable insight into the options available for simplifying two-loop Yang-Mills amplitudes.

\renewcommand{\chaptername}[1]{Appendix A.}
\renewcommand{\chaptermark}[1]{\markboth{\chaptername \ #1}{}}

\appendix 

\chapter{\texorpdfstring{$D$}{D}-dimensional Dirac traces}\label{app:traces}

Writing down Dirac traces containing $D$-dimensional loop momenta $\ell_i$ necessitates the use of a $D$-dimensional Clifford algebra.  Our conventions follow those of 't Hooft and Veltman \cite{Collins:1984xc,tHooft:1972tcz}, which are further elaborated by Bern and Morgan~\cite{Bern:1995db}.  The $D$-dimensional Clifford algebra is formed by matrices $\gamma^\alpha$ satisfying
\begin{align}\label{eq:DDirac}
\{\gamma^\alpha,\gamma^\beta\}=2\eta^{\alpha\beta},
\end{align}
where the $D$-dimensional metric is $\eta^{\alpha\beta}=\text{diag}(+,-,-,-,-,\ldots)$.  There is a four-dimensional subalgebra formed by the first four matrices $\gamma^\mu$ with its own $\gamma_5$ matrix $\gamma_5=i\gamma^0\gamma^1\gamma^2\gamma^3$.  This matrix anticommutes with the four-dimensional gamma matrices as usual, but commutes with the rest:
\begin{align}
\{\gamma_5,\gamma^\mu\}=0,&&
[\gamma_5,\gamma^\alpha]=0\qquad\alpha>3.
\end{align}

We decompose $D$-dimensional Dirac traces into four-dimensional ones as follows.  Recalling that $D$-dimensional loop momenta can be expressed as $\ell_i=\bar{\ell}_i+\mu_i$,
\begin{align}
&\tr(i_1\cdots i_k\ell_x\ell_yi_{k+1}\cdots i_n)\\
&\qquad=
\tr(i_1\cdots i_k\bar{\ell}_x\bar{\ell}_yi_{k+1}\cdots i_n)+
\tr(i_1\cdots i_k\bar{\ell}_x\mu_yi_{k+1}\cdots i_n)\nn\\
&\qquad\qquad+
\tr(i_1\cdots i_k\mu_x\bar{\ell}_yi_{k+1}\cdots i_n)+
\tr(i_1\cdots i_k\mu_x\mu_yi_{k+1}\cdots i_n),\nn
\end{align}
where $n$ is even.  The external momenta $p_i$ live entirely in the four-dimensional subspace, so in the second two terms the extra-dimensional loop momenta $\mu_x$ and $\mu_y$ can be freely anticommuted around the traces.  They pick up an overall minus sign, so must both be zero.  As for the last term, anticommuting $\mu_x$ around the trace gives
\begin{align} 
\tr(i_1\cdots i_k\mu_x\mu_yi_{k+1}\cdots i_n)
&=\tr(i_1\cdots i_k\mu_y\mu_xi_{k+1}\cdots i_n)\nn\\
&=\frac{1}{2}\tr(i_1\cdots i_k\{\mu_x,\mu_y\}i_{k+1}\cdots i_n)
\nn\\
&=-\mu_{xy}\tr(i_1\cdots i_n)
\end{align}
where to get the last line we used eq.~(\ref{eq:DDirac}).  We are left with
\begin{align}
\tr(i_1\cdots i_k\ell_x\ell_yi_{k+1}\cdots i_n)=
\tr(i_1\cdots i_k\bar{\ell}_x\bar{\ell}_yi_{k+1}\cdots i_n)
-\mu_{xy}\tr(i_1\cdots i_n).
\end{align}
As $\gamma_5$ commutes with extra-dimensional gamma matrices, this argument applies equally well to $\trfive$.  So,
\begin{align}\label{eq:apptrdecomp}
\text{tr}_\pm(i_1\cdots i_k\ell_x\ell_yi_{k+1}\cdots i_n)
=\text{tr}_\pm(i_1\cdots i_k\bar{\ell}_x\bar{\ell}_yi_{k+1}\cdots i_n)
-\mu_{xy}\text{tr}_\pm(i_1\cdots i_n).
\end{align}

We can also establish that, when $k$ is odd,
\begin{align}
&\trp(i_1\cdots i_k\bar{\ell}_x\bar{\ell}_yi_{k+1}\cdots i_n)
\nn\\&\qquad=
[i_1,i_2]\braket{i_2i_3}\cdots\braket{i_{k-1}i_k}[i_k|\bar{\ell}_x\bar{\ell}_y|i_{k+1}]
\braket{i_{k+1}i_{k+2}}\cdots\braket{i_{n}i_1}
\nn\\&\qquad=
\frac{[i_1,i_2]\braket{i_2i_3}\cdots\braket{i_{n}i_1}}{[i_{k}i_{k+1}]\braket{i_{k+1}i_k}}
[i_k|\bar{\ell}_x\bar{\ell}_y|i_{k+1}]\braket{i_{k+1}i_k}
\nn\\&\qquad=
\frac{\trp(i_1\cdots i_n)}{s_{i,i+1}}\trp(i_k\bar{\ell}_x\bar{\ell}_yi_{k+1}),
\end{align}
and similarly for $\trm$.  When combined with (\ref{eq:apptrdecomp}) this gives
\begin{align}\label{eq:apptrstacking}
\text{tr}_\pm(i_1\cdots i_k\ell_x\ell_yi_{k+1}\cdots i_n)
=\frac{\text{tr}_\pm(i_1\cdots i_n)}{s_{i_k,i_{k+1}}}\text{tr}_\pm(i_k\ell_x\ell_yi_{k+1}),
\end{align}
as expected.  So it works equally well for $D$-dimensional loop momenta.

\chapter{Local integrands in \texorpdfstring{$\cN=4$}{N=4} SYM}\label{sec:N4localintegrands}

In this appendix we show how momentum-twistor-based expressions for
planar $\mathcal{N}=4$ local integrands up to two loops,
given in ref.~\cite{ArkaniHamed:2010kv}, may be converted to a form applicable in $D\neq4$.
For instance,
\begin{subequations}\label{eq:arkaniexamples}
\begin{align}
\usegraph{13}{pent23s}
&=\int\displaylimits_{AB}\frac{\braket{1456}\braket{AB|(612)\cap(345)}}
{\braket{AB12}\braket{AB34}\braket{AB45}\braket{AB56}\braket{AB61}},\\
\usegraph{10}{441s}
&=\!\!\!\!\int\displaylimits_{(AB,CD)}\!\!\!\!
\frac{\braket{1346}\braket{AB|(612)\cap(234)}\braket{CD|(345)\cap(561)}}
{\braket{AB61}\cdots\braket{AB34}\braket{CD34}\cdots\braket{CD61}\braket{ABCD}}.
\end{align}
\end{subequations}

Momentum twistors are defined with respect to dual-space coordinates $x_i$,
themselves introduced using $p_i=x_i-x_{i-1}$.
A dual-space point $x$ is identified with a line of twistors, $Z=(\lambda,\mu)$,
satisfying the incidence relation $\mu_{\dot{a}}=\lambda^a x_{a\dot{a}}$.
These twistors form a projective line in $\mathbb{C}\mathbb{P}^3$.
To identify a point $x$ in dual space
it therefore suffices to specify two momentum twistors, $Z_i$ and $Z_j$:
\begin{align}\label{eq:dualspace}
x_{a\dot{a}}=\frac{\lambda_{i,a}\mu_{j,\dot{a}}-\lambda_{j,a}\mu_{i,\dot{a}}}{\braket{ij}}
=x_{i,a\dot{a}}+\frac{\lambda_{i,a}\lambda_j^b(p_{i+1,\ldots,j})_{b\dot{a}}}{\braket{ij}},
\end{align}
where $\braket{ij}\equiv\epsilon_{ab}\lambda_i^a\lambda_j^b$.
The latter identity, taken together with the massless Weyl equation $\lambda_i^ap_{i,a\dot{a}}=0$,
implies that when $j=i+1$, $x=x_i$.

As for the loop momenta, for $n$-point scattering at one loop we introduce
$y=\ell+x_n$ associated with the line in $(Z_A,Z_B)$.
At two loops we introduce $y_1=\ell_1+x_n$ and $y_2=-\ell_2+x_n$
associated with the lines $(Z_A,Z_B)$ and $(Z_C,Z_D)$ respectively.\footnote{For a complete introduction see ref.~\cite{ArkaniHamed:2010gh}.}

The basic building block to evaluate is the twistor four-bracket,
\begin{align}\label{eq:twistor4bracket}
\braket{i,j,k,l}\equiv\epsilon_{IJKL}Z_i^IZ_j^JZ_k^KZ_l^L
=\braket{ij}\braket{kl}(x-y)^2,
\end{align}
where $x$ and $y$ are associated with the lines $(Z_i,Z_j)$ and $(Z_k,Z_l)$ respectively.
Using the dual-space definition (\ref{eq:dualspace}) it follows that
\begin{subequations}\label{eq:4bracketexamples}
\begin{align}
\braket{ijkl}&=\braket{ij}\braket{kl}
\left(p_{i+1,\ldots,k}^\mu-\frac{\braket{i|\gamma^\mu p_{i+1,\ldots,j}|j}}{2\braket{ij}}
+\frac{\braket{k|\gamma^\mu p_{k+1,\ldots,l}|l}}{2\braket{kl}}\right)^2,\label{eq:4bracketgeneral}\\
\braket{ABij}&=\braket{AB}\braket{ij}
\left(\ell_1^\mu-p_{1,\ldots,i}^\mu-\frac{\braket{i|\gamma^\mu p_{i+1,\ldots,j}|j}}{2\braket{ij}}\right)^2,
\label{eq:4bracketl1}\\
\braket{CDij}&=\braket{CD}\braket{ij}
\left(\ell_2^\mu+p_{1,\ldots,i}^\mu+\frac{\braket{i|\gamma^\mu p_{i+1,\ldots,j}|j}}{2\braket{ij}}\right)^2,
\label{eq:4bracketl2}\\
\braket{ABCD}&=\braket{AB}\braket{CD}(\ell_1+\ell_2)^2.\label{eq:4bracketl1l2}
\end{align}
\end{subequations}
where $i<j<k<l$ (this can always be ensured using antisymmetry of the 4-bracket).
Some frequently-encountered examples are
\begin{subequations}\label{eq:4bracketeasycases}
\begin{align}
\braket{i,i+1,j,j+1}
&=\braket{i,i+1}\braket{j,j+1}s_{i+1,\ldots,j},\label{eq:4bracket1}\\
\braket{A,B,i,i+1}
&=\braket{AB}\braket{i,i+1}(\ell_1-p_{1,\ldots,i})^2,\label{eq:4bracket2}\\
\braket{C,D,i,i+1}
&=\braket{CD}\braket{i,i+1}(\ell_2+p_{1,\ldots,i})^2,\label{eq:4bracket3}
\end{align}
\end{subequations}
where the massless Weyl equation is again used to make these simplifications.

When evaluating four-brackets involving intersections of planes in momentum-twistor space
one should use the twistor intersection formula,
\begin{align}\label{eq:twistorintersection}
\braket{AB|(abc)\cap(def)}
=\braket{cdef}\braket{ABab}+\braket{adef}\braket{ABbc}+\braket{bdef}\braket{ABca}.
\end{align}
From this we have established the general pattern that
\begin{align}
\usegraph{13}{pentschemes}
&\sim\braket{AB|(i-1,i,i+1)\cap(j-1,j,j+1)}\nonumber\\
&\qquad
=-\braket{AB}\braket{i-1,i}\braket{i,i+1}\braket{j-1,j}\braket{j,j+1}\left[i|\ell_x\ell_y|j\right],
\end{align}
which is manifestly local.
Application of $\trp(i\ell_x\ell_yj)=\left[i|\ell_x\ell_y|j\right]\braket{ji}$ gives
agreement with the integral definition we made in eq.~(\ref{eq:squiggledef}).
For instance, in the two examples given in eqs.~(\ref{eq:arkaniexamples}),
\begin{subequations}
\begin{align}
\braket{AB|(612)\cap(345)}
&=\frac{\braket{AB}\braket{61}\braket{12}\braket{34}\braket{45}}{\braket{14}}
\trp(1(\ell-p_1)(\ell-p_{123})4),\\
\braket{AB|(612)\cap(234)}
&=\frac{\braket{AB}\braket{61}\braket{12}\braket{23}\braket{34}}{\braket{13}}
\trp(1(\ell_1-p_1)(\ell_1-p_{12})3),\\
\braket{CD|(345)\cap(561)}
&=\frac{\braket{CD}\braket{34}\braket{45}\braket{56}\braket{61}}{\braket{46}}
\trp(4(\ell_2-p_{56})(\ell_2-p_6)6).
\end{align}
\end{subequations}

When calculating complete integrands in $D\neq4$ one should use the measure correspondences
\begin{align}\label{eq:twistormeasure}
\int\displaylimits_{AB}\frac{1}{\braket{AB}^4}\sim\int \d^4\ell,&&
\int\displaylimits_{(AB,CD)}\frac{1}{\braket{AB}^4\braket{CD}^4}
\sim\int \d^4\ell_1\d^4\ell_2,
\end{align}
at one and two loops repectively.
This accounts for all factors of $\braket{AB}$ and $\braket{CD}$.
An overall factor of the tree amplitude $\mathcal{A}^{(0),[\mathcal{N}=4]}_\text{MHV}$ then contributes both a Parke-Taylor denominator
$\braket{12}\cdots\braket{n-1,n}\braket{n1}$ and the supersymmetric delta function $\delta^8(Q)$.

\chapter{Two-loop rational integrals}\label{app:2lintegrals}

This appendix contains all the integrals required to compute the
rational parts of the five- and six-point all-plus amplitudes given in chapter~\ref{ch:localintegrands}.  These expressions are also available in an ancillary file included with the \texttt{arXiv} submission of ref.~\cite{Badger:2016ozq}.  For the sake of compactness, in this Appendix only we switch to the two-loop integral convention:
\begin{align}
I^D_T\left[\mathcal{P}(p_i,\ell_i,\mu_{ij})\right]
\equiv-(4\pi)^D\int\!\frac{\d^D\ell_1\d^D\ell_2}{(2\pi)^{2D}}
\frac{\mathcal{P}(p_i,\ell_i,\mu_{ij})}
{\prod_{\alpha\in T}\cQ_{\alpha}(p_i,\ell_1,\ell_2)},
\end{align}
where once again $\{\cQ_\alpha\}$ is the set of propagators of the topology $T$.  All integrals are taken in the limit $\eps\to0$.

\section{Five-point integrals}

\begingroup
\allowdisplaybreaks
\begin{subequations}
\begin{align}
I^D\bigg(\usegraph{13}{430s}\bigg)[\mu_{11}\mu_{22}]
\to{}&\frac{s_{13}}{4},\\
I^D\bigg(\usegraph{13}{430s}\bigg)[\mu_{11}\mu_{22}(\ell_1+\ell_2)^2]
\to{}&\frac{\trm(1345)}{36}\\
&\,-\frac{s_{13}(6s_{45}-2s_{13}-s_{34}-s_{15})}{36},\nn\\
I^D\bigg(\usegraph{13}{3305L}\bigg)[\mu_{11}\mu_{22}]
\to{}&\frac{1}{4},\\
I^D\bigg(\usegraph{13}{3305L}\bigg)[\mu_{11}\mu_{22}\text{tr}_+(123\ell_1\ell_2345)]
\to{}&\frac{\trp(123(2p_1\!+\!p_2)(p_4\!+\!2p_5)345)}{36},\\
I^D\bigg(\usegraph{13}{3305L}\bigg)[\mu_{11}\mu_{22}(\ell_1+\ell_2)^2]
\to{}&\frac{(2p_1\!+\!p_2)\cdot(p_4\!+\!2p_5)}{18}.
\end{align}
\end{subequations}
\endgroup

\section{Six-point integrals}

\begingroup
\allowdisplaybreaks
\begin{subequations}
\begin{align}
I^D\bigg(\!\usegraph{9}{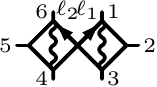}\!\bigg)[\mu_{11}\mu_{22}]
\to{}&\frac{s_{13}s_{46}}{4},\\
I^D\bigg(\!\usegraph{9}{440s}\!\bigg)[\mu_{11}\mu_{22}(\ell_1+\ell_2)^2]
\to{}&-\frac{s_{13}s_{46}(s_{23}+s_{45})}{12}\\&
\,+\frac{s_{13}s_{46}(2p_1\!+\!p_2\!+\!p_3)\!\cdot\!(p_4\!+\!p_5\!+\!2p_6)}{18}\nn\\&
\,+\frac{s_{46}\trp(12(p_4\!+\!p_5\!+\!2p_6)3)}{36}\nn\\&
\,+\frac{s_{13}\trp(4(2p_1\!+\!p_2\!+\!p_3)56)}{36}\nn\\&
\,+\frac{\trp(125643)}{36},\nn\\
I^D\bigg(\usegraph{9}{530}\!\bigg)[\mu_{11}^2\mu_{22}]
\to{}&0,\\
I^D\bigg(\usegraph{9}{530}\!\bigg)[\mu_{11}^2\mu_{22}(\ell_1+\ell_2)^2]
\to{}&-\frac{1}{12},\\
I^D\bigg(\usegraph{9}{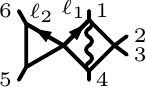}\!\bigg)[\mu_{11}\mu_{22}]
\to{}&\frac{s_{14}}{4},\\
I^D\bigg(\usegraph{9}{430M2s}\!\bigg)[\mu_{11}\mu_{22}(\ell_1+\ell_2)^2]
\to{}&\frac{\trm(1456)}{36}\\&
\,-\frac{s_{14}(5s_{23}\!+\!6s_{56}\!-\!2s_{14}\!-\!s_{45}\!-\!s_{16})}{36},\nn\\
I^D\bigg(\usegraph{9}{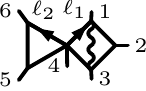}\!\bigg)[\mu_{11}\mu_{22}]
\to{}&\frac{s_{13}}{4},\\
I^D\bigg(\usegraph{9}{4306Ls}\!\bigg)[\mu_{11}\mu_{22}(\ell_1+\ell_2)^2]
\to{}&-\frac{s_{13}s_{23}}{12}+\frac{\trp(12(p_5\!+\!2p_6)3)}{36}\nn\\&
\,+\frac{s_{13}(2p_1\!+\!p_2\!+\!p_3)\cdot(p_5\!+\!2p_6)}{18},\\
I^D\bigg(\usegraph{9}{4306Ls}\!\bigg)[\mu_{11}\mu_{22}\ell_1^\mu\ell_2^\nu]
\to{}&\frac{2s_{13}(2p_1\!+\!p_2\!+\!p_3)^\mu(p_5\!+\!2p_6)^\nu}{72}\nn\\&
\,+\frac{\trp(12\gamma^\mu3)(p_5\!+\!2p_6)^\nu}{72},\\
I^D\bigg(\usegraph{9}{3305LM}\!\bigg)[\mu_{11}\mu_{22}]
\to{}&\frac{1}{4},\\
I^D\bigg(\usegraph{9}{3305LM}\!\bigg)[\mu_{11}\mu_{22}(\ell_1+\ell_2)^2]
\to{}&\frac{(2p_1\!+\!p_2)\cdot(p_5\!+\!2p_6)}{18},\\
I^D\bigg(\usegraph{9}{3305LM}\!\bigg)[\mu_{11}\mu_{22}\ell_1^\mu\ell_2^\nu]
\to{}&\frac{(2p_1\!+\!p_2)^\mu(p_5\!+\!2p_6)^\nu}{36},\\
I^D\bigg(\usegraph{13}{3306L}\!\bigg)[\mu_{11}\mu_{22}]
\to{}&\frac{1}{4},\\
I^D\bigg(\usegraph{13}{3306L}\!\bigg)[\mu_{11}\mu_{22}(\ell_1+\ell_2-p_1-p_5)^2]
\to{}&\frac{(p_1\!-\!p_2)\cdot(p_5\!-\!p_4)}{18}-\frac{s_{12}+s_{45}}{12},\\
I^D\bigg(\usegraph{13}{3306L}\!\bigg)[\mu_{11}\mu_{22}(\ell_1-p_1)^\mu]
\to{}&\frac{p_2^\mu-p_1^\mu}{12},\\
I^D\bigg(\usegraph{13}{3306L}\!\bigg)[\mu_{11}\mu_{22}(\ell_2-p_5)^\mu]
\to{}&\frac{p_4^\mu-p_5^\mu}{12},\\
I^D\bigg(\usegraph{13}{3306L}\!\bigg)[\mu_{11}\mu_{22}(\ell_1-p_1)^\mu(\ell_2-p_5)^\nu]
\to{}&\frac{(p_1^\mu-p_2^\mu)(p_5^\nu-p_4^\nu)}{36},\\
I^D\bigg(\usegraph{13}{3306L}\!\bigg)[\mu_{11}\mu_{22}(\ell_1-p_1)^2]
\to{}&-\frac{s_{12}}{12},\\
I^D\bigg(\usegraph{13}{3306L}\!\bigg)[\mu_{11}\mu_{22}(\ell_2-p_5)^2]
\to{}&-\frac{s_{45}}{12}.
\end{align}
\end{subequations}
\endgroup

\chapter{Five-point two-loop soft divergences}\label{app:2lsoftintegrals}

The following integrals are required for the check on soft divergences in chapter~\ref{ch:allplusfull}.  They have all been checked numerically using the sector decomposition methods implemented in \textsc{FIESTA}~\cite{Smirnov:2013eza} and \textsc{SecDec}~\cite{Borowka:2015mxa}.  Some of the integrals were computed long ago in $D$ dimensions and can be used to write the full integrals including finite terms via the dimensional reduction identities implemented in \textsc{LiteRed}~\cite{Lee:2012cn} and IBP relations from \textsc{FIRE5}~\cite{Smirnov:2014hma}.\footnote{We thank Claude Duhr for providing his own computation of the integrals for $e^+e^-\to3j$ \cite{Gehrmann:2000zt,Gehrmann:2001ck}.}

\begingroup
\allowdisplaybreaks
\begin{align}
I^D\bigg(\usegraph{9}{431i}\!\!\!\:\bigg)[F_1]&=
\frac{1}{(4\pi)^4} \frac{D_s-2}{3 s_{12}s_{23}\eps^2}+\cO(\eps^{-1}),\\
I^D\bigg(\usegraph{9}{431}\!\!\!\:\bigg)[F_1\,(\ell_1\!\cdot\!p_5)]&=
\frac{1}{(4\pi)^4}\frac{(D_s-2)(2 s_{15}+s_{25})}{12 s_{12}s_{23}\eps^2}
+\cO(\eps^{-1}),\\
I^D \bigg(\usegraph{9}{332i}\bigg)[F_1]&=
\cO(\eps^{-1}),\\
I^D \bigg(\usegraph{9}{332}\bigg)[F_1\,(\ell_1\!\cdot\!(p_5-p_4))]&=
\cO(\eps^{-1}),\\
I^D \bigg(\usegraph{9}{332}\bigg)[F_1\,((\ell_1-\ell_2)\!\cdot\!p_3)]&=
\cO(\eps^{-1}),\\
I^D \bigg(\!\!\!\:\usedelta{422i}\!\!\!\:\bigg)[F_1]&=
\frac{1}{(4\pi)^4} \frac{D_s-2}{3s_{12}s_{23}\eps^2} + \cO(\eps^{-1}),\\
I^D\bigg(\!\!\!\:\usegraph{9}{422}\!\!\!\:\bigg)[F_1 (\ell_1\!\cdot\!(p_5-p_4))]&=
\frac{1}{(4\pi)^4} \frac{(D_s-2)(s_{15}-s_{14}+s_{34}-s_{35})}{12s_{12}s_{23}\eps^2}\nn\\
&\qquad+\cO(\eps^{-1}),\\
I^D \bigg(\usegraph{13}{3315Li}\bigg)[F_1]&=
\frac{1}{(4\pi)^4} \frac{D_s-2}{6\eps^2}\left(\frac{1}{s_{12}}+\frac{1}{s_{45}}\right)
+\cO(\eps^{-1}),\\
I^D \bigg(\!\!\!\;\usegraph{16}{3225Li}\bigg)[F_1]&=
\frac{1}{(4\pi)^4} \frac{D_s-2}{6s_{12}\eps^2}+ \cO(\eps^{-1}),\\
I^D \bigg(\usegraph{9}{331M1i}\bigg)[F_1]&=
\frac{1}{(4\pi)^4} \frac{D_s-2}{6s_{45}\eps^2} + \cO(\eps^{-1}) ,\\
I^D\bigg(\usegraph{9}{232i}\!\bigg)[F_1]&=
\frac{1}{(4\pi)^4} \frac{D_s-2}{6s_{45}\eps^2} + \cO(\eps^{-1}) ,\\
I^D \bigg(\!\!\!\:\usegraph{8.8}{322M1i}\bigg) [F_1]&=
\cO(\eps^{-1}) .
\end{align}
\endgroup
\small{
\singlespacing

\addcontentsline{toc}{chapter}{Bibliography}
\bibliographystyle{JHEP}
\bibliography{refs}


}

\end{document}